\documentclass[preprint,print,5p,times,10pt]{elsarticle}

\usepackage{graphicx,graphics,epsfig}
\usepackage{amsmath,amsfonts,amssymb}
\usepackage{siunitx}
\graphicspath{{}}
\usepackage{algorithm}
\usepackage{algpseudocode}
\usepackage{url}
\usepackage{xcolor}
\usepackage{listings}
\usepackage{multirow}
\usepackage{booktabs}
\usepackage{hyperref}
\usepackage{microtype}
\definecolor{codegreen}{rgb}{0,0.6,0}
\definecolor{codegray}{rgb}{0.5,0.5,0.5}
\definecolor{backcolour}{rgb}{0.95,0.95,0.92}
\lstdefinestyle{mystyle}{
    language=Matlab,
    backgroundcolor=\color{backcolour},
    commentstyle=\color{codegreen},
    keywordstyle=\color{magenta},
    numberstyle=\tiny\color{codegray},
    basicstyle=\ttfamily\footnotesize,
    breakatwhitespace=false,
    breaklines=true,
    captionpos=b,
    keepspaces=true,
    numbers=left,
    numbersep=5pt,
    showspaces=false,
    showstringspaces=false,
    showtabs=false,
    tabsize=2,
}
\biboptions{comma}

\journal{Computer Physics Communications}

\begin{document}

\begin{frontmatter}

\title{DeHNSSo-v2: A compressible, curvilinear, and efficient extension of the Delft Harmonic Navier--Stokes Solver\tnoteref{t1}}
\tnotetext[t1]{Source code open and available under the terms of the GNU General Public License v3 on \url{https://github.com/pcboldini/DeHNSSo}.}

\author[inst1]{Pietro Carlo Boldini\corref{author}}
\affiliation[inst1]{organization={Department of Flow Physics and Technology, Delft University of Technology},
            addressline={Kluyverweg 1},
            city={Delft},
            postcode={2629 HS},
            country={The Netherlands}}

\author[inst1]{Yifu Chen}
\author[inst1]{Marios Kotsonis}
\cortext[author]{Corresponding author.\\\textit{E-mail address:} pietro.c.boldini@gmail.com (P.~C.~Boldini)}

\begin{abstract}
We present DeHNSSo-v2, a major update of DeHNSSo-v1 (Westerbeek et al., Comput.~Phys.~Commun.\ 302, 2024), the open-source Delft Harmonic Navier--Stokes Solver for nonlinear stability analysis of shear flows. The new release adds a compressible formulation on general two-dimensional body-fitted geometries, automatic grid generation, linear-cost sparse assembly, reusable per-mode factorizations, and Newton pseudo-arclength continuation for high-amplitude nonlinear solutions. The curvilinear formulation preserves the fourth-order accuracy of the discretization on strongly distorted grids, and the compressible solver is confirmed to reproduce DeHNSSo-v1 in the incompressible limit. Validation spans published benchmarks for linear instability of compressible flat-plate boundary layers from $\mathrm{Ma}=0.5$ to $\mathrm{Ma}=4.5$, nonlinear oblique-mode breakdown at $\mathrm{Ma}=1.6$, nonlinear swept-wing crossflow compared with direct numerical simulations, and linear Tollmien--Schlichting wave interaction with a modeled surface-embedded Helmholtz resonator compared with experimentally validated linearized Navier--Stokes simulations. On the nonlinear test cases of DeHNSSo-v1, the computational developments result in a speed-up of approximately one order of magnitude. By broadening the scope and reducing the cost of DeHNSSo-v1, DeHNSSo-v2 provides the boundary-layer-transition and flow-control community with an accessible tool for studying receptivity, nonlinear disturbance evolution, and instability interactions with complex surface geometry and metamaterials across compressibility regimes. \\

\noindent{\bf New version program summary} \\
\begin{small}
\noindent
{\em Program Title:} Delft Harmonic Navier--Stokes Solver (DeHNSSo)\\
{\em CPC Library link to program files:}  \\
{\em Developer's repository link:} \url{https://github.com/pcboldini/DeHNSSo} \\
{\em Licensing provisions:} GPLv3 \\
{\em Programming language:} MATLAB, C \\
{\em Journal Reference of previous version:} Westerbeek S., Hulshoff S., Schuttelaars H., Kotsonis M., DeHNSSo: The Delft Harmonic Navier--Stokes Solver for nonlinear stability problems with complex geometric features, Computer Physics Communications 302 (2024) 109250. \\
{\em Does the new version supersede the previous version?:} Yes \\
{\em Reasons for the new version:} Release of new features, extension to compressible regimes, and performance improvements. \\
{\em Summary of revisions:}\\
\noindent$\bullet$~general curvilinear perturbation formulation and automatic body-fitted stability-grid generation;\\
\noindent$\bullet$~nonuniform streamwise grids with fourth-order finite differences;\\
\noindent$\bullet$~compressible primitive-variable formulation with thermal and far-field boundary conditions, freestream forcing, and parabolized outflow treatment;\\
\noindent$\bullet$~linear-cost sparse assembly and reduced operator storage;\\
\noindent$\bullet$~reusable per-mode factorizations with sparse LU, banded LU, and out-of-core block-Thomas back-ends;\\
\noindent$\bullet$~Newton pseudo-arclength continuation in the inflow disturbance amplitude.\\
{\em Nature of problem:} DeHNSSo-v2 computes the linear and nonlinear spatial evolution of temporally and spanwise periodic disturbances in incompressible and compressible laminar boundary layers over general two-dimensional body-fitted geometries. It is designed for studies of receptivity, nonlinear modal interactions, transition, and interactions with localized surface features or flow-control devices. \\
{\em Solution method:} Perturbations to a steady base flow are expanded in temporal and spanwise Fourier modes and discretized using streamwise finite differences and wall-normal finite differences or Chebyshev collocation. The resulting sparse systems, which couple the whole two-dimensional domain, are solved by direct factorization, with nonlinear coupling treated by fixed-point iteration or Newton pseudo-arclength continuation.
\end{small}

\end{abstract}


\end{frontmatter}

\section{Introduction}
\label{sec:intro}

Reducing drag on transport aircraft and limiting the aerodynamic heating of high-speed vehicles are central objectives of aerodynamic design. Both depend strongly on whether the boundary layer over the vehicle surface remains laminar or becomes turbulent: a turbulent boundary layer produces substantially greater skin friction and wall heat flux than a laminar one. Predicting the transition location therefore underlies two distinct design problems. On subsonic transport aircraft, skin friction accounts for a large fraction of the total drag, so delaying transition is among the most effective aerodynamic means of reducing drag \citep{Schrauf2005,Joslin1998}. At high speed, transition governs the heat loads that size the thermal-protection system and is often driven by the first and second Mack modes \citep{Fedorov2011}. Both regimes ultimately pose the same problem: accurately simulating how small disturbances are amplified within a laminar boundary layer and how that amplification is modified by wall geometry, thermal conditions, and Mach number, among other factors.

The subsonic swept wing illustrates the sensitivity of this process to surface geometry. At the low freestream-turbulence levels typical of cruise flight, transition is commonly driven by stationary crossflow instability \citep{Saric2003}. Laminar-flow-control strategies targeting this instability can be compromised by the geometric imperfections carried by real wings, including steps and gaps at panel junctions and the bumps and waviness produced by manufacturing and structural loading \citep{Holmes1985,Worner2003}. Improvements in manufacturing have reduced distributed surface waviness, increasing the relative importance of localized steps and gaps, which modify the incoming instability and may displace the transition front upstream \citep{Crouch2006,WangGaster2005,RiusVidales2021}. Conversely, a deliberately shaped smooth hump has recently been shown to delay swept-wing transition by approximately $14\%$ of the chord when encountered by a low-amplitude crossflow disturbance, but to advance transition when the incoming amplitude is larger \citep{RiusVidales2025Hump,Westerbeek2026Hump}. The effect of the same surface feature can thus change sign according to the amplitude of the incoming disturbance and hence the nonlinear state of the perturbation field. Predicting either response requires resolving the nonlinear interaction between the instability and the surface feature at the scale of the boundary layer. Closely related interactions underlie wall-based control devices, both active, such as suction and plasma actuators, and passive, such as Helmholtz resonators \citep{Michelis2023} and other surface metamaterials \citep{Avallone2026}. Through localized forcing, these devices can damp or reinforce selected instability components.

Several numerical frameworks are available for predicting disturbance evolution, as reviewed by \citet{Westerbeek2024DeHNSSo}. Local stability theory treats the flow at successive streamwise stations and does not retain the global streamwise coupling of the disturbance field. The parabolized stability equations (PSE) \citep{Bertolotti1992,Herbert1997}, including their extensions to compressible boundary layers and supersonic oblique breakdown \citep{BertolottiHerbert1991,ChangMalik1991,ChangMalik1994}, retain nonparallel development while neglecting the second-order streamwise derivatives of a slowly varying disturbance shape function. The regularization required to obtain a stable downstream march accurately retains a single slowly varying spatial component at each temporal frequency, but may distort other instability mechanisms, nonmodal responses, and acoustic components \citep{Towne2019}. The slowly varying approximation itself may lose validity where localized geometry or forcing produces rapid streamwise variation, as occurs near steps and humps \citep{Cooke2019,Casacuberta2022}, during the scattering of high-speed boundary-layer modes by wall roughness \citep{Zhao2019}, and near surface indentations, which have been treated with adaptive harmonic linearized formulations \citep{Franco2018,FrancoSumariva2020}.

The harmonic Navier--Stokes (HNS) framework \citep{Streett1998,Paredes2019} avoids the parabolizing approximation by retaining all streamwise derivatives and solving the perturbation equations as a global boundary-value problem. It can therefore represent upstream and downstream influence and the elliptic interaction between a disturbance and a localized surface feature. HNS exploits the periodic structure of the disturbance by expanding it into a finite set of temporal and spanwise Fourier harmonics. Its computational cost exceeds that of a parabolized marched method but can remain substantially below that of direct numerical simulation when the disturbance spectrum is represented accurately by a limited number of harmonics \citep{Westerbeek2024DeHNSSo}.

Recent HNS developments include both linear and nonlinear formulations. \citet{Paredes2024} solved the harmonic linearized Navier--Stokes (HLNS) equations about spanwise- or azimuthally varying base flows using a block-pentadiagonal direct factorization, while \citet{ZhaoDong2025} coupled HLNS to a fitted bow shock to study hypersonic receptivity to freestream vortical, entropy, and acoustic disturbances. Both formulations remain linear, as do the open-source global tools BROADCAST \citep{Poulain2023} and LinStab2D \citep{MartiniSchmidt2024}, which provide stability and resolvent analysis of compressible flows. \citet{Scholten2024} extended HNS to nonlinear compressible boundary-layer disturbances, closing the coupled harmonic system implicitly \citep{Rigas2021}. A complementary route is provided by the one-way Navier--Stokes equations \citep{TowneColonius2015}. Recursive and carrier-wave-factored formulations reduce the cost of resolving multiple downstream-traveling components in wall-bounded and compressible flows \citep{ZhuTowne2023,BadcockMughal2026}, and \citet{Sleeman2025} extended the approach to nonlinear harmonic interactions, reporting greater robustness than nonlinear PSE under strong forcing. One-way methods exclude upstream influence by construction; global HNS remains appropriate when such influence or the elliptic interaction with localized geometry is significant.

The Delft Harmonic Navier--Stokes Solver (DeHNSSo) was introduced by \citet{Westerbeek2024DeHNSSo}, hereafter DeHNSSo-v1, as an open-source MATLAB implementation of nonlinear HNS for spanwise-invariant laminar base flows and temporally and spanwise periodic disturbances. It discretizes the wall-normal direction by Chebyshev collocation and the streamwise direction by high-order finite differences, assembles the resulting global sparse systems, and closes the nonlinear interactions by harmonic balancing. The original release was validated for Tollmien--Schlichting (TS) waves, swept-wing crossflow instability, including the interaction of the latter with a smooth hump and a forward-facing step, all within a strictly incompressible framework. To the authors' knowledge, it remains the only openly available nonlinear HNS implementation for this class of boundary-layer problems. Open-source nonlinear alternatives based on parabolized formulations include \texttt{glimPSE} for incompressible flows \citep{Gonzalez2024} and \texttt{Krypton} for transonic flows \citep{Lacombe2022}.

Three problem classes fall outside the scope of the original release of DeHNSSo. The first concerns geometry. The swept wing considered above has a curved surface, as do airfoils, leading edges, and concave walls supporting G\"ortler-type instabilities \citep{Saric1994}. DeHNSSo-v1 applied its user-tuned algebraic grids only to localized humps on flat plates, without the mixed second-order metrics, and provided neither a standalone preprocessing pipeline for external CFD data nor mesh smoothing. The second concerns the fluid model. The incompressible formulation excludes high-speed regimes, in which instabilities such as Mack modes can dominate, and the low-speed flows in which non-adiabatic wall conditions can be used to control boundary-layer stability \citep{Barahona2026}. Both regimes require the energy equation, which adds a fifth unknown variable, and temperature-dependent transport properties. The third concerns computational cost. DeHNSSo-v1 refactorizes each per-mode operator at every nonlinear iteration and solves the modes sequentially, and at larger resolutions the fill-in of the sparse factorization sets the peak memory. The fixed-point iteration that closes the nonlinear coupling also fails to converge at sufficiently high disturbance amplitudes. These limitations restrict practical use to modest grids and disturbance amplitudes.

DeHNSSo-v2 addresses these three limitations. It generalizes the curvilinear formulation to include the complete metric terms required by the compressible operator and introduces a standalone grid-generation and preprocessing pipeline for base flows imported from external CFD solvers, with optional elliptic smoothing for curved and concave surfaces. It introduces a compressible primitive-variable formulation with an energy equation and temperature-dependent transport properties, following \citet{Scholten2024}. It combines $O(N_\xi)$ operator assembly and reusable per-mode factorizations with memory-scalable direct solvers and optional parallel execution, and adds Newton pseudo-arclength continuation, which closes the nonlinear coupling at disturbance amplitudes beyond the convergence range of the fixed-point iteration. Together, these developments broaden the physical and geometrical scope of DeHNSSo and substantially reduce preprocessing effort and computational cost.

The present release considers steady, spanwise-invariant base flows on body-fitted curvilinear meshes, with three-dimensional disturbances represented through a spanwise Fourier expansion. The compressible formulation assumes a calorically perfect ideal gas at constant Prandtl number. Fully three-dimensional base flows, nonideal-gas and high-temperature thermochemical effects, and the direct treatment of shocks remain outside its scope.

The remainder of the manuscript is organized as follows. Section~\ref{sec:equations} presents the governing equations, the harmonic-balanced operator, and its boundary conditions. Section~\ref{sec:gridgen} describes the stability-grid generation and the spatial discretization, and Sec.~\ref{sec:efficiency} details the linear-algebra and nonlinear-solver developments. Sections~\ref{sec:verification} and~\ref{sec:validation} verify and validate the new capabilities against DeHNSSo-v1 \citep{Westerbeek2024DeHNSSo}, published PSE \citep{Bertolotti1992,ChangMalik1994,Zanus2021}, HNS \citep{Scholten2024}, and DNS \citep{Moniripiri2026} results, and the experimentally validated Helmholtz-resonator computations of \citet{Michelis2023}. Section~\ref{sec:performance} reports computational performance, and Sec.~\ref{sec:conclusions} summarizes the principal findings. The coefficient matrices of the compressible operator are given in \ref{sec:app_compressible_matrices}, the elliptic smoother in \ref{sec:app_elliptic}, and the parameters of all calculations in \ref{app:params}.

\section{Governing equations}
\label{sec:equations}

This section formulates the harmonic Navier--Stokes operator discretized by DeHNSSo-v2. The formulation begins with the fully compressible Navier--Stokes equations in primitive variables, closed by the ideal-gas equation of state and temperature-dependent transport properties. The flow is then decomposed into a steady base flow and a perturbation, transformed to body-fitted coordinates, and expanded into temporal and spanwise Fourier harmonics. The resulting curvilinear operator and its boundary conditions are presented in Secs.~\ref{subsec:operator}--\ref{subsec:bcs}.

\begin{figure*}[t]
  \centering
  \includegraphics[width=\textwidth]{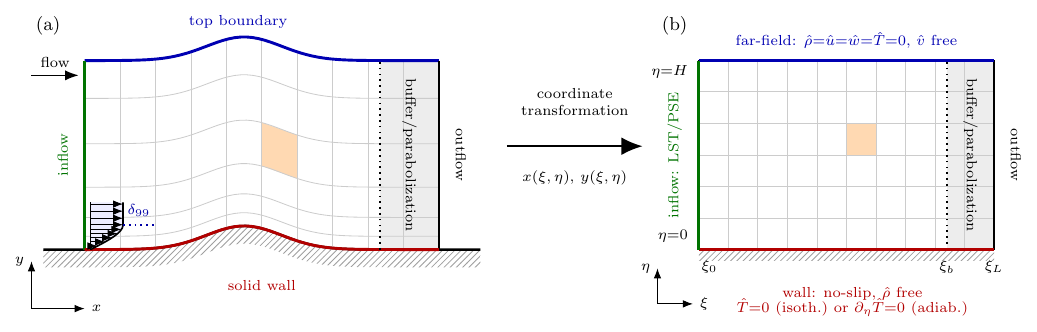}
\caption{Physical-to-computational mapping used by DeHNSSo-v2.
  		(a)~Body-fitted grid in the physical $(x,y)$ plane, showing the
  		boundary-layer edge $\delta_{99}$ and the outflow buffer.
  		(b)~Corresponding rectangular domain in the computational coordinates
  		$(\xi,\eta)$, with the wall, far-field, inflow, and outflow boundaries.
  		Matching colors identify corresponding boundaries, and the shaded cell
  		illustrates the mapping between the two coordinate systems.}
  \label{fig:domain_schematic}
\end{figure*}

\subsection{Flow-conservation equations}
\label{subsec:flow_equations}

The nondimensional, fully compressible Navier--Stokes equations in primitive variables are
\begin{align}
	\partial_t \rho
	+ \nabla\!\cdot\!(\rho\mathbf{u})
	&= 0,
	\label{eq:cns_continuity}
	\\
	\rho\,\partial_t\mathbf{u}
	+ \rho(\mathbf{u}\!\cdot\!\nabla)\mathbf{u}
	&= -\nabla p
	+ \frac{1}{\mathrm{Re}}\nabla\!\cdot\!\boldsymbol{\tau},
	\label{eq:cns_momentum}
	\\
	\rho\,\partial_t T
	+ \rho(\mathbf{u}\!\cdot\!\nabla)T
	&= -(\gamma-1)\rho T\,\nabla\!\cdot\!\mathbf{u}
	+ \frac{\gamma}{\mathrm{Re}\,\mathrm{Pr}}
	\nabla\!\cdot\!(\kappa\nabla T)
	\nonumber\\
	&\quad
	+ \frac{\gamma(\gamma-1)\mathrm{Ma}^2}{\mathrm{Re}}\,\Phi .
	\label{eq:cns_energy}
\end{align}
The physical coordinates $(x,y,z)$ are Cartesian, with $z$ denoting the homogeneous spanwise direction, and $\mathbf{u}=(u,v,w)^\top$ contains the corresponding velocity components. The remaining variables are the density $\rho$, pressure $p$, and temperature $T$. Density and temperature are selected as the independent thermodynamic variables, giving the primitive state vector $\mathbf{q}=(u,v,w,\rho,T)^\top$.

The nondimensional viscous stress tensor and dissipation function are
\begin{equation}
	\boldsymbol{\tau}
	= \mu\!\left[
	\nabla\mathbf{u}
	+ (\nabla\mathbf{u})^\top
	- \tfrac{2}{3}
	(\nabla\!\cdot\!\mathbf{u})\mathbf{I}
	\right],
	\qquad
	\Phi
	= \boldsymbol{\tau}:\nabla\mathbf{u},
	\label{eq:viscous_stress}
\end{equation}
where $\mu$ is the dynamic viscosity, $\kappa$ the thermal conductivity, and $\mathbf{I}$ the identity tensor. The stress definition invokes Stokes' hypothesis for the second viscosity coefficient, $\lambda=-2\mu/3$.

The reference length, velocity, and time scales, and consequently the nondimensional frequency $\omega$, spanwise wavenumber $\beta$, and Reynolds number $\mathrm{Re}$, follow \citet{Westerbeek2024DeHNSSo}. The additional compressible variables are nondimensionalized according to
\begin{equation}
	\begin{aligned}
		\rho
		&= \frac{\rho^*}{\rho^*_\infty},
		&\qquad
		T
		&= \frac{T^*}{T^*_\infty},
		&\qquad
		p
		&= \frac{p^*}
		{\rho^*_\infty u_\infty^{*2}},
		\\
		\mu
		&= \frac{\mu^*}{\mu^*_\infty},
		&\qquad
		\kappa
		&= \frac{\kappa^*}{\kappa^*_\infty}.
	\end{aligned}
	\label{eq:nondim_refs}
\end{equation}
A superscript $(\cdot)^*$ denotes a dimensional quantity, while the subscript $(\cdot)_\infty$ identifies the freestream reference state. The associated nondimensional parameters are
\begin{equation}
	\mathrm{Ma}
	= \frac{u^*_\infty}{c^*_\infty},
	\qquad
	\mathrm{Pr}
	= \frac{c^*_{p,\infty}\mu^*_\infty}
	{\kappa^*_\infty},
	\qquad
	\gamma
	= \frac{c^*_{p,\infty}}
	{c^*_{v,\infty}}.
	\label{eq:char_numbers}
\end{equation}
The parameters $\mathrm{Ma}$ and $\mathrm{Pr}$ are the freestream Mach and Prandtl numbers, respectively, and $\gamma$ is the ratio of specific heats. The freestream speed of sound is denoted by $c^*_\infty$, while $c^*_{p,\infty}$ and $c^*_{v,\infty}$ are the freestream specific heats at constant pressure and constant volume.

\subsection{Equation of state and transport properties}
\label{subsec:eos}

DeHNSSo-v2 assumes a calorically perfect ideal gas. With the scaling in Eq.~\eqref{eq:nondim_refs}, the thermal equation of state becomes
\begin{equation}
	p
	= \frac{\rho T}
	{\gamma\mathrm{Ma}^2}.
	\label{eq:eos}
\end{equation}
The ratio of specific heats is constant, and the dimensional caloric equation of state is $e^*=c_v^*T^*$.

The nondimensional dynamic viscosity is evaluated using Sutherland's law,
\begin{equation}
	\mu(T)
	=
	T^{3/2}
	\frac{1+S^*/T^*_\infty}
	{T+S^*/T^*_\infty},
	\label{eq:sutherland}
\end{equation}
where $S^*$ is the dimensional Sutherland constant, equal to $110.6\,\mathrm{K}$ for air, which can be set by the user for other gases. Under the calorically perfect and constant-Prandtl-number assumptions, the nondimensional thermal conductivity satisfies $\kappa(T)=\mu(T)$.

\subsection{Perturbation decomposition}
\label{subsec:compressible_eqs}

The flow is decomposed into a steady, spanwise-invariant base flow and a three-dimensional, time-dependent perturbation,
\begin{equation}
	\mathbf{q}(x,y,z,t)
	=
	\bar{\mathbf{q}}(x,y)
	+
	\mathbf{q}'(x,y,z,t).
	\label{eq:perturbation_decomp}
\end{equation}
The pressure is similarly decomposed as $p=\bar{p}+p'$, with $\bar{p}=\bar{\rho}\bar{T}/(\gamma\mathrm{Ma}^2)$. Substitution of $\rho=\bar{\rho}+\rho'$ and
$T=\bar{T}+T'$ into Eq.~\eqref{eq:eos} gives the exact pressure perturbation
\begin{equation}
	p'
	=
	\frac{
		\bar{T}\rho'
		+\bar{\rho}T'
		+\rho'T'
	}{\gamma\mathrm{Ma}^2}.
	\label{eq:eos_perturbation}
\end{equation}
The first two terms are linear in the perturbations and enter the HNS operator, whereas the product $\rho'T'$ contributes to the nonlinear forcing.

The temperature dependence of the transport properties is linearized about the base-flow temperature,
\begin{equation}
	\mu(\bar{T}+T')
	\approx
	\bar{\mu}
	+\bar{\mu}_T T',
	\qquad
	\kappa(\bar{T}+T')
	\approx
	\bar{\kappa}
	+\bar{\kappa}_T T',
	\label{eq:transport_linearisation}
\end{equation}
where $\bar{\mu}=\mu(\bar{T})$ and $\bar{\kappa}=\kappa(\bar{T})$ are the base-flow transport properties, and
\begin{equation}
	\bar{\mu}_T
	=
	\left.
	\frac{\mathrm{d}\mu}{\mathrm{d}T}
	\right|_{\bar{T}},
	\qquad
	\bar{\kappa}_T
	=
	\left.
	\frac{\mathrm{d}\kappa}{\mathrm{d}T}
	\right|_{\bar{T}}.
\end{equation}
The equality $\kappa=\mu$ of Sec.~\ref{subsec:eos} gives $\bar{\kappa}=\bar{\mu}$ and $\bar{\kappa}_T=\bar{\mu}_T$. Their spatial derivatives follow from the chain rule, $\nabla\bar{\mu}=\bar{\mu}_T\nabla\bar{T}$ and $\nabla\bar{\mu}_T=\bar{\mu}_{TT}\nabla\bar{T}$, where $\bar{\mu}_{TT}=\mathrm{d}^2\mu/\mathrm{d}T^2|_{\bar{T}}$, and the same relations hold for $\bar{\kappa}$.

Substituting Eq.~\eqref{eq:perturbation_decomp} into
Eqs.~\eqref{eq:cns_continuity}--\eqref{eq:cns_energy} and subtracting the steady equations, which the base flow is assumed to satisfy, gives
\begin{equation}
	\mathcal{L}(\bar{\mathbf{q}})\mathbf{q}'
	=
	\mathbf{r}^{(2)}(\mathbf{q}',\mathbf{q}')
	+
	\mathbf{r}^{(3)}(\mathbf{q}',\mathbf{q}',\mathbf{q}'),
	\label{eq:perturbation_equations}
\end{equation}
where $\mathcal{L}$ contains the terms linear in the perturbations. The terms $\mathbf{r}^{(2)}$ and $\mathbf{r}^{(3)}$ collect the quadratic and cubic interactions among the perturbations, and are resolved into a per-mode forcing in Sec.~\ref{subsec:operator}.

\subsection{Coordinate transformation and first-order inverse metrics}
\label{subsec:coord_transform}
\label{subsec:first_order_metrics}

The Cartesian derivatives appearing in both the linear operator and the nonlinear forcing are transformed to body-fitted coordinates. The transformation acts in the $(x,y)$ plane, while the spanwise coordinate $z$ remains Cartesian.

Consider the forward mapping $x=x(\xi,\eta)$, $y=y(\xi,\eta)$, illustrated in Fig.~\ref{fig:domain_schematic}. Its first derivatives are denoted by $a=x_\xi$, $b=x_\eta$, $c=y_\xi$, and $d=y_\eta$, and the Jacobian determinant is $J=ad-bc$.
A valid mapping requires $J$ to remain nonzero and retain the same sign throughout the computational domain.

The first-order inverse metrics are
\begin{equation}
	\xi_x=\frac{d}{J},
	\qquad
	\xi_y=-\frac{b}{J},
	\qquad
	\eta_x=-\frac{c}{J},
	\qquad
	\eta_y=\frac{a}{J}.
	\label{eq:first_order_metrics}
\end{equation}
These metrics are the coefficients of the chain rule that expresses the Cartesian derivatives of the perturbation in terms of computational ones,
\begin{equation}
	\partial_x
	=
	\xi_x\partial_\xi
	+
	\eta_x\partial_\eta,
	\qquad
	\partial_y
	=
	\xi_y\partial_\xi
	+
	\eta_y\partial_\eta.
	\label{eq:chain_first}
\end{equation}

\subsection{Second-order inverse metrics}
\label{subsec:second_order_metrics}

The second-order inverse metrics are obtained by differentiating the inverse-coordinate identities. In compact form, the physical Hessian of either inverse coordinate
$s\in\{\xi,\eta\}$ satisfies
\begin{equation}
	\mathbf{H}_{x,y}(s)
	=
	-\mathbf{J}_f^{-\top}
	\left[
	s_x\mathbf{H}_{\xi,\eta}(x)
	+
	s_y\mathbf{H}_{\xi,\eta}(y)
	\right]
	\mathbf{J}_f^{-1},
	\label{eq:inverse_hessian}
\end{equation}
where $\mathbf{H}_{\xi,\eta}$ is the Hessian with respect to the computational coordinates and $\mathbf{J}_f$ is the Jacobian matrix of the forward mapping, with entries $a$, $b$, $c$, and $d$ and determinant $J$. This relation gives
$\xi_{xx}$, $\xi_{xy}$, $\xi_{yy}$,
$\eta_{xx}$, $\eta_{xy}$, and $\eta_{yy}$ from the first- and second-order derivatives of the forward mapping.

For illustration purposes, the second $x$-derivative of $\xi$ is expressed as:
\begin{equation}
	\xi_{xx}
	=
	\frac{
		-d^3x_{\xi\xi}
		+bd^2y_{\xi\xi}
		-c^2d\,x_{\eta\eta}
		+bc^2y_{\eta\eta}
		+2cd^2x_{\xi\eta}
		-2bcd\,y_{\xi\eta}
	}{J^3}.
	\label{eq:xixx}
\end{equation}
The implementation evaluates the pure derivatives
$\xi_{xx}$, $\xi_{yy}$, $\eta_{xx}$, and $\eta_{yy}$ from their analytical expressions. The mixed metrics, which enter only the compressible operator, are instead obtained by differentiating the first-order inverse metrics,
\begin{equation}
	\xi_{xy}
	=
	\xi_y\partial_\xi\xi_x
	+
	\eta_y\partial_\eta\xi_x,
	\qquad
	\eta_{xy}
	=
	\xi_y\partial_\xi\eta_x
	+
	\eta_y\partial_\eta\eta_x.
	\label{eq:mixed_inverse_metrics}
\end{equation}

The second Cartesian derivatives transform according to
\begin{align}
	\partial_{xx}^{2}
	&=
	\xi_x^2\partial_{\xi\xi}^{2}
	+2\xi_x\eta_x\partial_{\xi\eta}^{2}
	+\eta_x^2\partial_{\eta\eta}^{2}
	+\xi_{xx}\partial_\xi
	+\eta_{xx}\partial_\eta,
	\label{eq:chain_second_xx}\\
	\partial_{yy}^{2}
	&=
	\xi_y^2\partial_{\xi\xi}^{2}
	+2\xi_y\eta_y\partial_{\xi\eta}^{2}
	+\eta_y^2\partial_{\eta\eta}^{2}
	+\xi_{yy}\partial_\xi
	+\eta_{yy}\partial_\eta,
	\label{eq:chain_second_yy}\\
	\partial_{xy}^{2}
	&=
	\xi_x\xi_y\partial_{\xi\xi}^{2}
	+(\xi_x\eta_y+\xi_y\eta_x)
	\partial_{\xi\eta}^{2}
	+\eta_x\eta_y\partial_{\eta\eta}^{2}
	+\xi_{xy}\partial_\xi
	+\eta_{xy}\partial_\eta.
	\label{eq:chain_second_xy}
\end{align}

\subsection{Harmonic ansatz and curvilinear operator}
\label{subsec:operator}

The perturbation is expanded into a truncated ensemble of temporal and spanwise Fourier harmonics,
\begin{equation}
	\mathbf{q}'(\xi,\eta,z,t)
	=
	\sum_{m=-M}^{M}
	\sum_{n=-N}^{N}
	\hat{\mathbf{q}}_{m,n}(\xi,\eta)
	\exp\!\left[
	\mathrm{i}
	\left(
	\beta_n z-\omega_m t
	\right)
	\right],
	\label{eq:harmonic_ansatz}
\end{equation}
where $\omega_m=m\omega$ and $\beta_n=n\beta$.
For a real-valued physical perturbation, the Fourier coefficients satisfy
\begin{equation}
	\hat{\mathbf{q}}_{-m,-n}
	=
	\hat{\mathbf{q}}_{m,n}^{\dagger},
	\label{eq:reality_condition}
\end{equation}
where $^\dagger$ denotes the complex conjugate.
The conjugate modes $(-m,-n)$ are therefore reconstructed rather than computed, reducing the number of independent systems from $(2M+1)(2N+1)$ to $\tfrac{1}{2}\left[(2M+1)(2N+1)+1\right]$. \citet{Westerbeek2024DeHNSSo} describe an additional reduction based on spanwise-reflection symmetry, but this requires $\bar{w}=0$ together with compatible boundary conditions and forcing, and is not exploited by DeHNSSo-v2.

For each harmonic, the Fourier expansion replaces the temporal and spanwise derivatives by
\begin{equation}
	\partial_t
	\rightarrow
	-\mathrm{i}\omega_m,
	\qquad
	\partial_z
	\rightarrow
	\mathrm{i}\beta_n,
	\qquad
	\partial_{zz}^{2}
	\rightarrow
	-\beta_n^2.
	\label{eq:fourier_derivatives}
\end{equation}
Combining these substitutions with the coordinate transformations in
Eqs.~\eqref{eq:chain_first} and
\eqref{eq:chain_second_xx}--\eqref{eq:chain_second_xy}
gives, for each harmonic, 
\begin{equation}
	\begin{aligned}
		\mathbf{A}_{m,n}\hat{\mathbf{q}}_{m,n}
		&+
		\mathbf{B}_{m,n}\,
		\partial_\eta\hat{\mathbf{q}}_{m,n}
		+
		\mathbf{C}\,
		\partial_{\eta\eta}^{2}\hat{\mathbf{q}}_{m,n}
		\\
		&+
		\mathbf{D}_{m,n}\,
		\partial_\xi\hat{\mathbf{q}}_{m,n}
		+
		\mathbf{E}\,
		\partial_{\xi\eta}^{2}\hat{\mathbf{q}}_{m,n}
		+
		\mathbf{F}\,
		\partial_{\xi\xi}^{2}\hat{\mathbf{q}}_{m,n}
		=
		\mathbf{r}_{m,n},
	\end{aligned}
	\label{eq:lhs_compressible_operator}
\end{equation}
where $\mathbf{r}_{m,n}$ is the nonlinear forcing defined in Eq.~\eqref{eq:nonlinear_forcing}. The inverse metrics are incorporated into the coefficient matrices, so Eq.~\eqref{eq:lhs_compressible_operator} is discretized directly on the body-fitted $(\xi,\eta)$ mesh, following the general-coordinate treatment of \citet{Paredes2014}. The base-flow gradients entering the matrices are precomputed in physical coordinates as described in Sec.~\ref{subsec:metric_computation}.

The operator separates the wall-normal, streamwise, and mixed derivatives. The locality of the streamwise stencil makes the discretized operator block banded, which the direct solvers of Sec.~\ref{subsec:factorization} exploit. The $5\times5$ block coefficient matrices depend on the base flow and its derivatives, the metric coefficients, and the thermodynamic and transport properties; $\mathbf{A}_{m,n}$ also depends on the frequency and spanwise wavenumber, and $\mathbf{B}_{m,n}$ and $\mathbf{D}_{m,n}$ on the spanwise wavenumber. Their explicit entries are given in \ref{sec:app_compressible_matrices}.

Spatial discretization yields the per-mode system
\begin{equation}
	\mathbb{M}_{m,n}\hat{\mathbf{q}}_{m,n}=\mathbf{r}_{m,n},
	\label{eq:permode_system}
\end{equation}
where $\mathbb{M}_{m,n}$ is the discretized per-mode operator. As in DeHNSSo-v1, the operator is assembled from mode-independent blocks:
\begin{equation}
	\mathbb{M}_{m,n}
	=
	\mathbb{M}^{(0)}
	+\omega_m\mathbb{M}^{(\omega)}
	+\beta_n\mathbb{M}^{(\beta)}
	+\beta_n^2\mathbb{M}^{(\beta^2)}
	+\beta_n\mathbb{M}^{(\beta\xi)}.
	\label{eq:mode_blocks}
\end{equation}
The final term contains the compressible $\mathrm{i}\beta_n\partial_\xi$ cross-viscous contribution and is absent from the incompressible operator. The imaginary unit and derivative matrices are absorbed into the blocks, and the mean-flow distortion (MFD), the $(0,0)$ mode, uses its own boundary rows (Sec.~\ref{subsec:bcs}). These blocks are constructed once and recombined for each retained harmonic, as described in Sec.~\ref{sec:efficiency}.

Harmonic balancing resolves $\mathbf{r}^{(2)}$ and $\mathbf{r}^{(3)}$ of Eq.~\eqref{eq:perturbation_equations} into a forcing for each retained harmonic,
\begin{equation}
	\mathbf{r}_{m,n}
	=
	\mathbf{r}^{(2)}_{m,n}
	+
	\mathbf{r}^{(3)}_{m,n}.
	\label{eq:nonlinear_forcing}
\end{equation}
Both are obtained by convolution of retained Fourier modes, while contributions outside the selected truncation ensemble are discarded.
The nonlinear forcing terms include the density--velocity coupling in continuity; the unsteady and convection terms weighted by the base-flow and perturbation densities in the momentum and energy equations; the product $\rho'T'/(\gamma\mathrm{Ma}^2)$ from the equation of state; the viscous stresses and heat conduction from the temperature-dependent transport properties; and the compression work and viscous dissipation in the energy equation. These terms use the same coordinate metrics as the linear operator and are consistent with the transport-property linearization in Eq.~\eqref{eq:transport_linearisation}.

As in DeHNSSo-v1 \citep{Westerbeek2024DeHNSSo}, the nonlinear system is solved by default using a fixed-point (i.e.,~Picard) iteration. At iteration $k$, the nonlinear forcing $\mathbf{r}^{(k)}_{m,n}$ is evaluated from the modal solution $\{\hat{\mathbf q}^{(k)}\}$ and held fixed on the right-hand side of Eq.~\eqref{eq:lhs_compressible_operator} while the individual harmonic systems are solved for $\{\hat{\mathbf q}^{(k+1)}\}$. The modes therefore remain uncoupled within each iteration. The inflow amplitude is ramped toward its target by an amplitude factor, which is increased each time the iteration converges, and the generated harmonics are activated as the solution converges. Convergence is measured, for each active mode, by the change in the modal amplitude between successive iterations,
\begin{equation}
\sigma^{(k)}_{m,n} = \frac{1}{N_\xi \, \max_i A^{(k)}_i} \sum_{i=1}^{i_b} \left| A^{(k)}_i - A^{(k-1)}_i \right|_{m,n},
\label{eq:conv_criterion}
\end{equation}
where $A_i$ is the peak streamwise perturbation velocity of the mode at station $i$, and both the sum and the maximum run over the stations upstream of the outflow buffer. The iteration stops once $\sigma^{(k)}_{m,n}$ falls below a prescribed threshold for every active mode. DeHNSSo-v2 adds two options to this iteration. Anderson acceleration \citep{WalkerNi2011} combines the last $m_{\mathrm{A}}$ iterates and their residuals through a regularized least-squares fit, restarted whenever the ramped amplitude changes (\texttt{Opt.anderson} $=m_{\mathrm{A}}$). Alternatively, the update is under-relaxed, $\hat{\mathbf q}^{(k+1)}_{m,n}=(1-\theta)\,\hat{\mathbf q}^{(k)}_{m,n}+\theta\,\mathbb{M}_{m,n}^{-1}\mathbf{r}^{(k)}_{m,n}$ with $\theta\in(0,1]$ (\texttt{Opt.relax}). Neither is applied by default; the choice for each calculation is given in Tab.~\ref{tab:calc_num}. Unlike DeHNSSo-v1, the per-mode factorizations are reused across iterations (Sec.~\ref{subsec:factorization}). The coupled Newton alternative for cases beyond the convergence range of Picard iteration is described in Sec.~\ref{subsec:newton}.

In the limit $\mathrm{Ma}\to0$, the compressible disturbance solution approaches that of the incompressible formulation of \citet{Westerbeek2024DeHNSSo}. This asymptotic consistency is verified numerically in Sec.~\ref{subsec:verif_incomp} using a Cartesian benchmark at $\mathrm{Ma}=10^{-3}$, for which the two solutions agree closely for all computed modes.

\subsection{Boundary conditions}
\label{subsec:bcs}

The treatment of no-slip and inhomogeneous wall conditions, freestream decay, inflow disturbances, outflow buffers, and embedded boundaries at sharp wall features follows \citet{Westerbeek2024DeHNSSo}. For the compressible state vector
\[
\hat{\mathbf{q}}_{m,n}
=
(\hat{u}_{m,n},\hat{v}_{m,n},\hat{w}_{m,n},
\hat{\rho}_{m,n},\hat{T}_{m,n})^{\top},
\]
DeHNSSo-v2 introduces thermal wall conditions, a density-based far-field condition, compressible and PSE inflow profiles, and a parabolized outflow treatment, as illustrated in Fig.~\ref{fig:domain_schematic}(b).

\paragraph{Velocity and wall conditions}
At the wall ($\eta=0$, superscript $\mathrm{w}$), the velocity perturbation is set to
\begin{equation}
	\hat{\mathbf{u}}_{m,n}(\xi,0)=\hat{\mathbf{u}}^{\mathrm{w}}_{m,n},
	\label{eq:bc_wall_velocity}
\end{equation}
where $\hat{\mathbf{u}}^{\mathrm{w}}_{m,n}$ is the prescribed wall velocity of each mode. Zero values recover the no-slip condition, whereas nonzero values represent wall actuation such as blowing and suction (Sec.~\ref{sec:gridgen}). The temperature perturbation satisfies either an isothermal or an adiabatic condition,
\begin{equation}
	\begin{aligned}
		\hat{T}_{m,n}(\xi,0)&=0 &&(\text{isothermal}),\\
		\partial_\eta\hat{T}_{m,n}(\xi,0)&=0 &&(\text{adiabatic}).
	\end{aligned}
	\label{eq:bc_wall_thermal}
\end{equation}
The density perturbation $\hat{\rho}_{m,n}$ is left unconstrained at the wall, in the same way as the pressure perturbation in DeHNSSo-v1 \citep{Westerbeek2024DeHNSSo}.

\paragraph{Density-based far-field condition}
The homogeneous Dirichlet condition of the pressure-based formulation of \citet{Westerbeek2024DeHNSSo} may reflect acoustic disturbances at finite Mach number. At the upper boundary $\eta=H$, DeHNSSo-v2 therefore employs the density-based condition of \citet{Scholten2024},
\begin{equation}
	\begin{aligned}
		\hat{\rho}_{m,n}
		=\hat{u}_{m,n}
		=\hat{w}_{m,n}
		=\hat{T}_{m,n}
		&=0,\\
		\hat{v}_{m,n}
		&\ \text{unconstrained},
	\end{aligned}
	\label{eq:bc_top_compressible}
\end{equation}
which admits a wall-normal mass-flux perturbation at the upper boundary. This generalizes to all $(m,n)$ modes the treatment applied by \citet{Westerbeek2024DeHNSSo} to the MFD, while the incompressible operator retains the original condition. The height $H$ is chosen sufficiently large for the boundary-layer perturbations to decay before reaching the upper boundary. In supersonic applications, this boundary must remain below any shock so that the base flow remains smooth throughout the stability domain; direct treatment of the shock would require a fitted-shock formulation such as that of \citet{ZhaoDong2025}. Alternatively, a homogeneous Neumann condition may be imposed at $\eta=H$, or nonzero values may be prescribed to introduce disturbances from the freestream.

\paragraph{Inflow condition}
\label{subsec:inflow}

Dirichlet conditions are imposed on all perturbation components at the inflow. User-defined primary modes are prescribed there, while the MFD and remaining higher harmonics are set to zero and generated downstream by nonlinear interactions. As in DeHNSSo-v1, the inflow profiles may instead be supplied externally (\texttt{LOAD}) or left undisturbed (\texttt{ZERO}).

An inflow profile $\hat{\mathbf{q}}_{m,n}(\eta)$ may also be obtained from a local stability analysis, as in DeHNSSo-v1 \citep{Westerbeek2024DeHNSSo}. The base flow is assumed parallel at the inflow station, and the streamwise derivatives are replaced according to
\begin{equation}
	\partial_\xi\rightarrow \mathrm{i}\alpha,
	\qquad
	\partial_{\xi\xi}^{2}\rightarrow-\alpha^2.
	\label{eq:local_stability_substitution}
\end{equation}
For prescribed frequency and spanwise wavenumber, these substitutions yield the local eigenvalue problem for the complex streamwise wavenumber $\alpha$ and its associated eigenfunction. While DeHNSSo-v1 offered only an incompressible stability solver, DeHNSSo-v2 adds a compressible one; the two are selected as \texttt{ILST} and \texttt{CLST}.

DeHNSSo-v2 additionally provides a self-contained linear PSE inflow treatment, selected as \texttt{PSE}, which accounts for the nonparallel development upstream of the HNS domain; the compressible version follows \citet{ChangMalik1993}. For each primary mode, the Fourier coefficient of Eq.~\eqref{eq:harmonic_ansatz} is factorized as
\begin{equation}
	\hat{\mathbf{q}}_{m,n}(\xi,\eta)
	=
	\check{\mathbf{q}}_{m,n}(\xi,\eta)
	\exp\!\left[
	\mathrm{i}
	\int_{\xi_0}^{\xi}
	\alpha_{m,n}(\xi')\,\mathrm{d}\xi'
	\right],
	\label{eq:pse_ansatz}
\end{equation}
where $\alpha_{m,n}(\xi)$ carries the rapidly varying streamwise phase and $\check{\mathbf{q}}_{m,n}(\xi,\eta)$ is a slowly varying shape function. The decomposition is closed using the density-weighted normalization condition of \citet{ChangMalik1993}, evaluated over the three velocity components. The calculation is initialized with a local eigenmode on a self-similar precursor and marched to the HNS inflow using a first-order backward-$\xi$ scheme, with $\alpha_{m,n}$ updated from the normalization condition at every step. The marching step is fixed, and its default satisfies the minimum-step restriction $\Delta\xi > 1/|\alpha_r|$ that suppresses the upstream-propagating component of the PSE operator \citep{LiMalik1996}.

\paragraph{Outflow parabolization}
The amplitude-damping buffer of \citet{Westerbeek2024DeHNSSo} is retained over $\xi\in[\xi_b,\xi_L]$ (cf.~\citet{Kloker1993}). As an alternative, Eq.~\eqref{eq:lhs_compressible_operator} is gradually parabolized within the buffer \citep{LiuLiu1993} by ramping down the streamwise second-derivative contribution,
\begin{equation}
	\mathbf{F}\,
	\partial_{\xi\xi}^{2}\hat{\mathbf{q}}_{m,n}
	\longrightarrow
	s(\xi)\,
	\mathbf{F}\,
	\partial_{\xi\xi}^{2}\hat{\mathbf{q}}_{m,n},
	\qquad
	\xi\in[\xi_b,\xi_L],
	\label{eq:parabolization}
\end{equation}
where the ramp $s(\xi)$ decreases smoothly from $s(\xi_b)=1$ at the start of the buffer to $s(\xi_L)=0$ at the outflow. At the downstream end of the buffer, the operator becomes:
\begin{equation}
	\begin{aligned}
		\mathbf{A}_{m,n}\hat{\mathbf{q}}_{m,n}
		&+
		\mathbf{B}_{m,n}\,
		\partial_\eta\hat{\mathbf{q}}_{m,n}
		+
		\mathbf{C}\,
		\partial_{\eta\eta}^{2}\hat{\mathbf{q}}_{m,n}
		\\
		&+
		\mathbf{D}_{m,n}\,
		\partial_\xi\hat{\mathbf{q}}_{m,n}
		+
		\mathbf{E}\,
		\partial_{\xi\eta}^{2}\hat{\mathbf{q}}_{m,n}
		=
		\mathbf{r}_{m,n}.
	\end{aligned}
	\label{eq:parabolized_operator}
\end{equation}
The buffer thus becomes first order in $\xi$, as in PSE, but since the system is solved globally rather than marched, the streamwise pressure gradient is kept in full without a step-size restriction. This option suits finite Mach numbers, where amplitude damping can reflect acoustic waves. The parabolized region is placed far enough downstream not to influence the retained solution.

\section{Stability-grid generation and numerical discretization}
\label{sec:gridgen}

DeHNSSo-v2 introduces a stability-grid generator through the \texttt{gridgen/} module. DeHNSSo-v1 already included user-tuned algebraic
grid construction, whereas the new module provides a solver-independent pipeline for importing, resampling, and differentiating base flows defined on Cartesian, structured-curvilinear, or unstructured grids. It constructs the body-fitted mesh on which the perturbation operator is assembled, with the wall-normal points clustered by the mapping of \citet{Malik1990}, and evaluates
the associated inverse metrics and base-flow derivatives with the selected discretization (Sec.~\ref{subsec:metric_computation}). The input may be obtained from a numerical simulation or from an analytical solution sampled on its native grid. The processed grid, base flow, metric coefficients,
derivatives, and forcing metadata are stored in a MATLAB \texttt{.mat} archive that can be read directly by the solver.

The grid-generation pipeline is summarized in Algorithm~\ref{algo:gridgen}. \textsc{ClassifyInput} first identifies the input as Cartesian, structured-curvilinear, or unstructured, and the corresponding branch resamples it. For a Cartesian grid, \textsc{ResampleCartesian} interpolates the base flow onto the prescribed streamwise stations and the clustered wall-normal points, after \textsc{DetectWallSolid} has identified any solids resting on the wall, such as steps and blocks, from the points of zero velocity; these are passed to the embedded-boundary treatment of \citet{Westerbeek2024DeHNSSo}. For a structured-curvilinear input, \textsc{ResampleCurvilinear} takes the wall from the boundary row of the native mesh and resamples the geometry and base-flow fields along the columns of the native mesh and along the wall arc length. For unstructured input, \textsc{ExtractWall} identifies the wall from the low-velocity portion of the boundary of the point cloud, smooths it, and parameterizes it by arc length, and \textsc{NormalMesh} constructs an algebraic mesh along the local wall normals. The automatic wall detection assumes a no-slip wall. The algebraic mesh may optionally be smoothed by \textsc{SmoothMesh} (Sec.~\ref{subsec:elliptic}) before \textsc{Interpolate} maps the base flow onto it. Finally, \textsc{BuildStabilityGrid} evaluates the differentiation operators and inverse metrics on the resampled grid, together with the physical derivatives of the base flow (Sec.~\ref{subsec:metric_computation}).

\begin{algorithm}
\caption{DeHNSSo-v2 stability-grid generation pipeline
(\texttt{gridgen/} module). Variable and function names follow the code; they are defined in the \texttt{gridgen} user guide distributed with the code.}
\label{algo:gridgen}
\begin{algorithmic}[1]
\Procedure{StabilityGridGen}{$\mathrm{BF}_{\mathrm{nat}},\mathrm{params}$}
  \State $\mathit{class}\gets\Call{ClassifyInput}{\mathrm{BF}_{\mathrm{nat}}}$
  \State $\hat f\gets\varnothing$
  \If{$\mathrm{params.forcing}$}
    \State $(x_c,w,\hat f)\gets\Call{PrepareForcing}{\mathrm{params}}$
    \State $\mathrm{params.mug}\gets x_c$, \ $\mathrm{params.spread}\gets w$
  \EndIf
  \If{$\mathit{class}=\textsc{Cartesian}$}
    \State $\mathrm{params.step}\gets\Call{DetectWallSolid}{\mathrm{BF}_{\mathrm{nat}}}$
    \State $\mathrm{BF}\gets\Call{ResampleCartesian}{\mathrm{BF}_{\mathrm{nat}},\mathrm{params}}$
  \ElsIf{$\mathit{class}=\textsc{Curvilinear}$}
    \State $\mathrm{BF}\gets\Call{ResampleCurvilinear}{\mathrm{BF}_{\mathrm{nat}},\mathrm{params}}$
  \Else
    \State $(x_w,y_w,\hat{\mathbf n})\gets\Call{ExtractWall}{\mathrm{BF}_{\mathrm{nat}}}$
    \State $(X,Y)\gets\Call{NormalMesh}{x_w,y_w,\hat{\mathbf n},\mathrm{params}}$
    \If{$\mathrm{params.elliptic}$}
      \State $(X,Y)\gets\Call{SmoothMesh}{X,Y,\mathrm{params}}$
    \EndIf
    \State $\mathrm{BF}\gets\Call{Interpolate}{\mathrm{BF}_{\mathrm{nat}},X,Y}$
  \EndIf
  \State $\mathrm{StabGrid}\gets\Call{BuildStabilityGrid}{\mathrm{BF},\mathrm{params}}$
  \State $\mathrm{StabGrid.forcing}\gets\hat f$
  \State \Call{SaveArchive}{$\mathrm{StabGrid}$}
\EndProcedure
\end{algorithmic}
\end{algorithm}

Wall forcing, namely the nonzero wall velocity $\hat{\mathbf{u}}^{\mathrm{w}}_{m,n}$ of Eq.~\eqref{eq:bc_wall_velocity}, is also prepared by the grid generator. A forcing patch is defined either by its center $x_c$ and half-width $w$ or by a distribution supplied along the wall. \textsc{PrepareForcing} uses the patch to refine the streamwise grid and returns a profile of unit peak, which is stored with the grid; the solver then scales this profile and applies it as the wall velocity. The same interface can also refine the grid around a wall-mounted device whose boundary condition is supplied separately, such as the resonator neck of Sec.~\ref{subsec:validation_resonator}.

\begin{figure*}[!t]
  \centering
  \includegraphics[width=\textwidth]
  {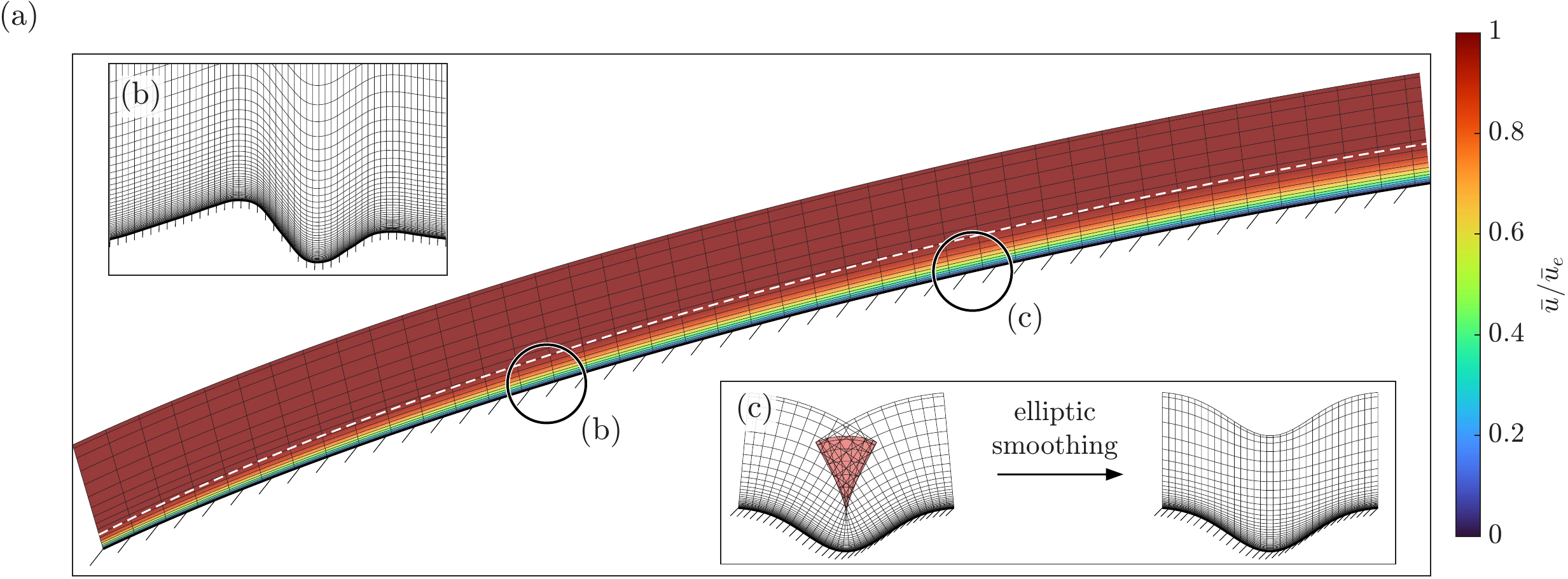}
\caption{Body-fitted stability grids generated by the \texttt{gridgen/} module. (a)~Grid for the clean M3J swept-wing geometry used in Sec.~\ref{subsec:validation_nonlin_curved}, with wall-normal distances enlarged four times, colored by the wall-tangential base-flow velocity normalized by its local edge value, $\bar u/\bar u_e$. The white dashed line marks the boundary-layer edge $\delta_{99}$, and the two circles mark example locations for the surface modifications shown in (b) and (c). For clarity, panel~(a) shows only every third wall-parallel and every sixtieth wall-normal grid line.}
\label{fig:gridgen_mesh}
\end{figure*}

Figure~\ref{fig:gridgen_mesh}(a) shows a grid generated for the M3J swept wing of the TU~Delft low-turbulence tunnel \citep{RiusVidales2025Hump}, using the base flow of \citet{Moniripiri2026}. The near-wall clustering resolves the boundary layer along the continuously curved surface. Panels~(b) and~(c) show grids over generic concave geometries, used to assess the robustness of the grid construction. Potential mesh singularities (i.e.,~folding) are mitigated using the optional elliptic smoothing described in Sec.~\ref{subsec:elliptic}.

\subsection{Elliptic mesh smoothing}
\label{subsec:elliptic}

Over a strongly concave wall, neighboring wall normals may converge and intersect, causing the algebraic normal-ray mesh to fold; an example is shown in Fig.~\ref{fig:gridgen_mesh}(c). A folded mesh contains cells for which the mapping Jacobian vanishes or changes sign, making the inverse metrics of Sec.~\ref{subsec:coord_transform} singular or incorrectly oriented. The optional smoothing procedure starts from the algebraic mesh and, in its default Laplace mode, solves the elliptic inverse-grid equations \citep{Thompson1999},
\begin{equation}
  \begin{aligned}
    g_{\xi\xi}x_{\eta\eta}
    -2g_{\xi\eta}x_{\xi\eta}
    +g_{\eta\eta}x_{\xi\xi}
    &=0,\\
    g_{\xi\xi}y_{\eta\eta}
    -2g_{\xi\eta}y_{\xi\eta}
    +g_{\eta\eta}y_{\xi\xi}
    &=0,
  \end{aligned}
  \label{eq:laplace_mesh}
\end{equation}
where
\begin{equation}
  g_{\xi\xi}
  =x_\xi^2+y_\xi^2,
  \qquad
  g_{\xi\eta}
  =x_\xi x_\eta+y_\xi y_\eta,
  \qquad
  g_{\eta\eta}
  =x_\eta^2+y_\eta^2.
  \label{eq:grid_metric_coefficients}
\end{equation}
The equations are solved on a uniformly spaced auxiliary computational grid using pointwise successive over-relaxation, with the algebraic mesh as the initial guess; the discretization and the iterative scheme are given in \ref{sec:app_elliptic}. To prevent the outer boundary itself from folding, it is translated along the mean outward wall normal rather than offset along the local normals. After convergence, the physical coordinates are redistributed onto the prescribed wall-normal points.

\subsection{Discretization, inverse metrics, and base-flow derivatives}
\label{subsec:metric_computation}

Two wall-normal schemes are available: Chebyshev collocation, as in DeHNSSo-v1, and finite differences of order $2$, $4$, or $6$. Both use the $N_\eta$ points of the mapped Chebyshev distribution of \citet{Malik1990}, which spans $\eta\in[0,H]$ and places half of the points below a prescribed height. The same operators are used for the perturbation operator, the metrics, and the base-flow derivatives. The only exception is the wall-normal derivative of the base flow at the wall and far-field boundaries, where local one-sided stencils reduce the sensitivity to the nonsmoothness of interpolated CFD data. Streamwise derivatives use fourth-order Fornberg finite-difference weights \citep{Fornberg1998}, which accommodate nonuniform station spacing, both in the perturbation operator and for the metrics and base-flow derivatives. These options are summarized in Tab.~\ref{tab:discretization}.

\begin{table}[!t]
\centering
\small
\caption{Spatial discretization in DeHNSSo-v1
\citep{Westerbeek2024DeHNSSo} and DeHNSSo-v2. FD denotes finite differences.}
\label{tab:discretization}
\vspace{2pt}
\begin{tabular}{@{}lll@{}}
\toprule
 & DeHNSSo-v1 & DeHNSSo-v2\\
\midrule
\multicolumn{3}{@{}l}{\textit{Wall-normal} ($\eta$)}\\
\quad scheme
  & Chebyshev
  & Chebyshev or FD\\
\quad FD order
  & ---
  & $2$, $4$, or $6$\\
\addlinespace
\multicolumn{3}{@{}l}{\textit{Streamwise} ($\xi$)}\\
\quad scheme
  & FD
  & Fornberg FD\\
\quad spacing
  & uniform
  & uniform or nonuniform\\
\quad FD order
  & $4$
  & $4$\\
\bottomrule
\end{tabular}
\end{table}

The fourth-order streamwise stencil couples each streamwise grid station to the two stations on either side of it, producing the block-pentadiagonal structure shown in Fig.~\ref{fig:lhs_sparsity}. Chebyshev collocation gives global wall-normal coupling within each streamwise station, whereas the finite-difference alternative produces sparse blocks with local interior stencils, although its boundary rows remain dense. The wall-normal scheme therefore sets the block density and the factorization memory of the direct solvers of Sec.~\ref{subsec:factorization} and of the Newton solver of Sec.~\ref{subsec:newton}.

With these operators, the grid generator evaluates the first- and second-order derivatives of the forward mapping and constructs the inverse metrics of Secs.~\ref{subsec:coord_transform} and~\ref{subsec:second_order_metrics}. The first-order metrics follow Eq.~\eqref{eq:first_order_metrics}. The pure second-order metrics are evaluated from the analytical relations represented by Eq.~\eqref{eq:inverse_hessian}, whereas the mixed metrics follow Eq.~\eqref{eq:mixed_inverse_metrics}. These coefficients enter the curvilinear chain rules that give the physical first and second derivatives of the base flow; the second derivatives and the mixed metrics are required only by the compressible operator. As in DeHNSSo-v1, compatible physical derivatives supplied with the input base flow may alternatively be retained.

\section{Efficient linear algebra and nonlinear continuation}
\label{sec:efficiency}

This section presents the linear-algebra and nonlinear-solver developments introduced in DeHNSSo-v2. They address four limitations of DeHNSSo-v1: the cost of assembling the global operators, the repeated factorization of the per-mode systems during Picard iteration, the memory required by the factors, and the loss of fixed-point convergence at large disturbance amplitudes.

In DeHNSSo-v1, the operator of each active mode is factorized again at every nonlinear iteration; for the nonlinear incompressible Blasius case of Sec.~\ref{subsec:perf_profile}, this repeated factorize-and-solve operation accounts for $92\%$ of the run time (see Fig.~\ref{fig:perf_linalg}a). In addition, the operator assembly grows super-linearly with $N_\xi$, and at larger resolutions the sparse fill-in sets the peak memory. DeHNSSo-v2 reduces these costs through $O(N_\xi)$ sparse assembly (Sec.~\ref{subsec:sparse_assembly}) and reusable per-mode factorizations (Sec.~\ref{subsec:factorization}). A banded-LU back-end provides predictable factor storage and a faster factorization (Sec.~\ref{subsec:lapack_band}), while an optionally out-of-core block-Thomas back-end permits calculations whose complete banded factor does not fit in memory (Sec.~\ref{subsec:block_thomas}). Finally, the Picard iteration inherited from DeHNSSo-v1 is complemented by a Newton continuation method for large disturbance amplitudes (Sec.~\ref{subsec:newton}).

\subsection{Sparse operator assembly}
\label{subsec:sparse_assembly}

Each mode-invariant block in Eq.~\eqref{eq:mode_blocks} has order $N_qN_\eta N_\xi$, where $N_q=4$ for the incompressible operator and $N_q=5$ for the compressible operator. DeHNSSo-v1 appended the sparse triplets $(\mathbf{i},\mathbf{j},\mathbf{v})$ of each streamwise station to growing arrays, so that repeated allocation and copying made the assembly time grow super-linearly, at a measured $O(N_\xi^{1.5})$ over $N_\xi=200$--$3200$ (see Fig.~\ref{fig:perf_linalg}b). DeHNSSo-v2 instead stores the triplets of each streamwise station separately and concatenates them once, after all stations have been processed. This reduces the assembly cost to $O(N_\xi)$ and changes only how the matrix is built, not its entries. The MFD operator is likewise stored as its difference from the common mode-invariant block rather than as a second complete operator. The assembly scaling is measured in Sec.~\ref{subsec:perf_profile}, and agreement with the DeHNSSo-v1 solution is shown in Sec.~\ref{subsec:verif_incomp}.

\subsection{Direct solvers}
\label{subsec:factorization}

For a fixed base flow and grid, the per-mode operator $\mathbb{M}_{m,n}$ of Eq.~\eqref{eq:mode_blocks} depends only on $(\omega_m,\beta_n)$ and therefore remains constant throughout the Picard iteration of Sec.~\ref{subsec:operator}, in which only the forcing $\mathbf{r}_{m,n}$ changes. Except for \texttt{seq} and \texttt{none}, the back-ends of Tab.~\ref{tab:opt} exploit this invariance by factorizing each operator once and reusing the factor at every iteration. The back-ends, selected through \texttt{Opt.lu\_mode}, differ in the factorization method and its storage. Their agreement is verified in Sec.~\ref{subsec:verif_lumode}, and their time and memory requirements are compared in Secs.~\ref{subsec:perf_backend} and~\ref{subsec:perf_scaling_hw}. The default \texttt{auto} is suitable for most incompressible calculations. For the compressible operator, \texttt{lapack\_band} is faster as long as its banded factors fit in memory; beyond that limit, \texttt{block\_thomas} can store the factor on disk, which allows finer grids.

\begin{table}[!t]
	\centering
	\small
	\caption{Direct-solver back-ends selected through
		\texttt{Opt.lu\_mode}. The first four settings use sparse LU and differ in the number of resident per-mode factors.}
	\label{tab:opt}
	\begin{tabular}{@{}ll@{}}
		\toprule
		\texttt{Opt.lu\_mode} & Factor handling \\
		\midrule
		\texttt{none}
		& Fresh factorization at every solve (DeHNSSo-v1) \\
		\texttt{seq}
		& One factor held temporarily and then discarded \\
		\texttt{auto}
		& Bounded cache of per-mode factors \\
		\texttt{full}
		& All per-mode factors retained \\
		\texttt{lapack\_band}
		& Compact banded factor retained per mode \\
		\texttt{block\_thomas}
		& Partitioned factor, optionally stored out of core \\
		\bottomrule
	\end{tabular}
\end{table}

\subsubsection{Cached sparse-LU factorization}
\label{subsec:lu_cache}

DeHNSSo-v1 passed each per-mode system to a new sparse direct solve at every Picard iteration, although $\mathbb{M}_{m,n}$ remains constant. DeHNSSo-v2 instead retains the factorization of $\mathbb{M}_{m,n}$, so that each subsequent solution reduces to a forward and a backward triangular substitution (see Fig.~\ref{fig:perf_linalg}c).

Sparse LU introduces fill-in: entries that are zero in $\mathbb{M}_{m,n}$ become nonzero in its factors, which may therefore require substantially more memory than the assembled operator. The amount of fill depends on the elimination order, so DeHNSSo-v2 permutes the operator before factorization. The solver applies the reverse Cuthill--McKee ordering \citep{CuthillMcKee1969} to the compressible operator, where it reduces the bandwidth of the block-banded structure, and the column approximate minimum degree ordering \citep{Davis2004Colamd}, which targets fill-in directly, to the incompressible operator, for which it is leaner and faster. Fill-in remains particularly costly for the five-variable compressible system and under Chebyshev collocation, which couples all wall-normal points within each station block.

The sparse-LU settings of Tab.~\ref{tab:opt} provide different memory--time compromises. The \texttt{full} setting retains every per-mode factor and \texttt{seq} only the current one. The default \texttt{auto} keeps up to \texttt{Opt.lu\_max\_cache} factors, a limit that is at least the number of computed modes unless set lower, and, when this limit is reached, removes the oldest factor not needed by the active modes. The \texttt{none} setting keeps no factor and solves each system directly, as in DeHNSSo-v1. With \texttt{full} or \texttt{auto}, the per-mode solves can be distributed over parallel workers through \texttt{Opt.parfor}, which requires the MATLAB Parallel Computing Toolbox; the resulting speed-up is measured in Sec.~\ref{subsec:perf_scaling_hw}.

The assembled blocks and cached factors may also be reused between solver calls through \texttt{Opt.rerun}, provided that all operator-defining quantities remain unchanged and only the right-hand side $\mathbf{r}_{m,n}$ differs between calls. This is useful in sweeps over an inflow or an actuator amplitude on a fixed grid and base flow, as measured in Sec.~\ref{subsec:perf_profile}. Changes to the grid, base flow, frequency, or spanwise wavenumber require the affected operators to be rebuilt.

\subsubsection{Banded LU factorization}
\label{subsec:lapack_band}

\begin{figure}[!t]
  \centering
  \includegraphics[width=\columnwidth]{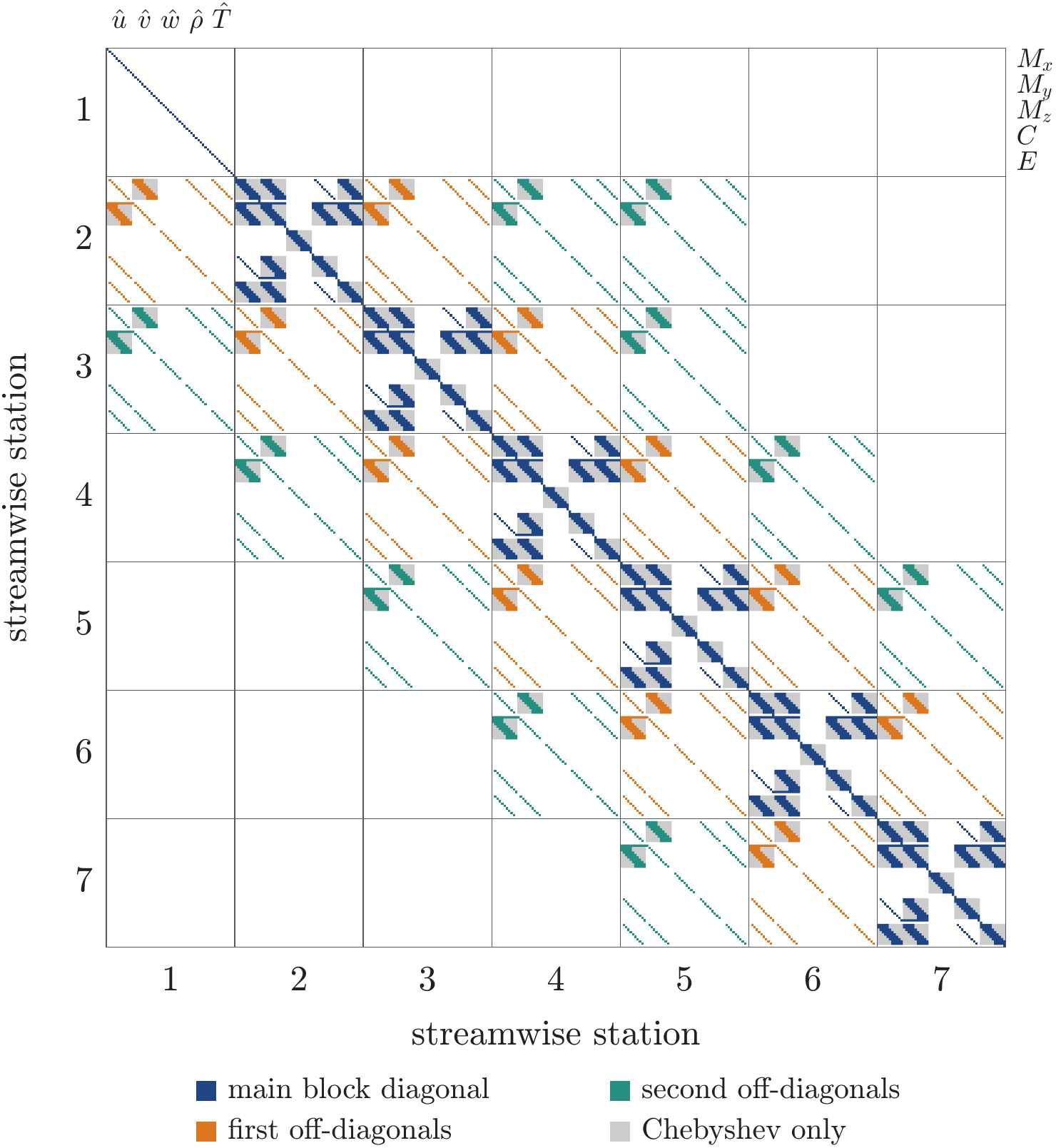}
\caption{Sparsity pattern of the mode-invariant block $\mathbb{M}^{(0)}$ of the compressible operator in Eq.~\eqref{eq:mode_blocks}, assembled on a grid with $N_\xi=7$, $N_\eta=11$, and fourth-order finite differences in $\xi$. Colored entries are those of fourth-order finite differences in $\eta$, distinguished by streamwise block diagonal; gray entries are added by Chebyshev collocation in $\eta$. Near the inflow and outflow, the biased streamwise stencils reach one station further. Block columns contain $(\hat u,\hat v,\hat w,\hat\rho,\hat T)$, and block rows contain the $x$-, $y$-, and $z$-momentum, continuity, and energy equations ($M_x$, $M_y$, $M_z$, $C$, and $E$).}
  \label{fig:lhs_sparsity}
\end{figure}

The locality of the streamwise stencil confines every nonzero entry of the per-mode operator to a band about the diagonal, giving the block-pentadiagonal structure of Fig.~\ref{fig:lhs_sparsity}. DeHNSSo-v2 stores only this band and factorizes it with the general-banded LU routines of LAPACK \citep{Anderson1990}, called through a C MEX interface. This avoids an additional MATLAB matrix representation and provides separate factorization and substitution calls, allowing the factor to be reused throughout the Picard iteration.

The band is set by the streamwise coupling and therefore has the same width for Chebyshev collocation and wall-normal finite differences, although it widens as $N_\eta$ grows. Within the band, finite-difference discretization of derivatives along $\eta$ produces far fewer nonzero entries than Chebyshev collocation (Fig.~\ref{fig:lhs_sparsity}). For a matrix of dimension $n$ and bandwidth $b$, banded storage scales as $O(nb)$ and factorization as $O(nb^2)$. At fixed grid resolution, both $n$ and $b$ are proportional to the number of variables $N_q$, so storage and factorization scale as $O(N_q^2)$ and $O(N_q^3)$, respectively (Sec.~\ref{subsec:perf_backend}). As banded storage allocates memory for every position within the band, including positions that contain zeros, it does not benefit from a sparse $\eta$ discretization, and its factor is larger than the corresponding sparse-LU factor in all cases examined in Sec.~\ref{subsec:perf_backend}. The advantages of the banded path are predictable memory requirements and a substantially faster factorization, which compensate for the larger factor when solving the five-variable compressible system. The routines apply partial pivoting \citep{GolubVanLoan2013}, which is necessary because the HNS operator is not diagonally dominant and, in the incompressible formulation, the continuity rows carry no diagonal entry.

\subsubsection{Partitioned block-Thomas factorization}
\label{subsec:block_thomas}

The banded back-end of Sec.~\ref{subsec:lapack_band} stores the complete factor in memory, which eventually limits the grid size. DeHNSSo-v2 therefore provides a block-Thomas back-end based on the dual-Schur \citep{Vanderstraeten1998} approach of \citet{Paredes2024}, selected through \texttt{Opt.lu\_mode = 'block\_thomas'}. The streamwise domain is divided into $n_p$ partitions separated by two-station interfaces, as required by the block-pentadiagonal stencil. Within each partition, the station blocks are eliminated by a block-Thomas recursion, which leaves a much smaller system on the interfaces. Following \citet{Paredes2024}, $n_p=2$ by default, since more partitions increase the operation count and data movement. The diagonal station blocks are LU-factorized with partial pivoting and applied by substitution \citep{Benkert2007}; storing their explicit inverses, as in \citet{Paredes2024}, is available as an option (\texttt{Opt.bt\_apply}). The factor is built once for each mode and reused across Picard iterations.

The factor can be kept in memory or streamed to disk with \texttt{Opt.bt\_ooc = 'on'}. In this out-of-core mode, the dense factor blocks are written to disk as they are formed, and only the interface system, the sparse operator blocks of the partitions, and the blocks of the current recursion remain in memory. Unlike banded storage, the resident memory therefore depends on the sparsity of the operator blocks and decreases with a finite-difference wall-normal discretization (Fig.~\ref{fig:lhs_sparsity}), as quantified in Sec.~\ref{subsec:perf_backend}. This allows resolutions whose in-core factor would exceed the available memory.

The recursion between stations is unpivoted and therefore requires nonsingular diagonal station blocks. This holds for the compressible operator but not for the incompressible one, so the back-end is restricted to compressible calculations.

\subsection{Newton continuation for high-amplitude nonlinear solutions}
\label{subsec:newton}

The Picard iteration inherited from DeHNSSo-v1 remains the default solver (see Sec.~\ref{subsec:operator}), but at high amplitudes the lagged nonlinear forcing may no longer define a convergent fixed-point iteration. Newton's method instead carries the coupling between harmonics in the Jacobian \citep{Rigas2021,Scholten2024}, and pseudo-arclength continuation \citep{Keller1977,Dijkstra2014,Busquet2021} provides successive initial guesses by advancing the disturbance field and the inflow amplitude $A_0$ together along the solution branch. The method is selected through \texttt{Opt.solver = 'newton'}.

Let $\hat{\mathbf{Q}}$ contain the coefficients $\hat{\mathbf{q}}_{m,n}$ retained as independent unknowns, so that the harmonic system of Eq.~\eqref{eq:permode_system} is written as
\begin{equation}
	\boldsymbol{\mathcal{F}}(\hat{\mathbf{Q}},\hat{\mathbf{Q}}^{\dagger},A_0)
	= \mathbb{M}\hat{\mathbf{Q}} - \mathbf{b}(A_0)
	- \mathbf{r}(\hat{\mathbf{Q}},\hat{\mathbf{Q}}^{\dagger}) = \mathbf{0},
	\label{eq:newton_residual}
\end{equation}
where $\mathbb{M}=\operatorname{diag}(\mathbb{M}_{m,n})$, $\mathbf{r}$ is the harmonic-balanced forcing, and $\mathbf{b}(A_0)$ represents the inflow condition, in which $A_0$ is the imposed amplitude of the fundamental mode. Since the reconstructed conjugate modes enter the nonlinear interactions, $\boldsymbol{\mathcal{F}}$ depends on $\hat{\mathbf{Q}}^{\dagger}$ as well as on $\hat{\mathbf{Q}}$. Newton's method is therefore applied to the real and imaginary parts of $\hat{\mathbf{Q}}$ and $\boldsymbol{\mathcal{F}}$, collected in the real vectors $\mathbf{x}$ and $\mathbf{f}$, with the real Jacobian $\mathbf{J}_R=\partial\mathbf{f}/\partial\mathbf{x}$ assembled analytically from the derivatives of $\mathbf{r}$ with respect to $\hat{\mathbf{Q}}$ and $\hat{\mathbf{Q}}^{\dagger}$.

Continuation treats $A_0$ as an additional unknown. At a converged point $(\mathbf{x}_c,A_{0,c})$ with normalized tangent $(\dot{\mathbf{x}},\dot{A}_0)$, the next solution satisfies
\begin{equation}
	\mathcal{S}(\mathbf{x},A_0)
	= \dot{\mathbf{x}}^{\top}(\mathbf{x}-\mathbf{x}_c)
	+ \dot{A}_0(A_0-A_{0,c}) - \Delta s = 0,
	\label{eq:newton_arclength}
\end{equation}
where $\Delta s$ is the continuation step. Two converged low-amplitude solutions provide the initial tangent, and a second-order predictor built from the two preceding tangents extrapolates along the branch. The corrector drives $\mathbf{f}$ and $\mathcal{S}$ to zero simultaneously. Defining $\mathbf{g}=\partial\mathbf{f}/\partial A_0$, the method of \citet{Dijkstra2014} avoids assembling and factorizing the bordered Jacobian by solving
\begin{equation}
	\begin{aligned}
		&\mathbf{J}_R\mathbf{z}_1=-\mathbf{f},
		\qquad
		\mathbf{J}_R\mathbf{z}_2=\mathbf{g},\\[2pt]
		&\Delta A_0
		=
		\frac{-\mathcal{S}-\dot{\mathbf{x}}^{\top}\mathbf{z}_1}
		     {\dot{A}_0-\dot{\mathbf{x}}^{\top}\mathbf{z}_2},
		\qquad
		\Delta\mathbf{x}=\mathbf{z}_1-\Delta A_0\mathbf{z}_2 .
	\end{aligned}
	\label{eq:newton_two_solves}
\end{equation}
The iteration continues until $(\lVert\mathbf{f}\rVert^{2}+\mathcal{S}^{2})^{1/2}<10^{-6}$, and $\mathbf{z}_2$ provides the next tangent, proportional to $[-\mathbf{z}_2^{\top},1]^{\top}$.

By default, $\mathbf{J}_{R}$ is held fixed during the corrector, and its factorization, computed with MUMPS \citep{Amestoy2001,Amestoy2006}, is carried across nearby continuation steps when the convergence history allows; it is refreshed after \texttt{Opt.newton\_reuse} steps or when the corrector exceeds \texttt{Opt.newton\_it\_thresh} iterations, as also done within the corrector by \citet{Busquet2021}. Most corrector iterations therefore require only a solve with the retained factor. The full Newton corrector and first-order predictor of \citet{Dijkstra2014} are available through \texttt{Opt.newton\_corrector} and \texttt{Opt.newton\_predictor}. The step $\Delta s$ follows the corrector iteration count, growing when convergence is quick and shrinking when it is slow or fails \citep{Beaume2017}. A final Newton solve at fixed amplitude returns the solution at the requested $A_0$. The cost of the Newton factorization and the saving from its reuse are assessed in Sec.~\ref{subsec:perf_newton}. 

\section{Verification}
\label{sec:verification}

This section verifies four aspects of DeHNSSo-v2: recovery of the incompressible limit and grid convergence (Sec.~\ref{subsec:verif_incomp}), independence of the solution from the direct-solver back-end (Sec.~\ref{subsec:verif_lumode}), accuracy and coordinate invariance of the curvilinear formulation (Sec.~\ref{subsec:verif_metric}), and high-amplitude Newton continuation (Sec.~\ref{subsec:verif_hns}).

Sections~\ref{subsec:verif_incomp}, \ref{subsec:verif_lumode}, and \ref{subsec:verif_hns} use the canonical nonlinear TS-wave development case of \citet{Bertolotti1992} and \citet{Westerbeek2024DeHNSSo} in a Blasius boundary layer. The local Reynolds number is $R=\sqrt{\mathrm{Re}_x}$, with $R_0=400$ at the inflow; the fundamental frequency is $\omega_1=0.0346$, corresponding to $F=86\times10^{-6}$, and $\beta=0$. Section~\ref{subsec:verif_metric} instead combines an analytical differentiation test with compressible flat-plate cases at $\mathrm{Ma}=0.5$ and $4.5$ of Sec.~\ref{subsec:validation_lin_compr}, comparing Cartesian and distorted grids. The parameters, grids, and discretizations of all calculations are summarized in \ref{app:params}.

\subsection{Incompressible-limit recovery and grid convergence}
\label{subsec:verif_incomp}

\begin{figure}[!t]
  \centering
  \includegraphics[width=\columnwidth]{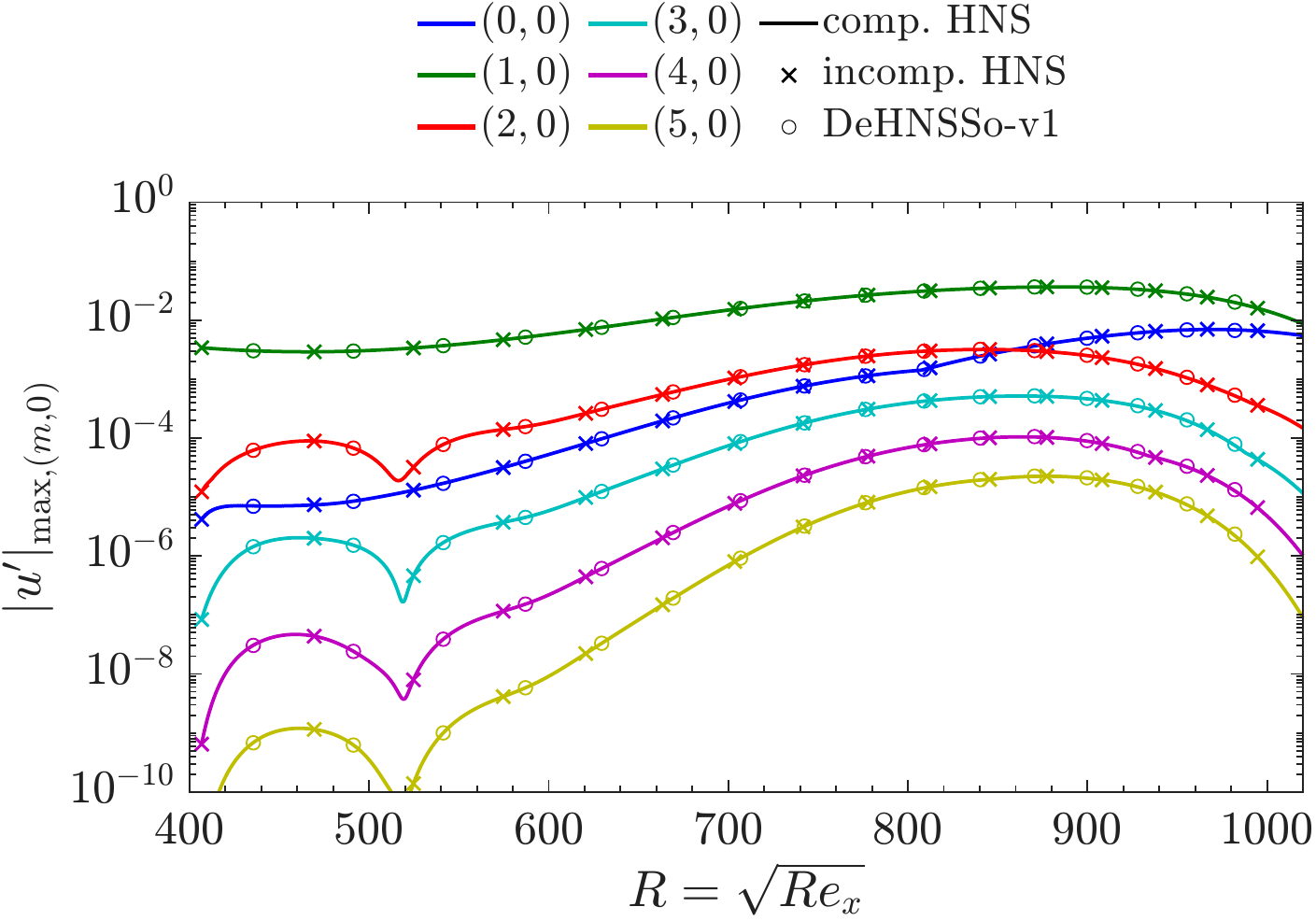}
    \caption{Incompressible-limit recovery: amplitude evolution $|u'|_{\max,(m,0)}$ of the MFD $(0,0)$ and harmonics $(1,0)$--$(5,0)$ for the Bertolotti TS case ($R_0=400$, $F=86\times10^{-6}$, $M=5$). DeHNSSo-v2 compressible at $\mathrm{Ma}=10^{-3}$ (solid lines) and incompressible (crosses) on the $1800\times140$ grid, and DeHNSSo-v1 on its $2674\times100$ grid (open circles); modes are distinguished by color.}
  \label{fig:verif_incomp}
\end{figure}

Figure~\ref{fig:verif_incomp} shows the amplitude evolution $|u'|_{\max,(m,0)}(R)$ against $R = \sqrt{\mathrm{Re}_x}$ for the MFD $(0,0)$ and the five temporal harmonics $(1,0)$--$(5,0)$ ($M = 5$, $N = 0$). The base flows are the steady solutions of the incompressible and compressible boundary-layer equations, respectively, and the primary mode $(1,0)$ is introduced at the inflow using the corresponding incompressible (ILST) or compressible (CLST) linear stability solution. Both calculations use the same refined grid and identical parabolized outflow treatment (\texttt{Opt.buffer = 'para'}, $\xi_b = 90\%$). The streamwise direction carries $1800$ equidistant points, corresponding to approximately $50$ points per wavelength of the primary mode, while the wall-normal direction carries $140$ points with the fourth-order finite-difference scheme.

The compressible and incompressible DeHNSSo-v2 solutions agree closely for all six modes, and both reproduce DeHNSSo-v1 computed on its own $2674\times100$ grid \citep{Westerbeek2024DeHNSSo}, previously verified against the NPSE reference case of \citet{Bertolotti1992}. This agreement verifies recovery of the established incompressible solver in the low-Mach limit. At $\mathrm{Ma}=10^{-3}$, the cubic forcing $\mathbf{r}^{(3)}_{m,n}$ of Eq.~\eqref{eq:nonlinear_forcing} is negligible and the quadratic forcing $\mathbf{r}^{(2)}_{m,n}$ reduces to its incompressible form. The nonlinear compressible forcing terms are validated at finite Mach number in Sec.~\ref{subsec:validation_nonlin_compr}.

Grid convergence is measured using the mode-summed perturbation kinetic energy over the buffer-free domain,
\begin{equation}
	E
	=
	\sum_{(m,n)\in\mathcal{A}}
	\int_{\xi_0}^{\xi_b}\!\int_0^{H}
	\frac{1}{2}
	\left(
	|\hat{u}_{m,n}|^2
	+|\hat{v}_{m,n}|^2
	+|\hat{w}_{m,n}|^2
	\right)
	|J|
	\,\mathrm{d}\eta\,\mathrm{d}\xi,
	\label{eq:verif_energy}
\end{equation}
where $\mathcal{A}$ is the active harmonic set, $\xi_b$ marks the beginning of the outflow buffer, and $J$ is the Jacobian of
Sec.~\ref{subsec:coord_transform}. On the Cartesian grids used here, $J$ equals unity. The corresponding local measure omits the streamwise integral and is evaluated at $R=825$.

\begin{figure}[!t]
  \centering
  \includegraphics[width=\columnwidth]{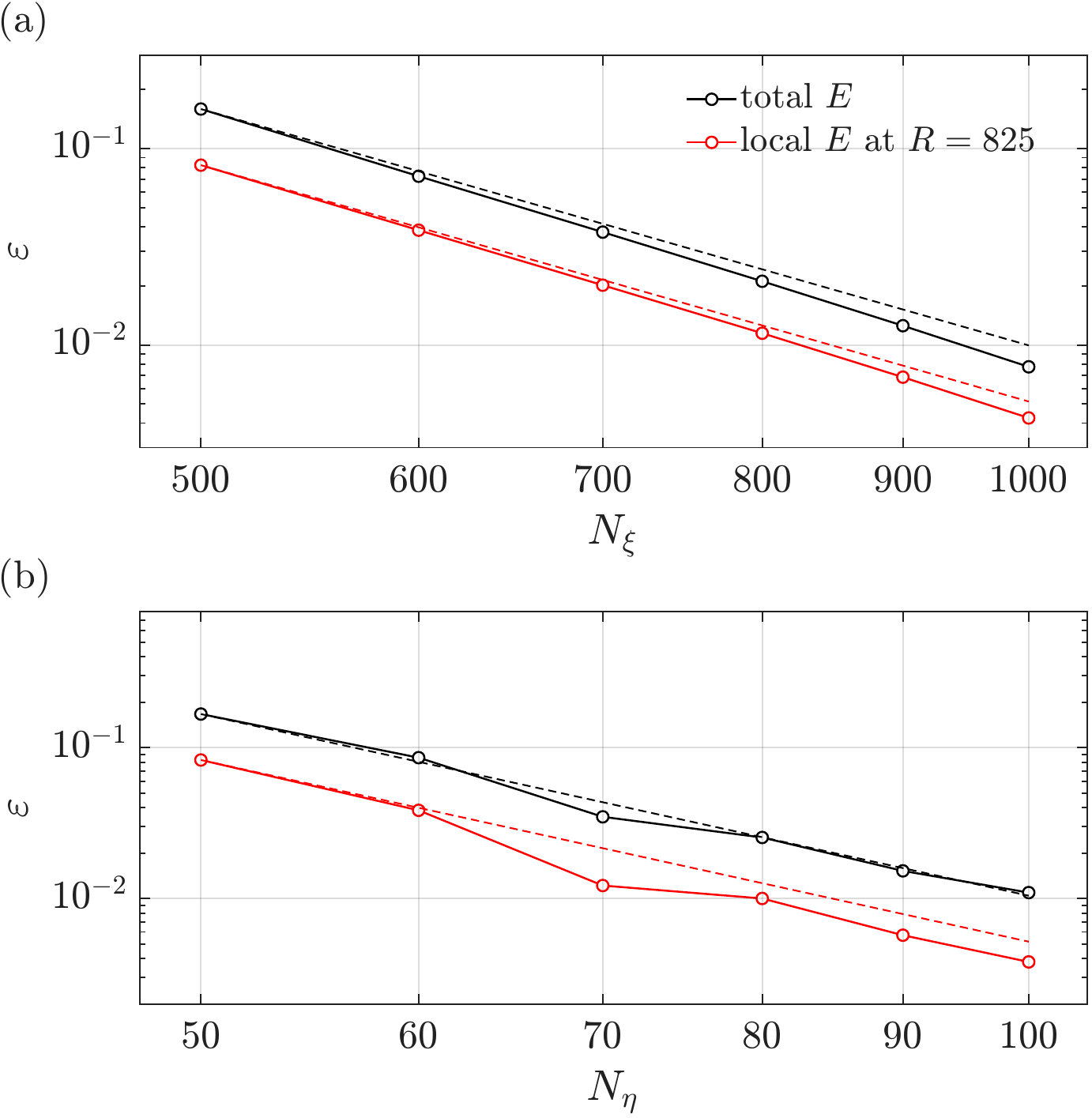}
  \caption{Grid convergence for the Blasius TS calculation: relative difference in the domain-integrated energy (black) and local energy at
	$R=825$ (red), referenced to the $1800\times140$ grid. (a)~Streamwise refinement at fixed $N_\eta=140$; (b)~wall-normal refinement at fixed $N_\xi=1800$. Dashed lines denote formal fourth-order slopes.}
  \label{fig:verif_grid}
\end{figure}

The resolution is assessed by varying $N_\xi=500$--$1000$ at fixed $N_\eta=140$ and $N_\eta=50$--$100$ at fixed $N_\xi=1800$, using the $1800\times140$ grid as reference. Each calculation is converged to $\sigma_{m,n}=10^{-5}$ according to Eq.~\eqref{eq:conv_criterion}. Figure~\ref{fig:verif_grid} reports the relative difference $\varepsilon=|E-E_{\mathrm{ref}}|/|E_{\mathrm{ref}}|$ in the domain-integrated energy and in the local energy at $R=825$. The streamwise sweep gives an observed order of $4.29$, with the $N_\xi=1800$ result within $0.02\%$ of the extrapolated value. The wall-normal sweep gives an order of $3.99$, with a small departure from the trend at $N_\eta=70$. Both sweeps therefore recover the formal fourth-order accuracy and support the adopted $1800\times140$ reference resolution.

\subsection{Direct-solver equivalence}
\label{subsec:verif_lumode}

A key requirement for robustness is that the converged solution must not depend on the direct-solver back-end. All settings of Tab.~\ref{tab:opt} are therefore applied to the Blasius TS case of Sec.~\ref{subsec:verif_incomp} on the reduced $N_\xi\times N_\eta=800\times40$ grid of \citet{Westerbeek2024DeHNSSo}, with only \texttt{Opt.lu\_mode} varied. The \texttt{none} setting, which factorizes each assembled operator at every solve as in DeHNSSo-v1, serves as the reference. The \texttt{block\_thomas} setting is restricted to the compressible operator, so it is applied to the corresponding compressible case at $\mathrm{Ma}=10^{-3}$ and compared with \texttt{full} on the same grid.

The back-ends are compared through the energy measures of Sec.~\ref{subsec:verif_incomp}. Every calculation is advanced for $80$ Picard iterations, beyond the point where the tolerance $\sigma_{m,n}=10^{-5}$ of Eq.~\eqref{eq:conv_criterion} is reached, so that the comparison is not affected by independent stopping decisions. Table~\ref{tab:lumode} lists the relative differences; the \texttt{seq}, \texttt{auto}, and \texttt{full} rows coincide because these settings use the same factorization and differ only in factor residency. For scale, continuing the \texttt{none} run from iteration $49$, when the stopping criterion is first satisfied, to iteration $80$ changes $E$ by a relative $9.5\times10^{-7}$, approximately $120$ times the largest back-end difference. The choice of back-end therefore has a negligible effect at the adopted convergence tolerance.

\begin{table}[!t]
	\centering
	\small
	\setlength{\tabcolsep}{4pt}
    \caption{Direct-solver equivalence for the Blasius TS case of Sec.~\ref{subsec:verif_incomp}. Relative difference $\varepsilon=|E-E_{\mathrm{ref}}|/|E_{\mathrm{ref}}|$ in the mode-summed perturbation kinetic energy of Eq.~\eqref{eq:verif_energy}, evaluated over the buffer-free domain and locally at $R=825$. The incompressible back-ends are referenced to the DeHNSSo-v1 approach \texttt{none}; \texttt{block\_thomas} is evaluated on the compressible case at $\mathrm{Ma}=10^{-3}$ and referenced to \texttt{full}.}
	\label{tab:lumode}
	\begin{tabular}{@{}lcc@{}}
		\toprule
		Back-end & $\varepsilon$ (domain) & $\varepsilon$ ($R=825$) \\
		\midrule
		\texttt{none} (DeHNSSo-v1) & reference & reference \\
		\texttt{seq} & $2\times10^{-9}$ & $9\times10^{-10}$ \\
		\texttt{auto} & $2\times10^{-9}$ & $9\times10^{-10}$ \\
		\texttt{full} & $2\times10^{-9}$ & $9\times10^{-10}$ \\
		\texttt{lapack\_band} & $3\times10^{-12}$ & $2\times10^{-12}$ \\
		\texttt{block\_thomas} & $8\times10^{-9}$ & $2\times10^{-9}$ \\
		\bottomrule
	\end{tabular}
\end{table}

\subsection{Curvilinear-operator accuracy and coordinate invariance}
\label{subsec:verif_metric}

The curvilinear formulation is verified in two stages. First, the transformed derivative operators are applied to an analytical field to establish their formal order of accuracy. Second, coordinate invariance of the linear HNS solution is tested by comparing Cartesian and distorted-grid solutions for the $\mathrm{Ma}=0.5$ and $4.5$ compressible flat-plate cases of Sec.~\ref{subsec:validation_lin_compr}. The distortion activates all metric terms, including $\xi_{xy}$ and $\eta_{xy}$, which enter only the compressible operator.

\begin{figure}[!t]
	\centering
	\includegraphics[width=\columnwidth]{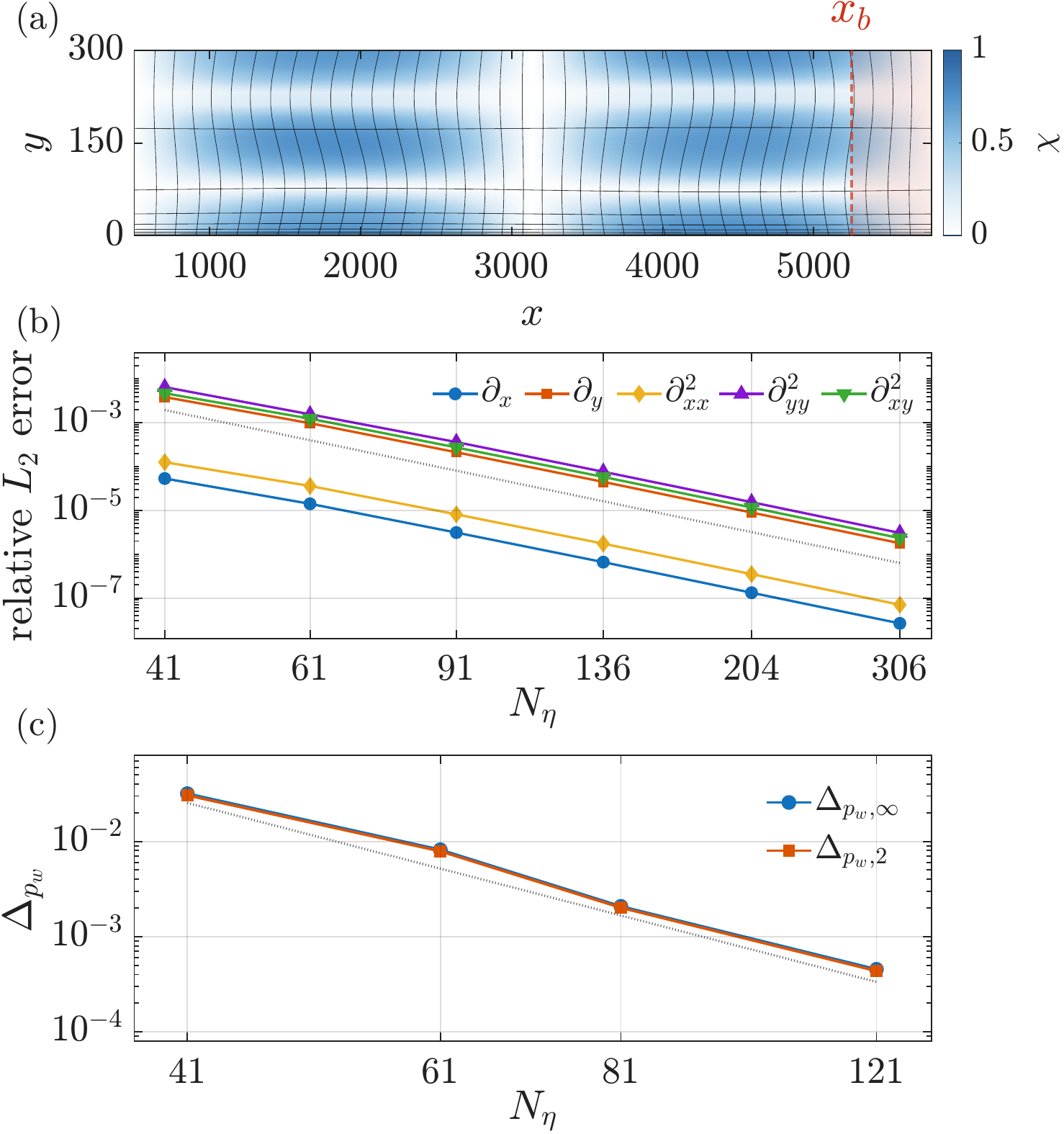}
	\caption{Curvilinear-operator accuracy and coordinate invariance at $\varepsilon=0.01$. (a)~Distorted grid superposed on the departure from orthogonality $\chi$; the shaded outflow buffer is excluded from the solution comparison, and the wall-normal scale is exaggerated by a factor of $4.1$. (b)~Relative $L_2$ errors of the transformed derivatives applied to an analytical field. (c)~Normalized $L_\infty$ and $L_2$ differences in the complex wall-pressure coefficient between the distorted- and Cartesian-grid $\mathrm{Ma}=0.5$ solutions. The finite-difference grids are refined in both directions with $N_\xi\approx10N_\eta$. Gray dotted lines in (b) and (c) show the formal fourth-order rate $N_\eta^{-4}$.}
	\label{fig:verif_metric}
\end{figure}

The grid is distorted with the sinusoidal mapping of \citet{Colella2011}, in the normalization of \citet{GrimmStrele2014}:
\begin{equation}
	\begin{aligned}
		x &= \xi_0
		+L\left[s+\varepsilon\sin(2\pi s)\sin(2\pi t)\right],\\
		y &= H\left[t+\varepsilon\sin(2\pi s)\sin(2\pi t)\right],
	\end{aligned}
	\label{eq:metric_distortion}
\end{equation}
with $s=(\xi-\xi_0)/L$ and $t=\eta/H$, where $L$ and $H$ are the streamwise and wall-normal extents of the domain and $\varepsilon$ sets the distortion amplitude. The mapping preserves all four boundaries and extends through the outflow buffer, which is excluded from the comparisons. Figure~\ref{fig:verif_metric}(a) shows the distorted $\mathrm{Ma}=0.5$ grid and $\chi=|\mathbf{r}_{\xi}\boldsymbol{\cdot}\mathbf{r}_{\eta}|/(|\mathbf{r}_{\xi}||\mathbf{r}_{\eta}|)$, the absolute cosine of the angle between its coordinate directions; $\chi=0$ denotes orthogonality. At $\varepsilon=0.01$, the Jacobian satisfies $0.937\leq J\leq1.063$ and $\chi_{\max}=0.76$. The grid is therefore nonsingular but strongly nonorthogonal.

For the analytical test, a smooth field with nonzero first, second, and mixed derivatives is differentiated using the discrete metrics and matrices employed in the assembled operator. For each derivative $\mathcal{D}$, Fig.~\ref{fig:verif_metric}(b) reports the relative error $\|\mathcal{D}_h f-\mathcal{D}f\|_2/\|\mathcal{D}f\|_2$, where $\mathcal{D}_h$ is its discrete transformed operator. All five operators in Eqs.~\eqref{eq:chain_first} and \eqref{eq:chain_second_xx}--\eqref{eq:chain_second_xy} converge at fourth order over six grids, from $401\times41$ to $3040\times306$, with finest-grid orders between $3.97$ and $3.99$. Setting $\xi_{xy}=\eta_{xy}=0$ leaves the other four errors unchanged, but the mixed-derivative error remains near $0.13$ when the resolution is tripled. Its fourth-order convergence therefore requires the terms in Eq.~\eqref{eq:mixed_inverse_metrics}.

The coordinate invariance of the complete HNS operator of Eq.~\eqref{eq:lhs_compressible_operator} is then tested using the $\mathrm{Ma}=0.5$ case of Sec.~\ref{subsec:validation_lin_compr}. At each resolution, the problem is solved on the Cartesian grid ($\varepsilon=0$) and the distorted grid ($\varepsilon=0.01$). Their common wall points allow direct comparison of the complex wall-pressure coefficients through $\Delta_{p_w,r}=\|\hat{p}_w^{(\varepsilon)}-\hat{p}_w^{(0)}\|_{L_r}/\|\hat{p}_w^{(0)}\|_{L_r}$, $r\in\{2,\infty\}$, with the norms evaluated along the buffer-free wall. Figure~\ref{fig:verif_metric}(c) plots both quantities against $N_\eta$. The $L_\infty$ norm, $\Delta_{p_w,\infty}$, measures the largest difference at any wall location, whereas the $L_2$ norm, $\Delta_{p_w,2}$, measures the accumulated difference along the buffer-free wall. Both decrease by a factor of $70$ over four grids, from $401\times41$ to $1201\times121$, and converge at order $3.98$. The two coordinate representations therefore converge to the same pressure amplitude and phase at the formal order of the scheme.

The coordinate-invariance test is repeated for the $\mathrm{Ma}=4.5$ second-Mack-mode case using Chebyshev collocation in the wall-normal direction. From $801\times101$ to $1201\times145$, the wall-pressure difference decreases by factors of $4.0$ and $4.5$ in the $L_\infty$ and $L_2$ norms, effective orders of $3.9$ and $4.2$. Omitting $\xi_{xy}$ and $\eta_{xy}$ changes the wall pressure six times more than at $\mathrm{Ma}=0.5$. Thus, the mixed metrics become considerably more relevant at high Mach number.

\subsection{High-amplitude Newton continuation}
\label{subsec:verif_hns}

\begin{figure*}[!t]
	\centering
	\includegraphics[width=\textwidth]{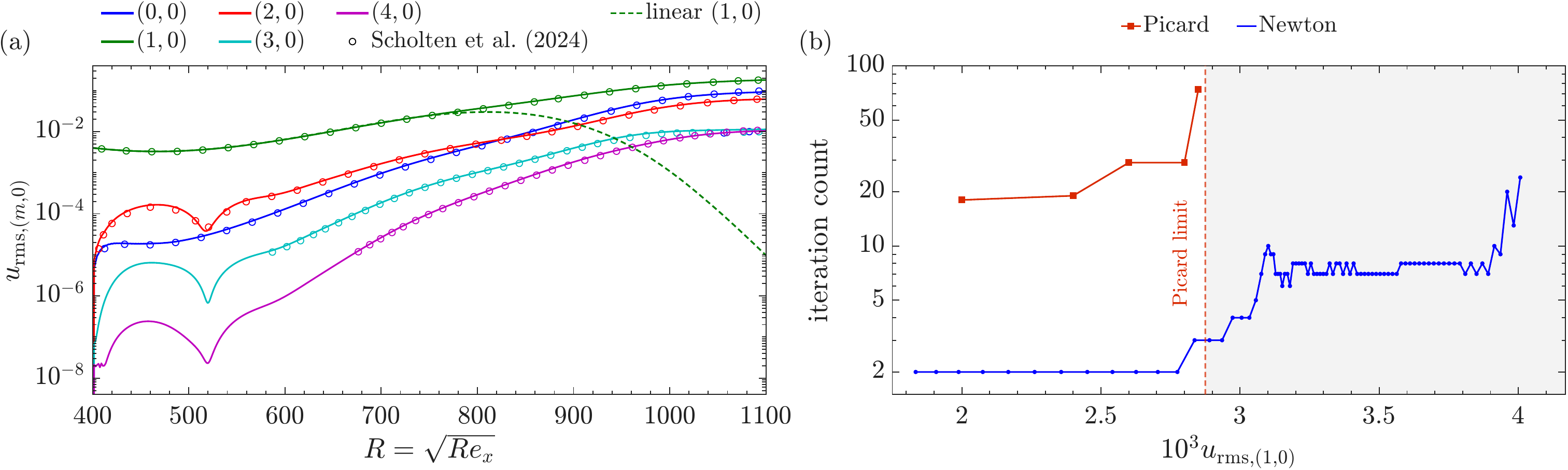}
    \caption{High-amplitude Newton continuation for the Blasius TS case ($M=4$, $N=0$, $N_\xi\times N_\eta=1050\times60$, fourth-order finite differences). (a) Modal amplitudes $u_{\mathrm{rms},(m,0)}$ for the inflow amplitude $u_{\mathrm{rms},(1,0)}=4\times10^{-3}$: DeHNSSo-v2 (solid lines), reference HNS data from \citet{Scholten2024} (circles), and the linear fundamental (dashed line). Colors identify the MFD, the fundamental $(1,0)$, and its harmonics $(2,0)$--$(4,0)$. (b) Picard and Newton corrector iteration counts versus the primary-wave inflow amplitude $u_{\mathrm{rms},(1,0)}$; the vertical dashed line marks the observed Picard limit, with shading beyond it.}
\label{fig:verif_hns}
\end{figure*}

The Newton continuation of Sec.~\ref{subsec:newton} is verified against the independent HNS implementation of \citet{Scholten2024} using the nonlinear Blasius benchmark of Sec.~\ref{subsec:verif_incomp}. Starting from two converged lower-amplitude solutions, the second-order predictor and modified Newton corrector, here with the Jacobian refactorized at every step, advance the disturbance field and inflow amplitude $A_0$ together along the solution branch. Since $A_0$ is the peak amplitude of the primary wave, its r.m.s.\ amplitude at the inflow is $u_{\mathrm{rms},(1,0)}=A_0/\sqrt{2}$, so the reference target $u_{\mathrm{rms},(1,0)}=4\times10^{-3}$ corresponds to $A_0=5.657\times10^{-3}$. This target amplitude lies beyond the convergence limit of the Picard iteration (Fig.~\ref{fig:verif_hns}b).

Figure~\ref{fig:verif_hns}(a) shows close agreement with the reference over the full range of the reference data for the MFD, the fundamental, and its three harmonics on a $1050\times60$ grid with fourth-order finite differences in both directions. Agreement remains very good downstream of $R\approx800$, where the linear fundamental decays, and beyond $R\approx875$, where the NPSE iterations reported by \citet{Scholten2024} fail to converge because of strong nonlinear coupling. DeHNSSo-v2 still converges in this regime, where the fundamental mode reaches $u_{\mathrm{rms},(1,0)}\approx0.18$ at $R=1100$, about $45$ times its inflow amplitude.

Figure~\ref{fig:verif_hns}(b) compares Picard iteration, supplemented by Anderson acceleration \citep{WalkerNi2011}, with Newton continuation along the same branch. Picard uses successive fixed-amplitude solves and converges up to an inflow amplitude $u_{\mathrm{rms},(1,0)}\approx2.9\times10^{-3}$, with its iteration count rising sharply before divergence at the next amplitude level. Chebyshev collocation and finite differences in the wall-normal direction give the same threshold within one amplitude increment. Newton continuation reaches the reference amplitude $u_{\mathrm{rms},(1,0)}=4\times10^{-3}$, approximately $40\%$ higher. The iteration counts indicate convergence behavior rather than computational cost, since the methods differ in convergence criteria and work per iteration (see Sec.~\ref{sec:performance}).

Although Chebyshev collocation closely reproduces the modal evolution with $N_\eta=40$ rather than $N_\eta=60$, it is not the cheaper choice. Its global wall-normal coupling makes the within-station blocks of the Newton Jacobian dense, whereas finite differences couple only nearby points (see Fig.~\ref{fig:lhs_sparsity}). Evaluated at zero disturbance on the $1050\times40$ Chebyshev and $1050\times60$ finite-difference grids, the Jacobian contains an average of $54$ nonzeros per row under Chebyshev collocation, compared with $13$ under finite differences. Consequently, the Chebyshev Jacobian contains $2.8$ times as many nonzeros overall despite having $33\%$ fewer unknowns. This greater density increases the memory and time required whenever the Jacobian is factorized; fourth-order finite differences are therefore the recommended choice for Newton-continuation calculations.

\section{Validation}
\label{sec:validation}

Four cases are used to validate the principal capabilities of DeHNSSo-v2 against independent reference data. The compressible formulation is assessed for linear and nonlinear disturbances in flat-plate boundary layers (Secs.~\ref{subsec:validation_lin_compr} and \ref{subsec:validation_nonlin_compr}), the curvilinear formulation for nonlinear stationary crossflow instability over a swept wing (Sec.~\ref{subsec:validation_nonlin_curved}), and the admittance wall condition for a TS wave interacting with a wall-embedded Helmholtz resonator (Sec.~\ref{subsec:validation_resonator}). Recovery of the incompressible limit is verified separately in Sec.~\ref{subsec:verif_incomp}.

\subsection{Linear compressible flat-plate flow}
\label{subsec:validation_lin_compr}
The linear compressible operator, compressible local-stability inflow, and far-field treatment are validated against the PSE results of \citet{Zanus2021} and an in-house PSE solver validated by \citet{Barahona2026}. Three adiabatic flat-plate cases span the subsonic, supersonic, and hypersonic regimes: $\mathrm{Ma}=0.5$ and $1.6$ at $F=40\times10^{-6}$, and $\mathrm{Ma}=4.5$ at $F=153\times10^{-6}$. Each calculation retains one two-dimensional mode ($M=1$, $N=0$), introduced at $A_0=2\times10^{-6}$ using the compressible local-stability solution of Sec.~\ref{subsec:inflow}. Chebyshev collocation is used on $800\times81$ grids at $\mathrm{Ma}=0.5$ and $1.6$ and an $800\times101$ grid at $\mathrm{Ma}=4.5$, with the density-based far-field condition of Sec.~\ref{subsec:bcs} in all three cases. Complete physical and numerical parameters are listed in Tabs.~\ref{tab:calc_phys} and \ref{tab:calc_num}.

\begin{figure}[!t]
	\centering
	\includegraphics[width=\columnwidth]{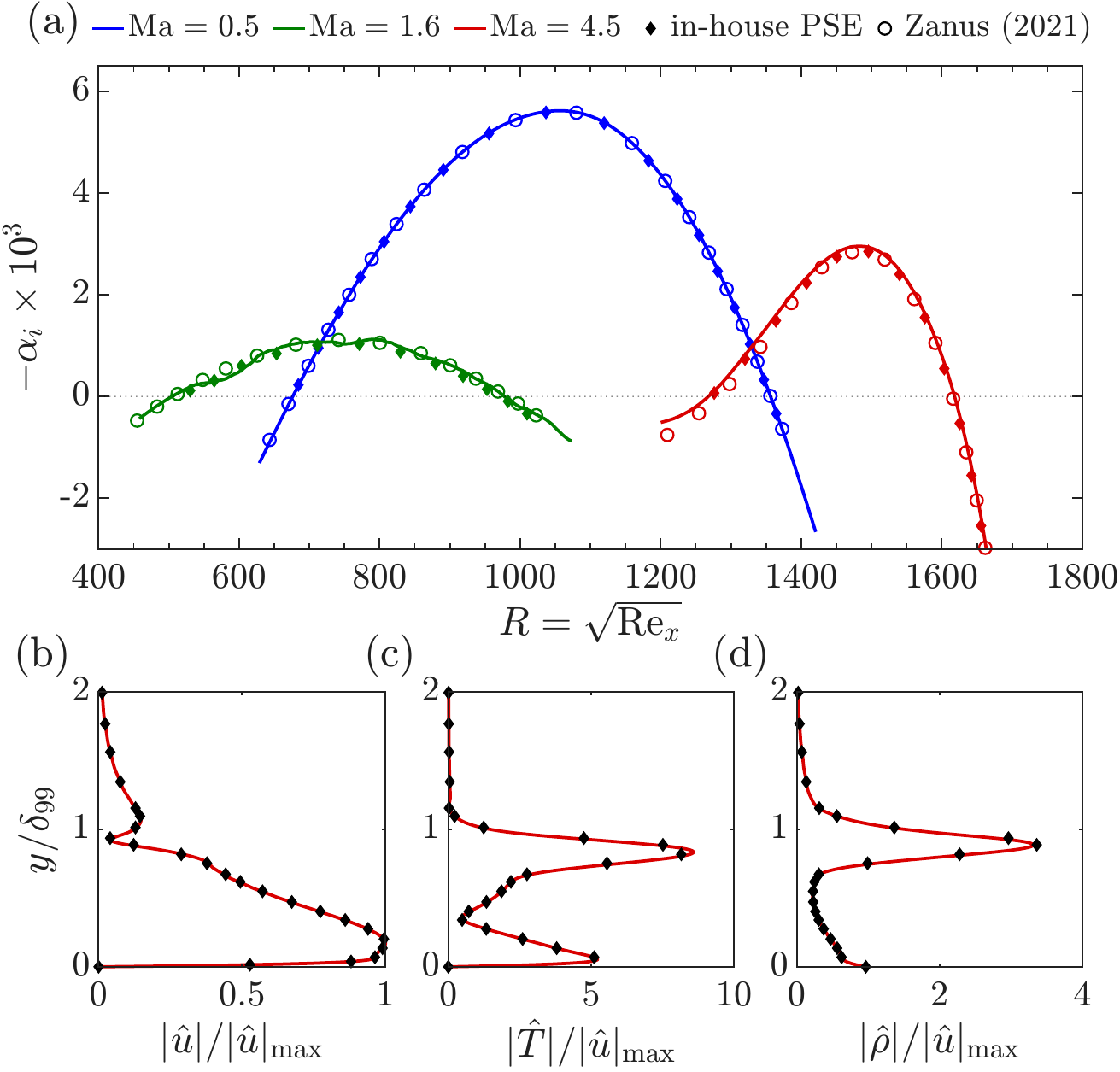}
	\caption{Linear compressible flat-plate validation. (a)~Spatial growth rate $-\alpha_i$ of two-dimensional instability modes, based on $(\rho u)'$ at $\mathrm{Ma}=0.5$ and $1.6$ and on $u'$ at $\mathrm{Ma}=4.5$, versus $R=\sqrt{\mathrm{Re}_x}$ at $\mathrm{Ma}=0.5$ and $1.6$ ($F=40\times10^{-6}$) and $\mathrm{Ma}=4.5$ ($F=153\times10^{-6}$): DeHNSSo-v2 (lines), the in-house PSE solver of \citet{Barahona2026} (filled diamonds), and the PSE data of \citet{Zanus2021} (circles). (b)--(d)~Wall-normal profiles of the streamwise-velocity, temperature, and density perturbations for the second Mack mode at $\mathrm{Ma}=4.5$ and $R\approx1486$, normalized by $|\hat{u}|_{\max}$ and plotted against $y/\delta_{99}$: DeHNSSo-v2 (lines) and results from the in-house PSE solver of \citet{Barahona2026} (black diamonds).}
	\label{fig:val_lin_compr}
\end{figure}

Figure~\ref{fig:val_lin_compr}(a) compares the spatial growth rate of the $(1,0)$ mode with both PSE solutions. Scaled with the local Blasius length, the growth rate is defined as
\begin{equation}
	-\alpha_i=\frac{1}{2\,|q'|_{\max}}\,\frac{\mathrm{d}|q'|_{\max}}{\mathrm{d}R},
	\label{eq:growth_rate}
\end{equation}
where $|q'|_{\max}$ is the wall-normal maximum of a scalar disturbance quantity: as in \citet{Zanus2021}, the streamwise mass flux $(\rho u)'=\bar{\rho}u'+\bar{u}\rho'$ at $\mathrm{Ma}=0.5$ and $1.6$, and the streamwise velocity $u'$ at $\mathrm{Ma}=4.5$. DeHNSSo-v2 closely reproduces the unstable region from branch~I to branch~II at all three Mach numbers. At $\mathrm{Ma}=4.5$ in Figs.~\ref{fig:val_lin_compr}(b)--(d), DeHNSSo-v2 also reproduces the wall-normal structure of the second Mack mode near its maximum growth rate at $R\approx1486$. Since \citet{Zanus2021} did not report these profiles, they are compared with the in-house PSE solver only. These comparisons validate both the streamwise amplification and the wall-normal structure of the compressible linear perturbations.

\subsection{Nonlinear compressible oblique-mode breakdown}
\label{subsec:validation_nonlin_compr}

The compressible nonlinear forcing and its harmonic balancing are validated against the nonlinear PSE calculation of \citet{ChangMalik1994} for oblique-mode breakdown over an adiabatic flat plate at $\mathrm{Ma}=1.6$, $T_\infty^*=\SI{300}{K}$, and $\mathrm{Pr}=0.72$. An oblique pair of first-mode waves $(1,\pm1)$ with equal frequency and opposite spanwise wavenumber is imposed at $R=\sqrt{\mathrm{Re}_x}=750$ with $F=2\times10^{-5}$ and $\beta/R=\pm0.83\times10^{-4}$, corresponding to a wave angle of approximately $60^\circ$, and an inflow amplitude $A_0=8\times10^{-4}$ each. The domain extends to $R=1550$, and the parabolizing buffer of Eq.~\eqref{eq:parabolization} begins at $\xi_b$, corresponding to $R=1500$. The calculation uses Chebyshev collocation on a $1500\times60$ grid, retains $M=N=4$, and uses the Picard iteration of Sec.~\ref{subsec:operator} with $\theta=0.4$ to $\sigma_{m,n}<7\times10^{-4}$. All parameters are listed in Tabs.~\ref{tab:calc_phys} and~\ref{tab:calc_num}.

The pair of oblique modes is generated by the linear PSE inflow of Sec.~\ref{subsec:inflow}, which develops both waves over a short upstream region and thereby avoids the transient that a locally parallel eigenfunction would introduce. The linear compressible operator was verified at these conditions beforehand: the $N$-factors based on the peak mass-flux fluctuation of the oblique first mode with $F=2\times10^{-5}$ and $\beta/R=0.96\times10^{-4}$ and of the two-dimensional mode with $F=4\times10^{-5}$ follow \citet{ChangMalik1994} (not shown). All other modes are set to zero at the inflow, so the MFD and the higher harmonics arise from the nonlinear coupling alone.

\begin{figure}[t]
	\centering
	\includegraphics[width=\columnwidth]{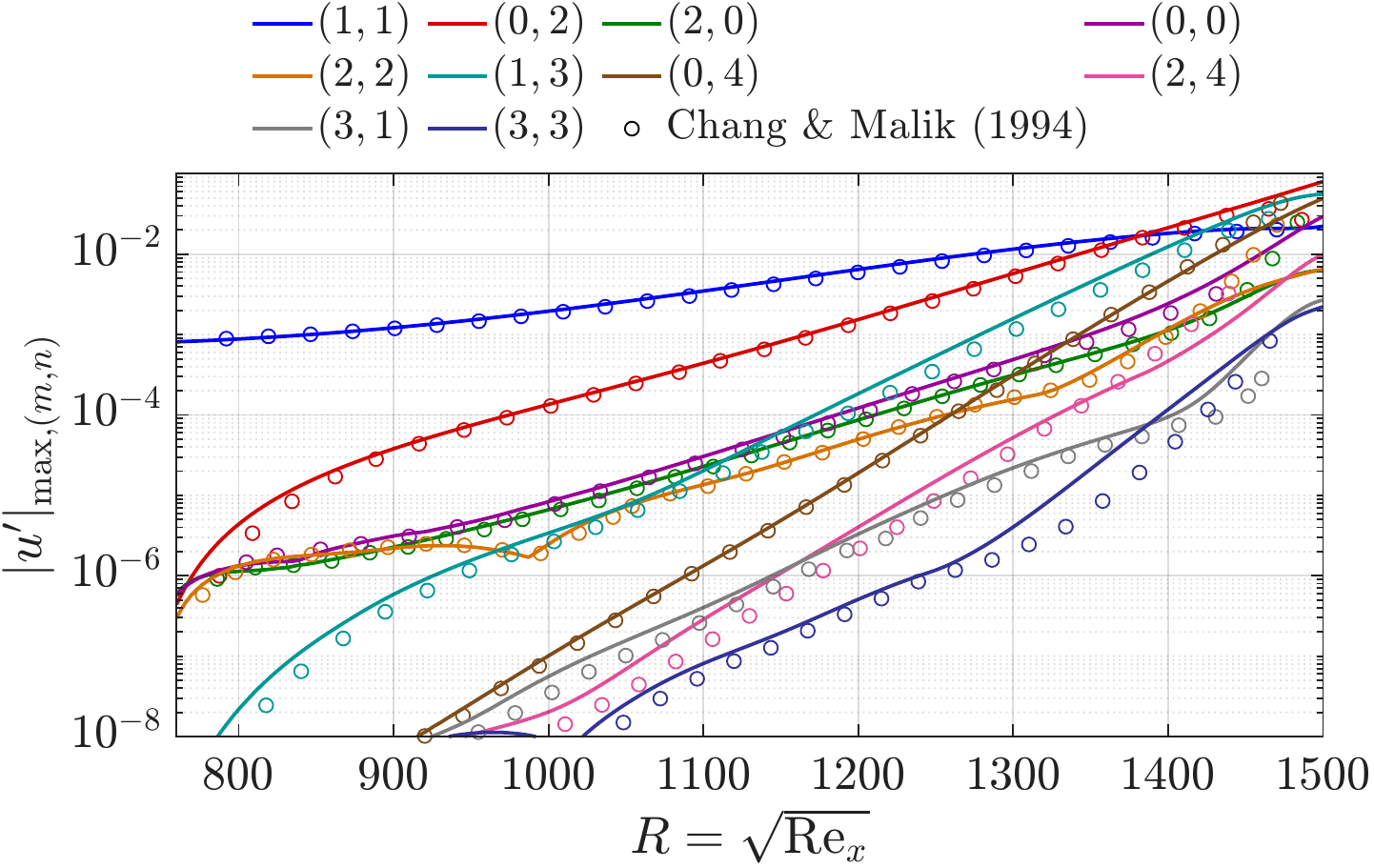}
	\caption{Nonlinear compressible validation for the oblique-mode breakdown of \citet{ChangMalik1994} at $\mathrm{Ma}=1.6$, $F=2\times10^{-5}$, and $A_0=8\times10^{-4}$. Wall-normal maxima of the streamwise-velocity harmonics are shown against $R=\sqrt{\mathrm{Re}_x}$: DeHNSSo-v2 with $M=N=4$ (lines) and the digitized nonlinear PSE results (circles). Modes are identified in the legend.}
	\label{fig:val_nonlin_compr}
\end{figure}

Figure~\ref{fig:val_nonlin_compr} shows the wall-normal maximum of the streamwise-velocity amplitude of every retained harmonic against $R$, compared with \citet{ChangMalik1994}. DeHNSSo-v2 reproduces the staggered harmonic cascade from the growing oblique pair through the MFD and the stationary streamwise streak $(0,2)$ to the onset of saturation. The cascade follows the wave--vortex triad of the reference: the oblique waves grow linearly, while their nonlinear interaction drives the rapid growth of the stationary streak $(0,2)$, and the interaction of the oblique waves with the streak then feeds the strong growth of mode $(1,3)$, which reaches the already saturated streak at the end of the computational domain. The weakest modes, $(3,1)$, $(3,3)$, and $(2,4)$, exceed the reference particularly beyond $R\approx1300$, consistent with the truncation at $M=N=4$ against $M=N=9$ in the reference.

\begin{figure}[t]
  \centering
  \includegraphics[width=\columnwidth]{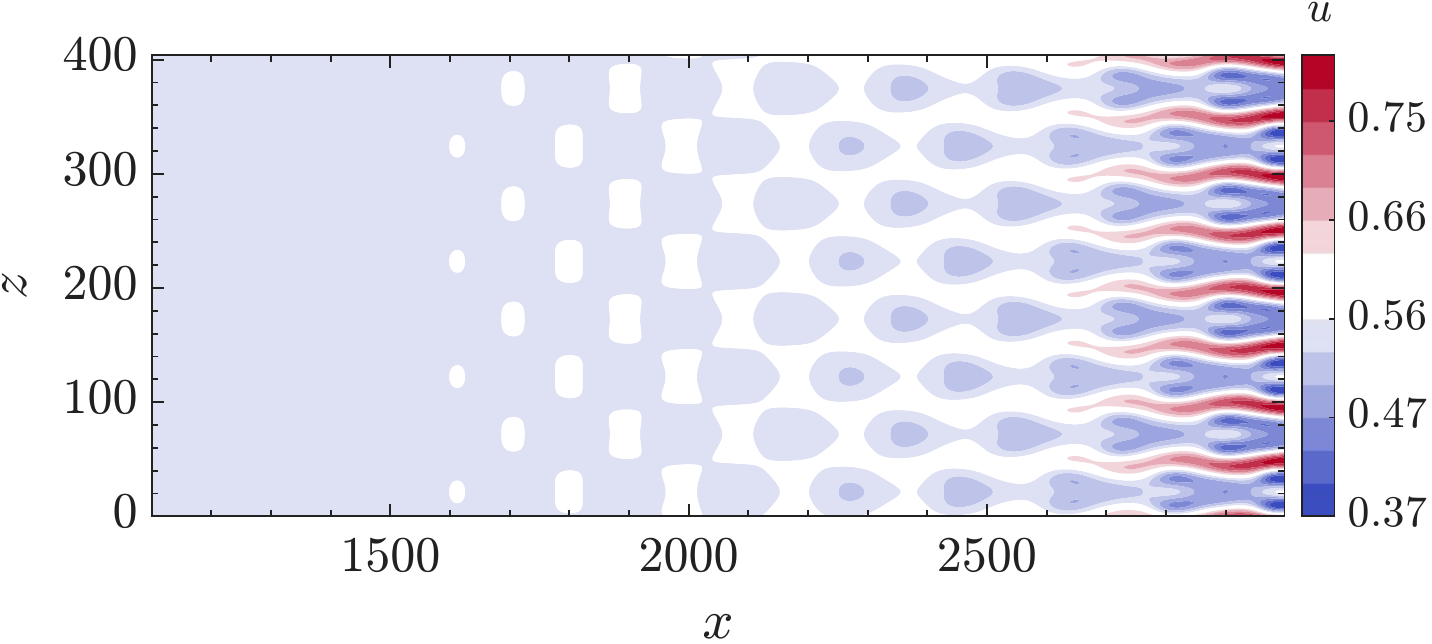}
	\caption{Instantaneous streamwise velocity reconstructed from the DeHNSSo-v2 solution (with $M=N=4$) at the constant scaled height $y/l_x=2.3$, i.e.,~$y=2.3\,R/R_0$, shown over four spanwise wavelengths. Here, $x=R^2/R_0$.}
  \label{fig:val_cm16_contour}
\end{figure}

Figure~\ref{fig:val_cm16_contour} shows the instantaneous streamwise velocity at the constant scaled height $y/l_x=2.3$, with $l_x$ the local Blasius length, i.e.,~at $y=2.3R/R_0$ in inflow units, for qualitative comparison with the corresponding visualization of \citet{ChangMalik1994}. As nonlinear interactions generate the stationary streamwise-streak mode $(0,2)$ and higher harmonics, the staggered modulation sharpens into high- and low-speed regions, with the high-speed streaks splitting and branching into a finger-like pattern. DeHNSSo-v2 recovers this characteristic breakdown structure with $M=N=4$, while the final breakdown near $x\approx3000$ at the onset of the outflow buffer remains sensitive to harmonic truncation, as also noted by \citet{ChangMalik1994}.

\subsection{Nonlinear crossflow instability over a curved swept wing}
\label{subsec:validation_nonlin_curved}

\begin{figure}[!ht]
	\centering
	\includegraphics[width=\columnwidth]{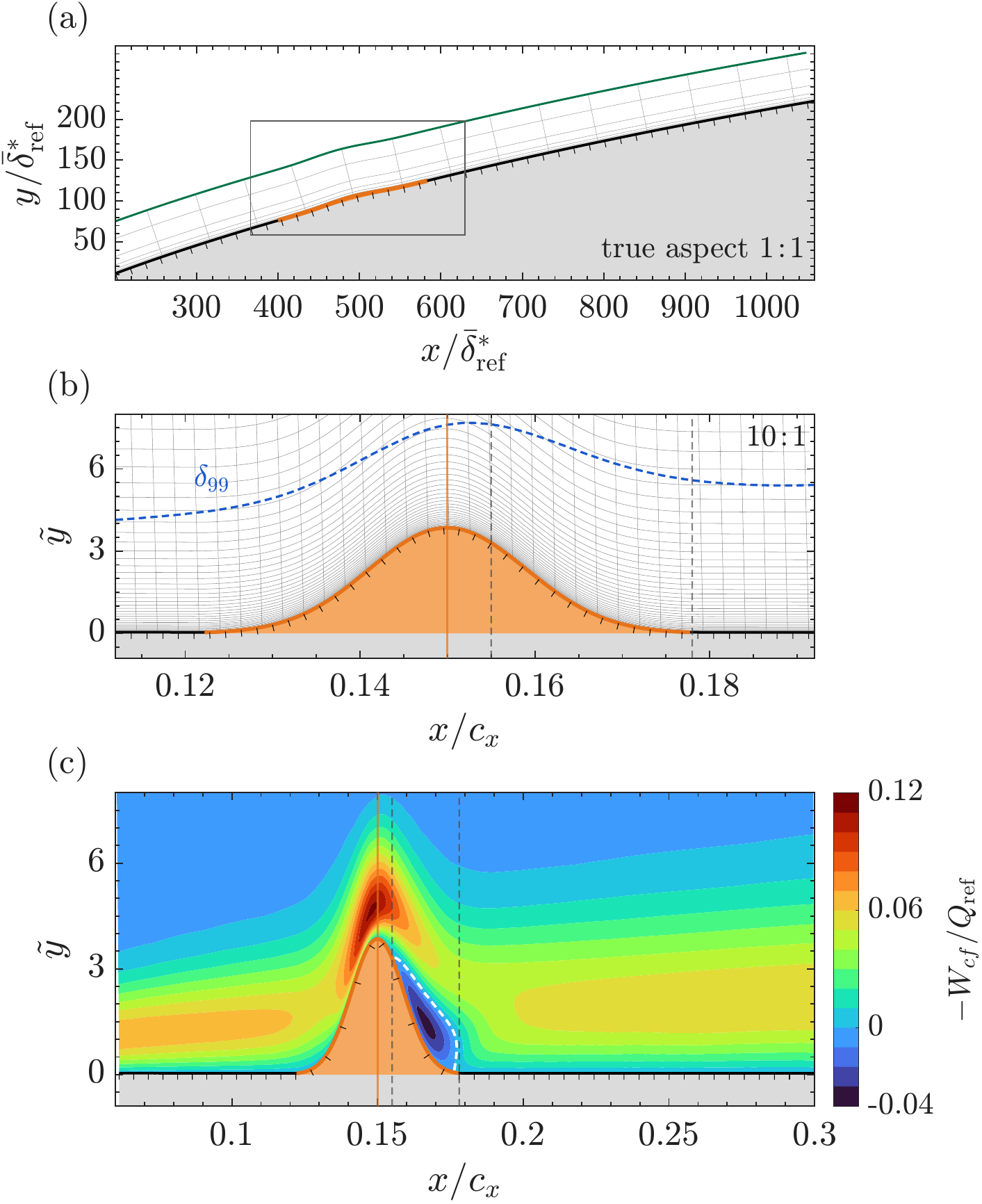}
	\caption{Stability domain, grid, and base flow for the hump configuration A1-H of the M3J swept wing of \citet{Moniripiri2026}. (a)~Body-fitted grid in the leading-edge-normal frame at true aspect ratio, with coordinates normalized by $\bar{\delta}^{*}_{\mathrm{ref}}$. The airfoil surface and hump are shown in black and orange, respectively, and the upper boundary $\eta=60$ in green; the box identifies the region enlarged in~(b). (b)~Near-wall grid in the wall-referenced coordinate $\tilde{y}=(y-y_{w,c})/\bar{\delta}^{*}_{\mathrm{ref}}$, where $y_{w,c}$ is the clean-wing wall at the same $x$, with the indicated vertical exaggeration and $\delta_{99}$ shown in blue. (c)~Normalized crossflow velocity $-W_{cf}/Q_{\mathrm{ref}}$; the white dashed contour denotes $W_{cf}=0$. In~(b,c), the orange line marks the hump apex and the gray dashed lines delimit the crossflow-reversal region. Grid lines are subsampled for clarity.}
	\label{fig:val_cfi_case}
\end{figure}

The curvilinear formulation of Secs.~\ref{subsec:coord_transform}--\ref{subsec:second_order_metrics} and the grid-processing pipeline of Sec.~\ref{sec:gridgen} are validated through a simulation of nonlinear stationary crossflow instability over the M3J swept wing tested in the TU~Delft low-turbulence tunnel \citep{RiusVidales2025Hump}. DeHNSSo-v2 simulations are tested against the DNS of \citet{Moniripiri2026}, itself designed to replicate the TU~Delft experiment. The airfoil is a modified NACA~66018 at $\Lambda = 45^\circ$, $3^\circ$ angle of attack, and $\mathrm{Re}_{c_X} = 2.38\times10^{6}$. The low-amplitude clean (A1-C) and hump (A1-H) configurations are considered; \citet{Moniripiri2026} reported good DNS agreement with the experimentally measured disturbance growth and transition to turbulence. Figure~\ref{fig:val_cfi_case} presents the DeHNSSo-v2 stability grid and imported base flow for the hump configuration.

\begin{figure*}[!t]
	\centering
	\includegraphics[width=\textwidth]{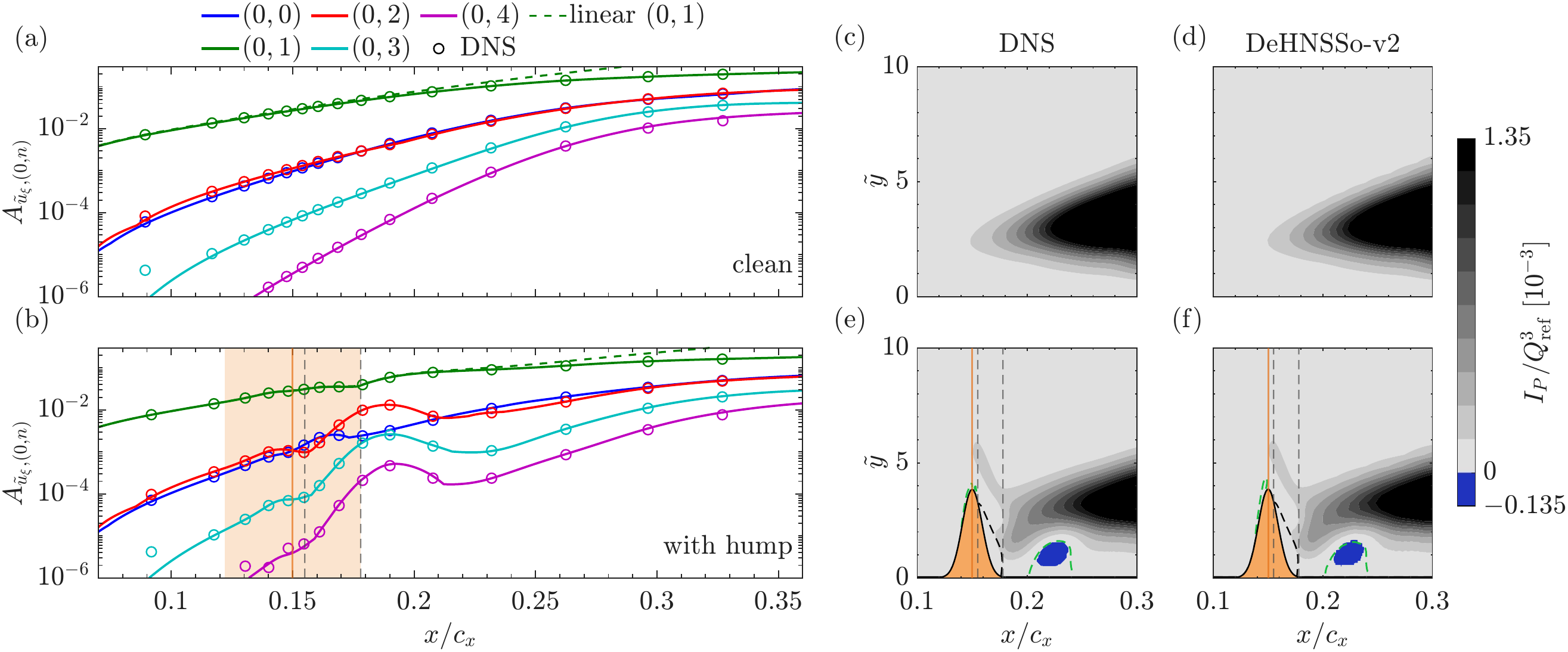}
	\caption{Nonlinear stationary crossflow over the clean (A1-C) and hump (A1-H) configurations of the M3J swept wing of \citet{Moniripiri2026}, retaining $M=0$ and $N=4$. (a)~Clean and (b)~hump configuration: amplitude of the $n$-th spanwise harmonic of the chordwise velocity perturbation $\tilde{u}_\xi$, $A=\max_{\eta}2|\tilde{u}_{\xi,(0,n)}|/Q_{\mathrm{ref}}$, with the factor two omitted for the MFD: DeHNSSo-v2 (lines), reference DNS (circles), and linear fundamental mode (dashed green). Only $(0,1)$ is imposed at the inflow, and the DeHNSSo-v2 curves terminate upstream of the buffer at $x/c_x=0.36$. (c--f)~Reynolds--Orr production integrand $I_P$ of the fundamental mode, defined as in \citet{Moniripiri2026} and normalized by $Q_{\mathrm{ref}}^3$: reference DNS in~(c,e) and DeHNSSo-v2 in~(d,f), for the clean~(c,d) and hump~(e,f) cases. The panels share the clipped color scale shown, with blue denoting negative production; the base-flow gradients are filtered over $51$ stations. In~(b), the shaded band marks the extent of the hump. In~(b,e,f), the orange line marks its apex and the gray dashed lines delimit the crossflow-reversal region; in~(e,f), the green and black dashed contours denote $I_P=0$ and $W_{cf}=0$, respectively.}
	\label{fig:val_cfi_combined}
\end{figure*}
As shown in Fig.~\ref{fig:val_cfi_case}(a), the local airfoil surface is inclined by up to $20^\circ$, so the metric terms of Sec.~\ref{subsec:second_order_metrics} are exercised in full. The spanwise-invariant hump in Fig.~\ref{fig:val_cfi_case}(b) is $\SI{1}{mm}$ high, has a height-to-width ratio of $1{:}50$, and is centered at $x/c_x=0.15$. Its aft slope produces near-wall crossflow reversal over $0.155\leq x/c_x\leq0.178$ (see Fig.~\ref{fig:val_cfi_case}(c)). The wall of the stability grid coincides with the DNS wall to within interpolation error. The wall-normal domain extends to $\eta=20$ on the clean wing and $\eta=60$ on the wing with hump to contain the outer-flow response induced by the hump. Lengths and velocities are scaled by $\bar{\delta}_{\mathrm{ref}}^*=\SI{2.687e-4}{m}$ and $Q_{\mathrm{ref}}=\SI{23.496}{m/s}$ at the DNS domain inflow, $x/c_x=0.055$, which fix the Reynolds number at $\mathrm{Re}_0=Q_{\mathrm{ref}}\bar{\delta}_{\mathrm{ref}}^*/\nu_\infty=451$.

In DeHNSSo-v2, only the stationary fundamental $(0,1)$ mode is imposed at the inflow, with $\beta_1=0.225$ ($\lambda_z=\SI{7.5}{mm}$) and its wall-normal profile prescribed by the DNS. The Newton continuation of Sec.~\ref{subsec:newton} advances its amplitude to $A_0=2.62\times10^{-3}$ in both configurations: the fundamental mode saturates at amplitudes of order $10^{-1}$ with the MFD of nearly the same order, the regime in which the Picard iteration fails to converge. The A1-C and A1-H cases retain $M=0$, $N=4$ on a fourth-order finite-difference grid of $2428\times60$, with $\Delta\xi \approx 0.51\,\bar\delta^{*}_{\mathrm{ref}}$, $56$ streamwise points per wavelength for $(0,1)$, and an amplitude-damping buffer over the final $15\%$ of the streamwise domain. Complete physical and numerical parameters are listed in Tabs.~\ref{tab:calc_phys} and \ref{tab:calc_num}.

Figure~\ref{fig:val_cfi_combined}(a,b) compares the chordwise development of the amplitude of every retained stationary mode, the MFD $(0,0)$ and the harmonics $(0,1)$ to $(0,4)$, with the DNS of \citet{Moniripiri2026}. On A1-C, all harmonics and the MFD follow the reference across the domain. On A1-H, the solver reproduces the fundamental mode stalling in the crossflow-reversal region (see Fig.~\ref{fig:val_cfi_combined}b) while growing less than on A1-C beyond $x/c_x\approx0.2$. The harmonics are significantly destabilized by the crossflow reversal up to $x/c_x\approx0.185$ before decaying. By $x/c_x \approx 0.35$, the fundamental mode reaches $A\approx22\%$ on A1-C and $18\%$ on A1-H, nearly reaching saturation. The main disagreement is near the inflow, particularly for the higher spanwise harmonics $(0,3)$ and $(0,4)$: the DNS superimposes the NPSE-derived disturbances there, whereas DeHNSSo-v2 imposes only $(0,1)$ and generates the harmonics through the nonlinear forcing; the DeHNSSo-v2 harmonics recover the reference values within a few stations downstream.

Figure~\ref{fig:val_cfi_combined}(c--f) indirectly compares the shape of the perturbation rather than its amplitude: the integrand of the Reynolds--Orr production of the fundamental mode (see \citet{Moniripiri2026}), whose negative values indicate the reverse lift-up effect. On A1-C, kinetic energy is transferred from the base flow to the perturbation field. By contrast, a region of negative production appears near the wall downstream of the reversal region in A1-H, responsible for the reduced disturbance growth downstream of the hump. DeHNSSo-v2 accurately reproduces this near-wall region at the same location and with the same extent. 

The A1-C and A1-H DeHNSSo-v2 calculations required \SI{15.2}{\hour} and \SI{11.5}{\hour} on system~E (\ref{app:hardware}), with four and two Jacobian factorizations, respectively. Each calculation used approximately $1.5\times10^{5}$ grid points and required at most $3\times10^{2}$ core hours. For context, the three-dimensional DNS, which also resolved laminar--turbulent transition, used up to $1.0\times10^{8}$ grid points and approximately $4\times10^{5}$ core hours (Moniripiri, private communication, 2026).

\subsection{Tollmien--Schlichting waves over a Helmholtz resonator}
\label{subsec:validation_resonator}

A passive wall-embedded Helmholtz resonator interacting with a linear TS wave is validated following the flow case investigated by \citet{Michelis2023}, whose linearized Navier--Stokes computations were validated against wind-tunnel particle image velocimetry measurements. A linear two-dimensional TS wave develops in a Blasius boundary layer at $\mathrm{Ma} = 0.06$ and $\mathrm{Re}_0 = 371$ ($\bar{U}_{\mathrm{ref}} = \SI{20.9}{m/s}$, $\bar{l}_{\mathrm{ref}} = \SI{2.70e-4}{m}$) and travels over a single-degree-of-freedom Helmholtz resonator whose \SI{1}{mm} neck opens at the wall at $x_r = \SI{0.55}{m}$. Cavity and TS wave are both tuned to the resonance frequency $f_r = \SI{295}{Hz}$ ($F=64.4\times10^{-6}$), where the resonator response is largest. Neither the cavity nor the neck is simulated: the resonator enters the calculation only through the coupled wall admittance condition described below, which replaces the wall condition of Sec.~\ref{subsec:bcs} over the neck. The baseline case with a solid wall is denoted \textrm{C} and the resonator case \textrm{R}. At this Mach number, the compressible and incompressible operators give nearly the same result, with $\Delta N$ differing by $1.4\%$.

\begin{figure}[!ht]
  \centering
  \includegraphics[width=\columnwidth]{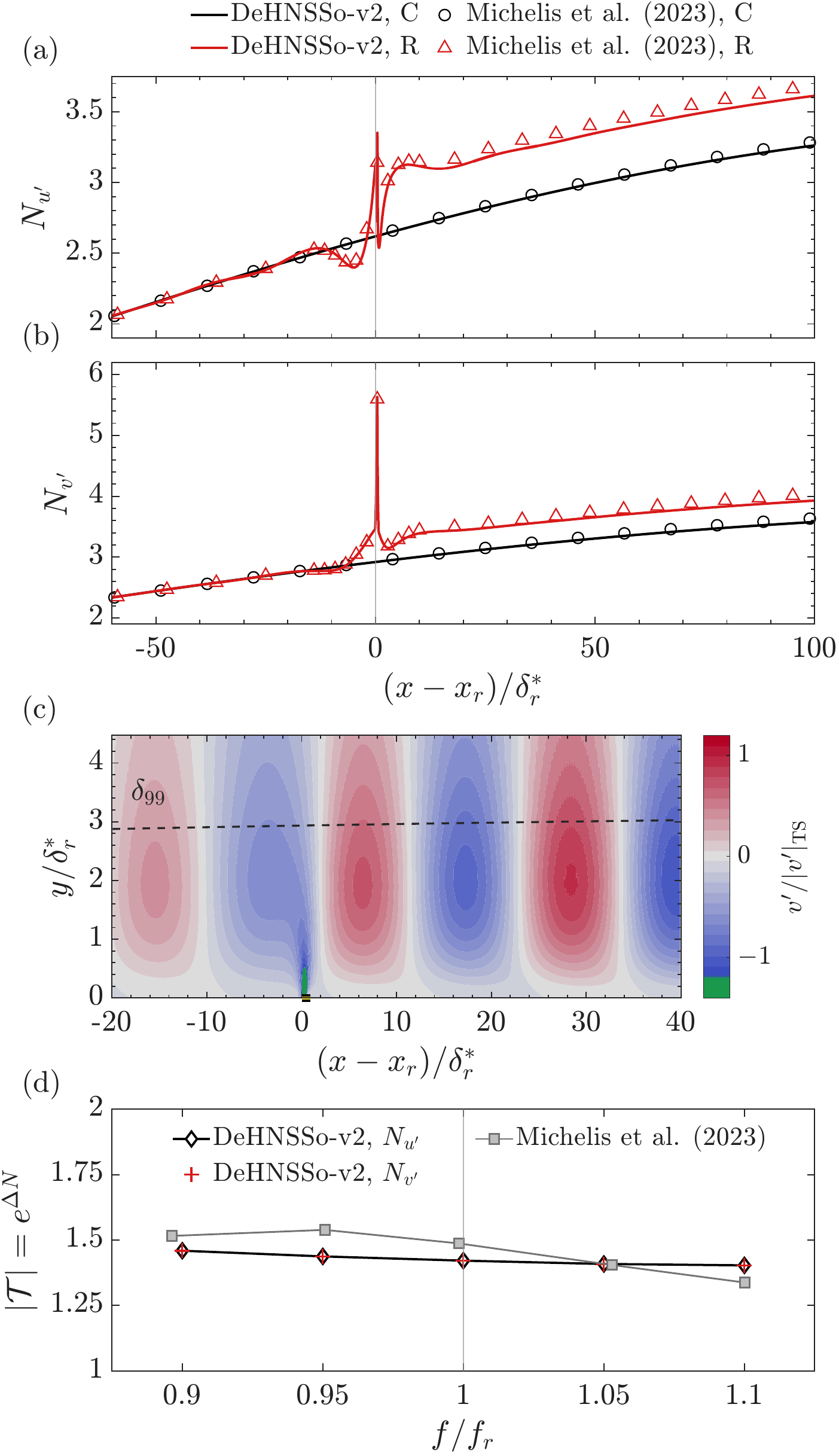}
	\caption{Tollmien--Schlichting wave interacting with a wall-embedded Helmholtz resonator at $\mathrm{Ma}=0.06$, $\mathrm{Re}_0=371$, and $f_{\mathrm{TS}}=f_r=\SI{295}{Hz}$ ($F=64.4\times10^{-6}$). Streamwise $N$-factors based on (a)~the streamwise and (b)~the wall-normal velocity perturbation; (c)~wall-normal velocity perturbation near the neck for the resonator case~\textrm{R}, normalized by the maximum of the TS wave, with amplitudes below $-1$ in green and the baseline $\delta_{99}$ dashed; (d)~transmission coefficient $|\mathcal{T}|=e^{\Delta N}$ across resonance. In~(a,b,d), lines denote DeHNSSo-v2 and symbols the reference results of \citet{Michelis2023}; \textrm{C} and \textrm{R} denote the baseline and resonator cases, respectively, and the two velocity components coincide in~(d). In~(a--c), the abscissa is the distance from the neck, $x-x_r$, scaled by the displacement thickness there, $\delta^{*}_{r}$, and the neck is marked at $x=x_r$; in~(d), it is the TS frequency relative to the resonance frequency.}
  \label{fig:val_resonator}
\end{figure}

Following \citet{Michelis2023}, the resonator is represented by the complex wall admittance $H_{\mathrm{BC}}$, the ratio of wall-normal velocity to pressure at the neck. Since the neck is about twenty-five times narrower than the TS wavelength, it experiences a nearly uniform pressure forcing from the grazing flow. Over the neck, the no-penetration condition $\hat{v}_{m,n}|_{\eta=0}=0$ of Sec.~\ref{subsec:bcs} is therefore replaced by the wall row $\hat{v}_{m,n} - H_{\mathrm{BC}}\,\hat{p}_{m,n}=0$ in the discrete system of Eq.~\eqref{eq:lhs_compressible_operator}, while the no-slip and isothermal conditions are retained. Unlike the prescribed wall forcing of \citet{Westerbeek2024DeHNSSo}, this condition specifies neither variable independently.

Figure~\ref{fig:val_resonator}(a,b) compares the $N$-factors of the streamwise and wall-normal velocity, $N_{u'} = \ln(A_{u'}/A_{u',\mathrm{I}})$ and $N_{v'} = \ln(A_{v'}/A_{v',\mathrm{I}})$, where $A$ is the peak amplitude of the perturbation and $A_{\mathrm{I}}$ its value at branch~I of the baseline case. DeHNSSo-v2 closely reproduces the resonator case~\textrm{R}: both integral amplifications exhibit a local dip followed by a sharp rise across the neck and settle a few displacement thicknesses downstream at $\Delta N\approx0.35$ above the baseline case~\textrm{C}. Their subsequent growth at the baseline rate indicates that the resonator changes the amplitude locally without altering the downstream modal growth. Figure~\ref{fig:val_resonator}(c) shows the wall-normal velocity perturbation of case~\textrm{R} near the neck, where the resonator drives $v'$ to $-6.6$ times the maximum of the TS wave; this peak is confined to the width of the neck and to about half the local displacement thickness $\delta^{*}_{r}$ above the wall.

Figure~\ref{fig:val_resonator}(d) compares the transmission coefficient $|\mathcal{T}| = \exp\{\Delta N\}$ over the investigated frequency range. DeHNSSo-v2 agrees with the reference to within approximately $7\%$ and reproduces the overall decrease with increasing $f_{\mathrm{TS}}/f_r$, although its frequency dependence is weaker. The comparison therefore validates the coupled admittance condition and the resulting resonator response without resolving the internal flow within the resonator, and the same condition applies to other surface metamaterials whose response can be expressed as a wall admittance.

\section{Computational performance}
\label{sec:performance}

The efficiency developments of Sec.~\ref{sec:efficiency} are first evaluated using the nonlinear Tollmien--Schlichting case, verified in Sec.~\ref{sec:verification}. On the same grid and hardware, DeHNSSo-v2 reaches convergence $18.9$ times faster than DeHNSSo-v1. Section~\ref{subsec:perf_profile} investigates the runtime distribution, LHS assembly, and factorization reuse. Section~\ref{subsec:perf_backend} compares the time--memory trade-offs of the direct-solver back-ends, and Sec.~\ref{subsec:perf_scaling_hw} examines scaling with resolution, hardware, and parallel workers. Finally, Sec.~\ref{subsec:perf_newton} assesses the cost of Newton continuation. The hardware is described in \ref{app:hardware}.

\subsection{Linear-algebra performance}
\label{subsec:perf_linalg}

\subsubsection{Runtime analysis}
\label{subsec:perf_profile}

\begin{figure*}[!t]
	\centering
	\includegraphics[width=\textwidth]{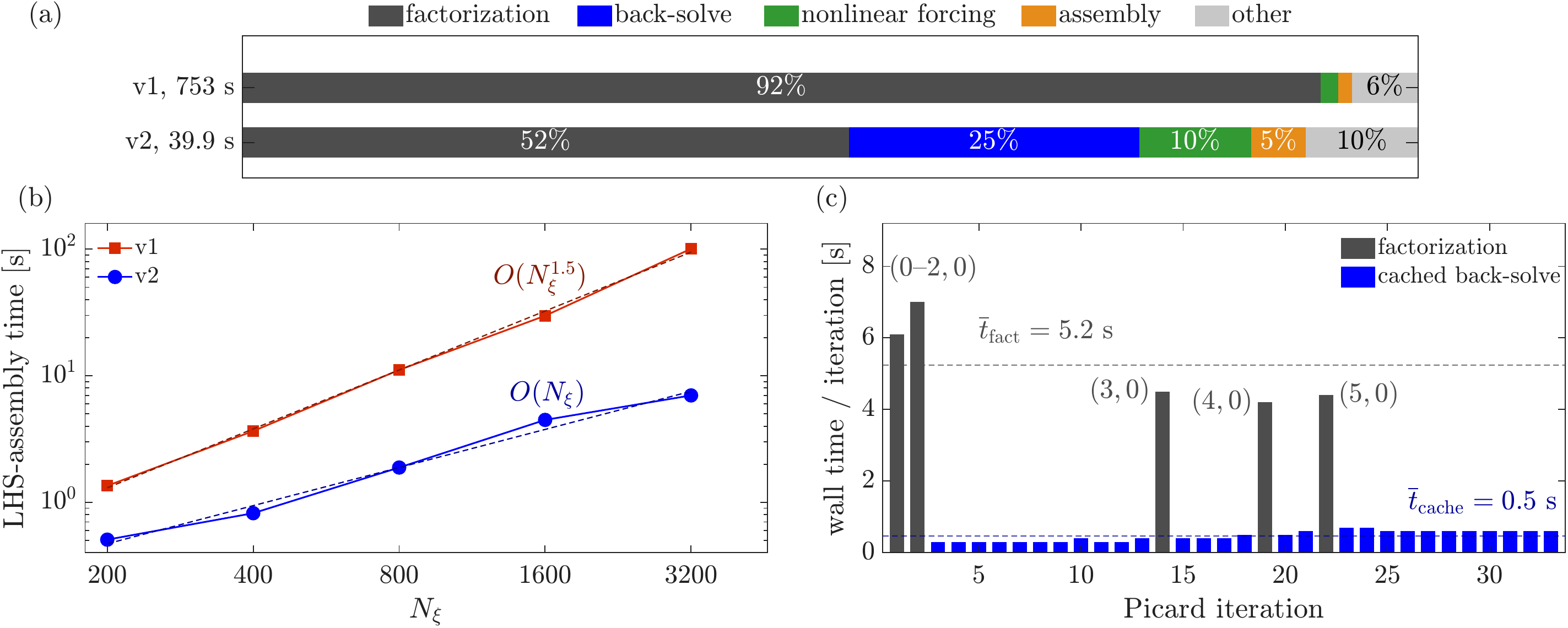}
	\caption{(a)~Share of the wall-clock time per task in DeHNSSo-v1 (v1) \citep{Westerbeek2024DeHNSSo} and DeHNSSo-v2 (v2) for the case of Tab.~\ref{tab:perf_blasius}, with the total wall-clock time of each calculation; in DeHNSSo-v1, factorization and solve form a single operation. (b)~LHS-assembly wall-clock time against the streamwise resolution $N_\xi$ ($N_\eta=40$, $M=1$, system~D); squares, DeHNSSo-v1; circles, DeHNSSo-v2; dashed lines, the fitted references. (c)~Per-iteration wall-clock time of the DeHNSSo-v2 Picard iteration for the case of~(a). Dark bars: iterations at which a mode is factorized as it enters the truncation; blue bars: iterations that reuse the cached factors through back-solves. Each factorization is labeled with the mode $(m,0)$ it factorizes, and the mean factorization and cached-solve times, $\overline{t}_{\rm fact}$ and $\overline{t}_{\rm cache}$, are marked.}
	\label{fig:perf_linalg}
\end{figure*}

Table~\ref{tab:perf_blasius} compares the wall-clock time and memory of DeHNSSo-v1 and DeHNSSo-v2 for the nonlinear TS case, computed with the same setup, harmonic truncation, grid, and hardware. Both solver versions converge the nonlinear terms with the Picard iteration of Sec.~\ref{subsec:operator}, so that the comparison isolates the linear algebra. Figure~\ref{fig:perf_linalg}(a) shows how the runtime of each version is divided among its tasks. In DeHNSSo-v1, the repeated factorize-and-solve operations dominate, as noted in Sec.~\ref{sec:efficiency}. In DeHNSSo-v2, factorization still takes the largest share, $52\%$, but it is performed only once per mode, when the mode enters the truncation, so that factorization takes about \SI{21}{\second}, against \SI{691}{\second} for the combined factorize-and-solve operations of DeHNSSo-v1. The two developments behind this reduction, linear-cost assembly and factorization reuse, are examined in turn.

\begin{table}[!htb]
	\centering
	\small
	\setlength{\tabcolsep}{3pt}
	\caption{Wall-clock time and memory of DeHNSSo-v1 \citep{Westerbeek2024DeHNSSo} and DeHNSSo-v2 for the nonlinear incompressible Blasius TS case ($M=5$, $N_\xi=800$, $N_\eta=40$) on system~D (\ref{app:hardware}), with parallel execution disabled.}
	\label{tab:perf_blasius}
	\begin{tabular}{lccc}
		\toprule
		& \shortstack{DeHNSSo-\\v1} & \shortstack{DeHNSSo-\\v2} & Ratio \\
		\midrule
		\multicolumn{4}{l}{\textit{Wall-clock time}}\\
		LHS assembly              & \SI{8.84}{\second} & \SI{2.20}{\second} & $4.0\times$  \\
		Solve per mode, first     & \SI{5.54}{\second} & \SI{3.60}{\second} & $1.5\times$  \\
		Solve per mode, repeat    & \SI{5.04}{\second} & \SI{0.10}{\second} & $50\times$   \\
		Nonlinear case ($M=5$)    & \SI{753}{\second}  & \SI{39.9}{\second} & $18.9\times$ \\
		\midrule
		\multicolumn{4}{l}{\textit{Memory}}\\
		LHS operator              & ---                & \SI{112}{MB}       & ---          \\
		\quad MFD block           & ---                & \SI{2}{MB}         & ---          \\
		Cached LU factors         & ---                & \SI{6.24}{GB}       & ---          \\
		\bottomrule
	\end{tabular}
\end{table}

The first development is the $O(N_\xi)$ assembly of Sec.~\ref{subsec:sparse_assembly}. Figure~\ref{fig:perf_linalg}(b) isolates the assembly over $N_\xi=200$--$3200$ with $M=1$. The assembly time of DeHNSSo-v2 follows the $O(N_\xi)$ reference, with a slope of $1.0$, whereas that of DeHNSSo-v1 follows $O(N_\xi^{1.5})$, because its sparse-index arrays are extended at every streamwise station. The speed-up therefore grows with resolution and reaches a factor of $14$ at $N_\xi=3200$.

The second development, the factorization reuse of Sec.~\ref{subsec:lu_cache}, provides the larger gain. DeHNSSo-v1 refactorizes every active per-mode system at each nonlinear iteration, whereas DeHNSSo-v2 factorizes each mode once and reduces every later solve to a back-solve with the cached factor, which is $50$ times faster (Tab.~\ref{tab:perf_blasius}). Figure~\ref{fig:perf_linalg}(c) shows when these factorizations occur: in the first two iterations for modes $(0,0)$--$(2,0)$, and later as modes $(3,0)$--$(5,0)$ enter the truncation. Such iterations take on average \SI{5.2}{\second}, whereas iterations that only reuse cached factors take \SI{0.5}{\second}. Together, the two developments make the complete nonlinear calculation $18.9$ times faster (Tab.~\ref{tab:perf_blasius}). For two stationary-crossflow cases of \citet{Westerbeek2024DeHNSSo}, the speed-ups are $9.3$ for nonlinear development and $2.0$ for linear development over a hump, where the single solve precludes factorization reuse.

This gain requires the factors to remain in memory. As listed in Tab.~\ref{tab:perf_blasius}, the operator occupies only \SI{112}{MB}, because its mode-invariant blocks are stored once and the MFD is retained as a \SI{2}{MB} correction, whereas the six cached LU factors occupy \SI{6.2}{GB}. Factorization reuse therefore saves time at the cost of memory, so that storing the factors, rather than computing them, becomes the main constraint.

The cache can also persist between solver calls through \texttt{Opt.rerun} (Sec.~\ref{subsec:lu_cache}). When only the inflow amplitude $A_0$ changes, the calculation of Tab.~\ref{tab:perf_blasius} repeats in \SI{15.3}{\second} instead of \SI{39.9}{\second}, as it skips operator assembly and all six factorizations.

\subsubsection{Direct-solver back-ends}
\label{subsec:perf_backend}

\begin{figure}[!t]
	\centering
	\includegraphics[width=\columnwidth]{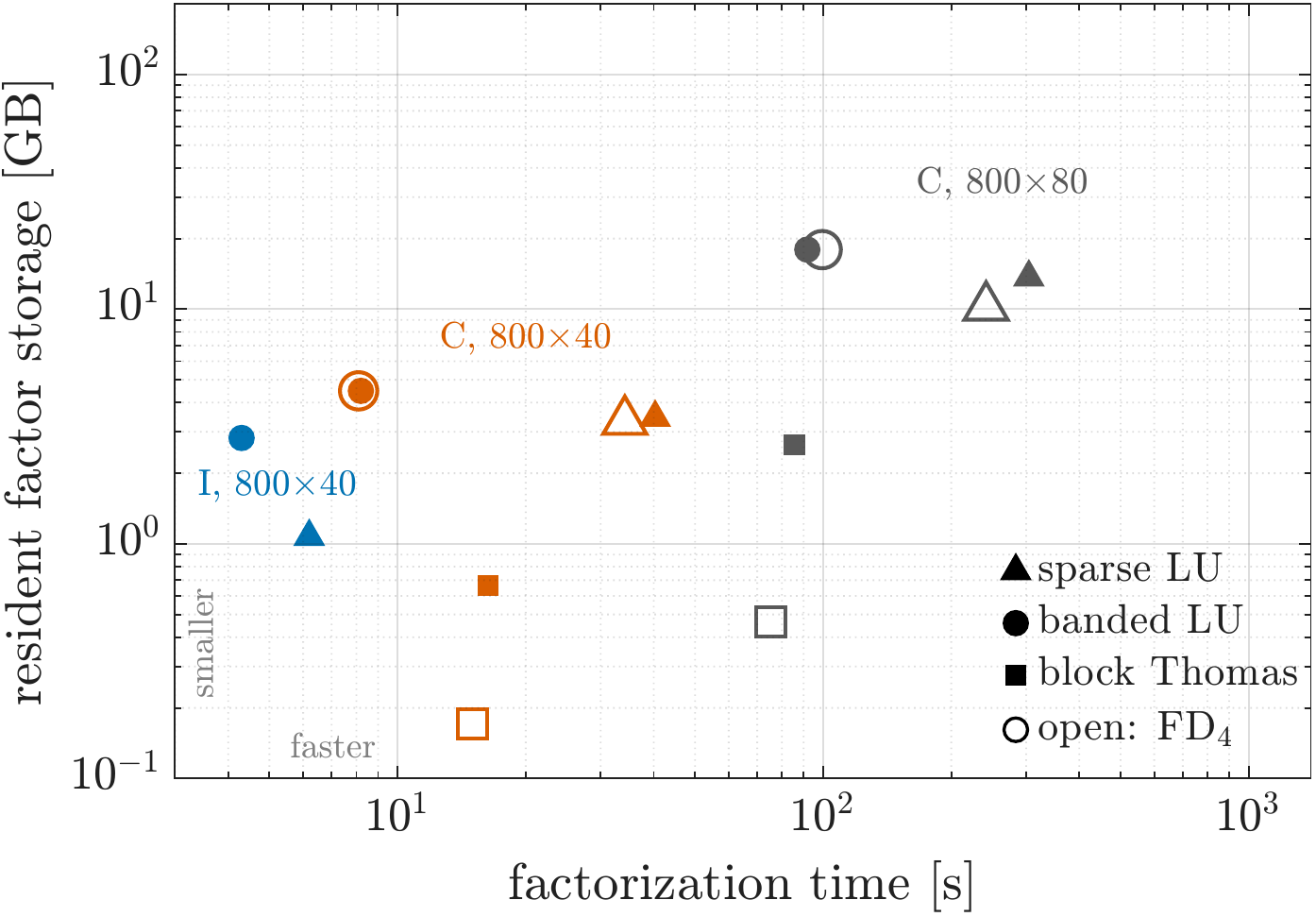}
	\caption{Resident factor storage versus factorization time for the direct-solver back-ends on one thread of system~D (\ref{app:hardware}). Marker shape denotes the back-end, color and label the operator and grid (I, incompressible; C, compressible), and fill the wall-normal discretization (filled, Chebyshev; open, fourth-order finite differences). For block Thomas, only the resident portion is shown; the remainder is streamed to disk.}
	\label{fig:backend_tradeoff}
\end{figure}

Figure~\ref{fig:backend_tradeoff} compares the cached sparse-LU, banded-LU, and block-Thomas back-ends of Secs.~\ref{subsec:lu_cache}--\ref{subsec:block_thomas} by their factorization time and resident factor storage. The baseline case uses the $800\times40$ Chebyshev grid of Sec.~\ref{subsec:perf_profile} and one thread of system~D, together with an $800\times80$ grid for a finer wall-normal resolution. Both the four-variable incompressible operator (I) and the five-variable compressible operator (C) at $\mathrm{Ma}=10^{-3}$ (Sec.~\ref{subsec:verif_incomp}) are included, so that the cost of the additional variable can be measured in the incompressible limit.

On the baseline grid, banded LU exchanges memory for speed (see Sec.~\ref{subsec:lapack_band}). For the incompressible operator, banded LU factorizes $1.4$ times faster than sparse LU, with factors $2.6$ times larger. For the compressible operator, the factorization is $4.9$ times faster and the factors are only $1.3$ times larger. The estimates of Sec.~\ref{subsec:lapack_band} predict that the fifth variable increases the banded storage and factorization time by $(5/4)^2=1.56$ and $(5/4)^3=1.95$, and the measured factors are $1.59$ and $1.91$. The sparse-LU factorization time instead increases by a factor of $6.5$.

Block Thomas, which is available only for the compressible operator and provides an out-of-core alternative (Sec.~\ref{subsec:block_thomas}), keeps a resident factor of approximately one fifth of the sparse-LU factor and one seventh of the banded-LU factor on the baseline grid, with a factorization time between the two.

Fourth-order wall-normal finite differences ($\mathrm{FD}_4$), shown by the open markers, test the dependence on operator sparsity described in Secs.~\ref{subsec:lapack_band} and~\ref{subsec:block_thomas}. They reduce the operator nonzeros by factors of $5$ and $9$ on the $800\times40$ and $800\times80$ grids. The banded factor is unchanged and the sparse-LU savings are limited by fill-in, whereas the block-Thomas resident factor decreases by factors of $4$ and $6$. On the $800\times80$ grid, block Thomas with finite differences is also faster than banded LU while using $40$ times less resident storage.

For the baseline compressible operator, one solve with a retained factor takes \SI{0.18}{\second} with sparse LU, \SI{0.36}{\second} with banded LU, and \SI{1.00}{\second} with out-of-core block Thomas, which does not alter the ordering of the back-ends for the iteration counts considered here. Banded LU is therefore generally fastest when sufficient memory is available, whereas block Thomas is preferable for memory-limited compressible calculations, particularly with finite differences. These measurements support the choice of back-end described in Sec.~\ref{subsec:factorization}.

\subsubsection{Scaling with resolution and hardware}
\label{subsec:perf_scaling_hw}

\begin{figure*}[!t]
	\centering
	\includegraphics[width=\textwidth]{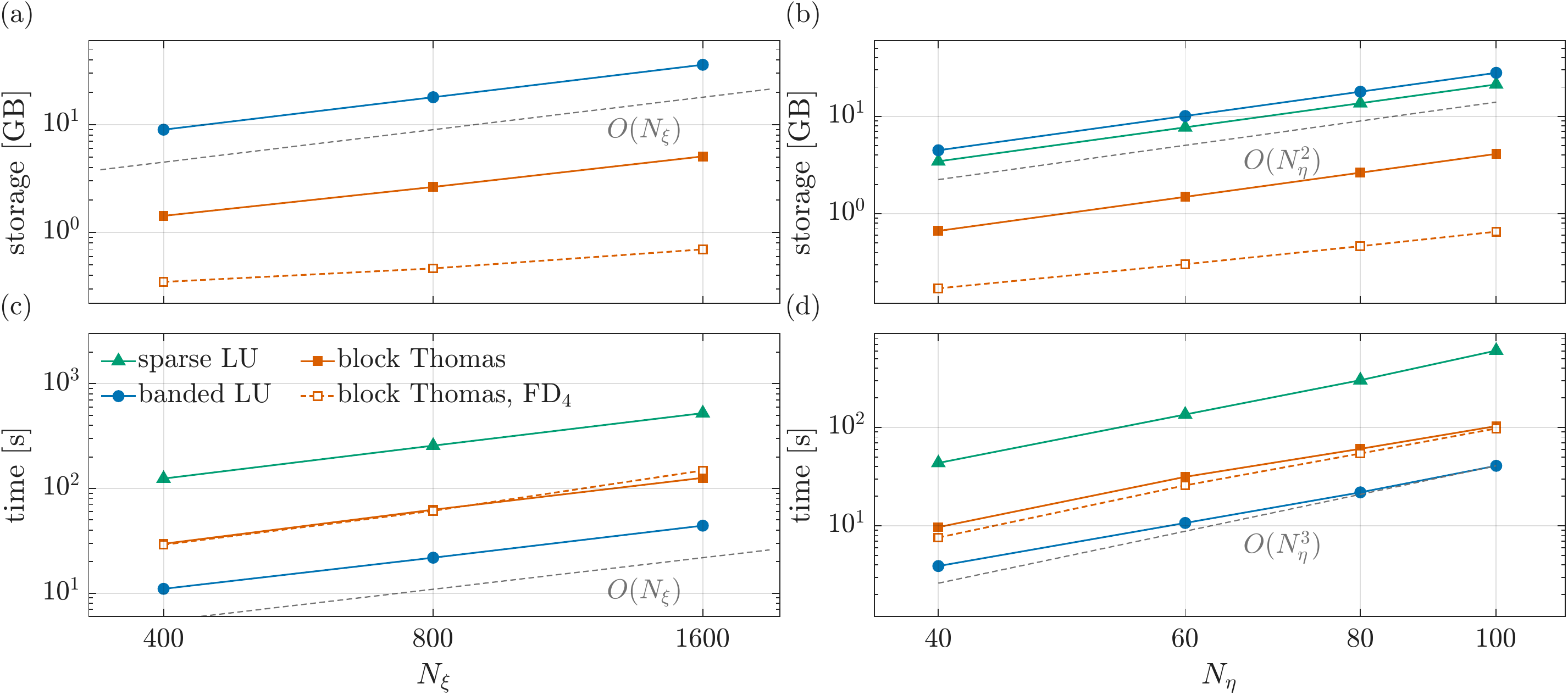}
	\caption{Resolution scaling for the compressible operator at $\mathrm{Ma}=10^{-3}$ using eight threads on system~E (\ref{app:hardware}). Resident factor storage against (a)~$N_\xi$ at $N_\eta=80$ and (b)~$N_\eta$ at $N_\xi=800$, and factorization time against (c)~$N_\xi$ and (d)~$N_\eta$. Sparse LU uses \texttt{seq} in~(a,c), so that no factor is retained and it is omitted from~(a), and \texttt{full} in~(b,d). Filled symbols denote Chebyshev grids and open squares block Thomas with $\mathrm{FD}_4$; for block Thomas, only the resident storage is shown. Gray dashed lines show the predicted scaling.}
	\label{fig:perf_scaling}
\end{figure*}

This subsection tests whether the conclusions of Sec.~\ref{subsec:perf_backend} persist for the larger systems required by practical nonlinear-transition calculations. Unless stated otherwise, all tests use the compressible operator at $\mathrm{Ma}=10^{-3}$ and eight threads of system~E. The three back-ends are compared on Chebyshev grids, and block Thomas is repeated with the $\mathrm{FD}_4$ wall-normal discretization, because its resident storage benefits most from wall-normal sparsity.

Figure~\ref{fig:perf_scaling} examines scaling with both grid directions. In panels~(a,c), increasing $N_\xi$ fourfold increases the resident factor storage of the Chebyshev back-ends by $3.56$--$4.00$ times and their factorization times by $4.0$--$4.3$ times. Both are therefore approximately linear in $N_\xi$, consistent with the fixed streamwise block bandwidth of Secs.~\ref{subsec:lapack_band} and~\ref{subsec:block_thomas}. With increasing $N_\eta$, their factor storage scales approximately as $N_\eta^2$ in panel~(b), while the factorization-time exponents in panel~(d) range from $2.55$ to $2.86$. The banded-LU exponent rises to $2.89$ on one thread, close to the predicted $O(N_\eta^3)$ cost. The ordering of the three Chebyshev back-ends remains unchanged over both sweeps.

In the $\mathrm{FD}_4$ block-Thomas series, shown by open squares, the resident factor remains between \SI{0.17}{GB} and \SI{0.71}{GB}, $3.9$--$7.3$ times smaller than with Chebyshev collocation, because the partition blocks inherit the operator sparsity. It also grows more slowly, approximately as $N_\xi^{0.5}$ and $N_\eta^{1.5}$, whereas its factorization time remains close to that of the Chebyshev block-Thomas series in panels~(c) and~(d).

Hardware affects the back-ends according to their parallelism and memory demand. Sparse LU is effectively serial, whereas banded LU and block Thomas benefit from additional threads, so that at eight threads banded LU is $2.6$ times faster on system~E than on system~D. Near the memory limit of the \SI{32}{GB} system~D, however, a solve with the \SI{17.9}{GB} banded factor takes \SI{7.5}{\second}, against \SI{1.1}{\second} on system~E, so that timings obtained near the memory limit are not transferable between machines.

Mode-level parallelism through \texttt{Opt.parfor} (Sec.~\ref{subsec:lu_cache}) distributes the back-solves over parallel workers. For the incompressible case of Sec.~\ref{subsec:perf_profile} on system~D, six workers accelerate the back-solve loop by $3.0$ times but reduce the total runtime by only $1.22$ times, because the loop accounts for about $29\%$ of the calculation and parallelism saturates at the number of active modes. Factorization reuse, which is also available in serial, therefore provides the larger reduction in runtime.

\subsection{Nonlinear-solver performance}
\label{subsec:perf_newton}

\begin{figure}[t]
    \centering
    \includegraphics[width=\columnwidth]{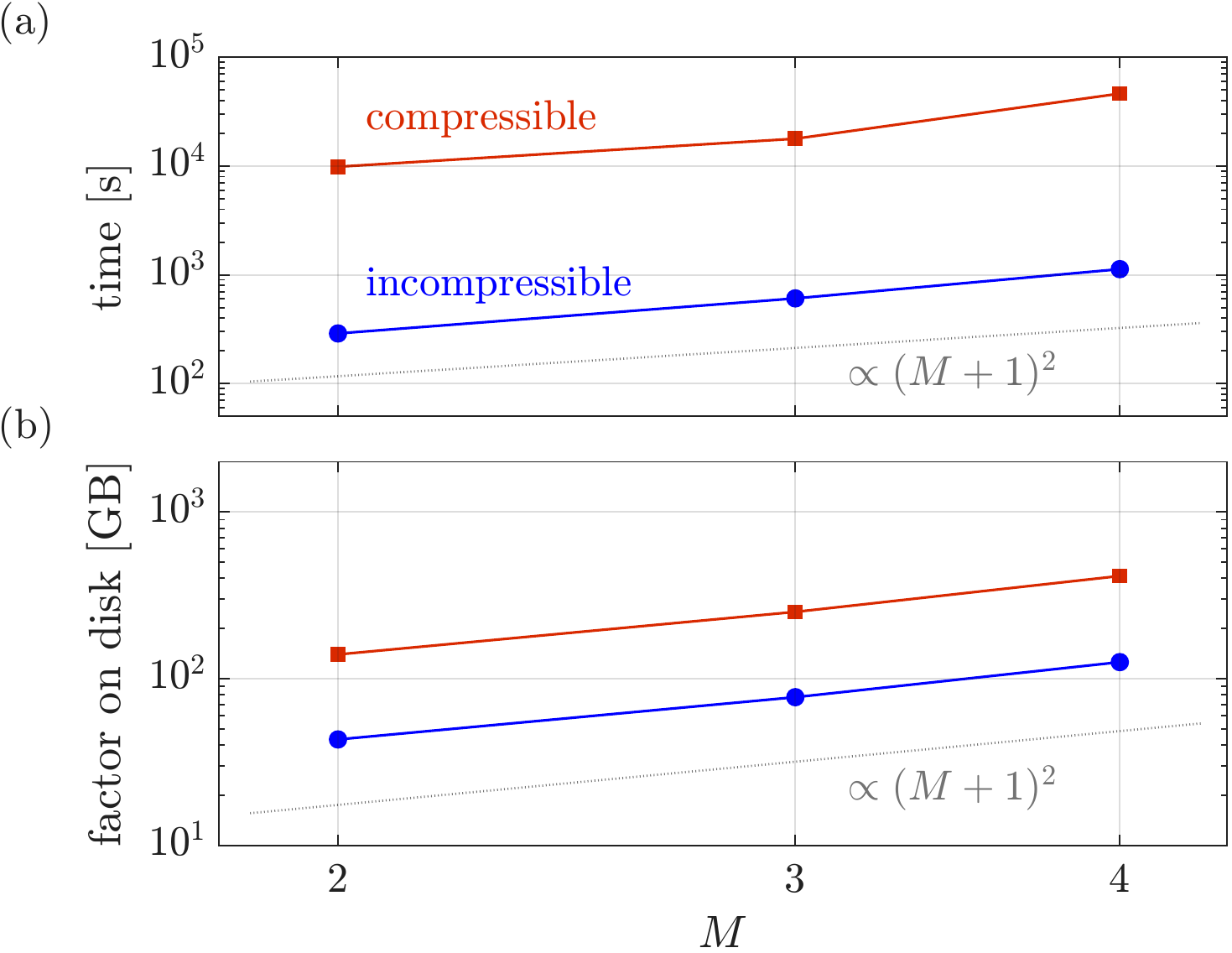}
    \caption{Newton Jacobian (a)~factorization time and (b)~factor size on disk versus harmonic truncation $M$ on a $1050\times60$ grid, for the incompressible and compressible operators. Tests use system~E, single-threaded MUMPS, out-of-core storage, and a \SI{120}{GB} workspace ceiling. Incompressible times are medians of three runs (two at $M=4$), and compressible times are single runs. Dotted lines indicate $(M+1)^2$ scaling.}
    \label{fig:perf_newton_cost}
\end{figure}

The Picard iteration of Sec.~\ref{subsec:operator} solves each harmonic mode separately, whereas Newton continuation includes their nonlinear coupling in the Jacobian of Sec.~\ref{subsec:newton}. This extends the convergence range, as verified in Sec.~\ref{subsec:verif_hns}, at the cost of a larger factorization. For the Blasius TS continuation of Fig.~\ref{fig:verif_hns} ($M=4$, $1050\times60$ grid), in which the Jacobian is refactorized at every continuation step, factorization accounts for approximately $70\%$ of the step time, and the $85$ steps from low amplitude to the target $u_{\mathrm{rms},(1,0)}=4\times10^{-3}$ require \SI{37}{\hour}. At the target amplitude, the out-of-core factor occupies \SI{126}{GB} on disk with \SI{8}{GB} resident.

Figure~\ref{fig:perf_newton_cost} shows how this cost grows with the harmonic truncation for both operators on the same $1050\times60$ grid. The Jacobians use modes $(0,0)$--$(M,0)$ of the converged $M=4$ solution at the target amplitude, with a nonzero temperature perturbation added at $\mathrm{Ma}=10^{-3}$ for the compressible operator. Over $M=2$--$4$, the factor sizes approximately follow the $(M+1)^2$ guide in panel~(b), whereas the factorization times in panel~(a) grow faster. The fifth variable is much more expensive here than in the per-mode systems of Sec.~\ref{subsec:perf_backend}, with $3.2$--$3.3$ times the factor storage and $29$--$42$ times the factorization time. Since the in-core and out-of-core factorizations of the compressible $M=2$ Jacobian take the same time to within $0.2\%$, this cost lies in the arithmetic rather than in disk access.

The modified Newton corrector of Sec.~\ref{subsec:newton} avoids repeated factorizations by retaining the Jacobian factor across continuation steps. Over the first ten steps of the Blasius continuation, taken with identical arclength increments, reusing the initial factor throughout reduces the runtime from \SI{299}{\minute}, with a factorization at every step, to \SI{57}{\minute}, a factor of $5.2$; the default reuse limit of eight steps gives \SI{82}{\minute}. With reuse, the corrector count rises from two to five iterations per step, whereas it remains at two without reuse. The final inflow amplitudes $u_{\mathrm{rms},(1,0)}$ of the runs differ by $1.5\%$, as each corrector lands at a slightly different point on the same branch.

The swept-wing cases of Sec.~\ref{subsec:validation_nonlin_curved} show the cost of repeated solves on a larger grid. The $N=4$ continuations on the $2428\times60$ grids require four and two factorizations over $20$ and $19$ steps, in \SI{15.2}{\hour} and \SI{11.5}{\hour}. Each corrector iteration takes approximately \SI{6.5}{\minute}, mostly for the out-of-core solve, and over ten to thirteen consecutive steps on the same factor the corrector count increases from one or two to ten. Factorizations therefore account for only $27\%$ and $17\%$ of the runtime. Holding a factor longer saves factorizations but adds corrector iterations.

\section{Conclusions}
\label{sec:conclusions}

We have presented DeHNSSo-v2, a new version of the open-source Delft Harmonic Navier--Stokes Solver \citep{Westerbeek2024DeHNSSo}. The present release combines major updates to the governing formulation, grid preprocessing, nonlinear solution, and linear algebra. These additions extend the scope of DeHNSSo-v1 to general two-dimensional body-fitted geometries, compressible flows, and higher disturbance amplitudes, targeting boundary-layer stability and flow-control problems in which nonparallel development, nonlinear interactions, or localized geometry require global streamwise coupling.

A general curvilinear formulation and an automatic grid-generation and preprocessing pipeline replace the problem-specific grid construction of DeHNSSo-v1 and support convex and concave wall geometries. A compressible formulation in the primitive perturbations $(u',v',w',\rho',T')$, including the required thermal and far-field boundary conditions, extends the operating range beyond incompressible flow. Finally, $O(N_\xi)$ sparse assembly, replacing the empirical $N_\xi^{1.5}$ scaling of DeHNSSo-v1, reusable per-mode factorizations, and complementary direct-solver back-ends reduce computational and memory requirements, while Newton pseudo-arclength continuation extends the reachable disturbance amplitude beyond the convergence range of the Picard iteration.

The curvilinear formulation preserves the fourth-order accuracy of the discretization on strongly distorted grids, and the compressible solver recovers the established incompressible solution in the low-Mach limit. Published comparisons span linear compressible instabilities up to $\mathrm{Ma}=4.5$, oblique-mode breakdown at $\mathrm{Ma}=1.6$, nonlinear stationary crossflow over clean and humped swept wings, and the interaction of a Tollmien--Schlichting wave with a wall-embedded Helmholtz resonator. On the matched nonlinear Blasius benchmark, the computational developments reduce the runtime from $\SI{753}{\second}$ to $\SI{39.9}{\second}$, a factor of $18.9$. The measured costs follow the scaling estimates of Sec.~\ref{sec:efficiency}, including their growth with the number of variables: banded LU is fastest when sufficient memory is available, whereas block Thomas with finite differences minimizes resident storage and supports out-of-core execution. Newton continuation extends the Picard limit by approximately $40\%$, reproduces the independent HNS solution of \citet{Scholten2024}, including the approximately $45$-fold amplification of the fundamental Tollmien--Schlichting wave, and provides the nonlinear crossflow solutions compared with direct numerical simulation. The coupled Newton Jacobian, particularly in compressible flows, remains the principal cost.

DeHNSSo-v2 provides a flexible software base for further development. Its present scope remains restricted to steady, spanwise-invariant base flows and, in the compressible formulation, to a calorically perfect ideal gas at constant Prandtl number. Future developments will target: (i) discrete adjoint and resolvent operators for sensitivity and receptivity analysis \citep{LuchiniBottaro2014}, (ii) an HNS--NPSE--LPSE coupling for efficient treatment of nonparallel effects \citep{Herbert1997}, (iii) nonideal and high-temperature gas models \citep{Boldini2025CUBENS,DiRenzo2020HTR}, and (iv) adapting scalable, memory-efficient iterative methods to three-dimensional HNS \citep{Cook2026}.

The open-source version of DeHNSSo-v2 is available at the following link: \url{https://github.com/pcboldini/DeHNSSo}.


\section*{Declaration of competing interest}
The authors declare that they have no known competing financial interests or personal relationships that could have appeared to influence the work reported in this paper.

\section*{Acknowledgements}
The European Research Council (grant no.~ERC-2023-CoG-101125132, MetaWing) has supported this study. P.~C.~Boldini acknowledges Dr.~M.~Moniripiri (KTH Royal Institute of Technology) for providing the swept-wing DNS data. The authors thank the Flow Stability and Flow Control group at TU Delft for testing and debugging DeHNSSo-v2.

\appendix

\section{Compressible LHS coefficient matrices}
\label{sec:app_compressible_matrices}

The $5\times 5$ coefficient matrices of the compressible operator~\eqref{eq:lhs_compressible_operator} are listed below. Rows correspond to the streamwise, wall-normal, and spanwise momentum equations, the continuity equation, and the energy equation; columns to the state components $(u',v',w',\rho',T')$. Derivatives of the perturbation are written in the computational coordinates, the inverse-metric factors of the chain rules~\eqref{eq:chain_first} and~\eqref{eq:chain_second_xx}--\eqref{eq:chain_second_xy} are absorbed into the coefficients through the building blocks defined below. Derivatives of the base flow are precomputed physical quantities and are therefore written with physical derivatives. The convention follows \citet{Scholten2024}. Only nonzero entries are listed.

\paragraph{Notation} The building blocks that collect the inverse-metric factors are defined as follows. The curvilinear advection velocities are
\begin{equation}
  \mathcal{U}_\xi \equiv \bar{u}\,\xi_x + \bar{v}\,\xi_y, \qquad
  \mathcal{U}_\eta \equiv \bar{u}\,\eta_x + \bar{v}\,\eta_y,
  \label{eq:app_Ucurv}
\end{equation}
and the metric weights that distribute a viscous Laplacian $a\,\partial^2_{xx} + b\,\partial^2_{yy}$ over the computational derivatives,
\begin{equation}
\begin{aligned}
  \Lambda^{\xi\xi}_{a,b} &= a\,\xi_x^2 + b\,\xi_y^2, &\quad
  \Lambda^{\eta\eta}_{a,b} &= a\,\eta_x^2 + b\,\eta_y^2, \\
  \Lambda^{\xi\eta}_{a,b} &= 2\,(a\,\xi_x\eta_x + b\,\xi_y\eta_y), &\quad
  \Lambda^{\xi}_{a,b} &= a\,\xi_{xx} + b\,\xi_{yy}, \\
  \Lambda^{\eta}_{a,b} &= a\,\eta_{xx} + b\,\eta_{yy}, &&
\end{aligned}
\label{eq:app_Lambda}
\end{equation}
such that $a\,\partial^2_{xx} + b\,\partial^2_{yy} = \Lambda^{\xi\xi}_{a,b}\,\partial^2_{\xi\xi} + \Lambda^{\eta\eta}_{a,b}\,\partial^2_{\eta\eta} + \Lambda^{\xi\eta}_{a,b}\,\partial^2_{\xi\eta} + \Lambda^{\xi}_{a,b}\,\partial_\xi + \Lambda^{\eta}_{a,b}\,\partial_\eta$. A single physical first derivative likewise splits as $\partial_x = \xi_x\partial_\xi + \eta_x\partial_\eta$ and $\partial_y = \xi_y\partial_\xi + \eta_y\partial_\eta$. The base-flow strain-rate components (without the $\bar{\mu}$ prefactor) are
\begin{align}
  \bar{S}_{xx} &= \tfrac{4}{3}\partial_x\bar{u} - \tfrac{2}{3}\partial_y\bar{v}, &
  \bar{S}_{yy} &= \tfrac{4}{3}\partial_y\bar{v} - \tfrac{2}{3}\partial_x\bar{u}, \nonumber \\
  \bar{S}_{xy} &= \partial_y\bar{u} + \partial_x\bar{v}, &
  \bar{S}_{zz} &= -\tfrac{2}{3}(\partial_x\bar{u} + \partial_y\bar{v}), \nonumber \\
  \bar{S}_{xz} &= \partial_x\bar{w}, &
  \bar{S}_{yz} &= \partial_y\bar{w}. \label{eq:app_strain}
\end{align}
Base-flow viscous Laplacians acting on $\bar{u}$, $\bar{v}$, $\bar{w}$, and $\bar{T}$ are:
\begin{align}
  \bar{\mathcal{L}}_u &= \tfrac{4}{3}\partial^2_{xx}\bar{u} + \partial^2_{yy}\bar{u} + \tfrac{1}{3}\partial^2_{xy}\bar{v}, \nonumber \\
  \bar{\mathcal{L}}_v &= \partial^2_{xx}\bar{v} + \tfrac{4}{3}\partial^2_{yy}\bar{v} + \tfrac{1}{3}\partial^2_{xy}\bar{u}, \nonumber \\
  \bar{\mathcal{L}}_w &= \partial^2_{xx}\bar{w} + \partial^2_{yy}\bar{w}, \nonumber \\
  \bar{\mathcal{L}}_T &= -\tfrac{\gamma\bar{\mu}_T}{\mathrm{Pr}\,\mathrm{Re}}\!\left(\partial^2_{xx}\bar{T} + \partial^2_{yy}\bar{T}\right) \nonumber \\
                &\quad - \tfrac{\gamma\bar{\mu}_{TT}}{\mathrm{Pr}\,\mathrm{Re}}\,|\nabla\bar{T}|^2. \label{eq:app_laplacians}
\end{align}
Base-flow normalized dissipation $\bar{\Phi}_0 \equiv \bar{\Phi}/\bar{\mu}$:
\begin{equation}
  \bar{\Phi}_0 = \bar{S}_{xx}\,\partial_x\bar{u} + \bar{S}_{yy}\,\partial_y\bar{v} + \bar{S}_{xy}^{\,2} + \bar{S}_{xz}\,\partial_x\bar{w} + \bar{S}_{yz}\,\partial_y\bar{w}.
  \label{eq:app_phib}
\end{equation}
The base-flow advective operator and divergence:
\begin{equation}
  \bar{\mathbf{u}}\!\cdot\!\nabla\bar{q}_i \equiv \bar{u}\,\partial_x\bar{q}_i + \bar{v}\,\partial_y\bar{q}_i, \qquad
  \bar{D}_{\mathbf{u}} \equiv \partial_x\bar{u} + \partial_y\bar{v}.
  \label{eq:app_advdiv}
\end{equation}
Dissipation prefactor:
\begin{equation}
  c_{\Phi} \equiv \frac{2\gamma(\gamma-1)\mathrm{Ma}^2}{\mathrm{Re}}.
  \label{eq:app_cphi}
\end{equation}
The shorthand $\bar{\kappa} = \bar{\mu}$, $\bar{\kappa}_T = \bar{\mu}_T$, $\bar{\kappa}_{TT} = \bar{\mu}_{TT}$ reflects the constant-Prandtl-number assumption of Sec.~\ref{subsec:eos}.

\paragraph{Viscous-diffusion matrices $\mathbf{F},\mathbf{C},\mathbf{E}$ (coefficients of $\partial^2_{\xi\xi},\partial^2_{\eta\eta},\partial^2_{\xi\eta}\,\hat{\mathbf{q}}$)} Writing $\mathbf{M}$ for any of $\mathbf{F},\mathbf{C},\mathbf{E}$ and $\Lambda^{\bullet}$ for the corresponding weight ($\Lambda^{\xi\xi}$ for $\mathbf{F}$, $\Lambda^{\eta\eta}$ for $\mathbf{C}$, $\Lambda^{\xi\eta}$ for $\mathbf{E}$), the diagonal viscous entries are
\begin{equation}
\begin{aligned}
  \mathbf{M}_{1,1}&=-\tfrac{\bar{\mu}}{\mathrm{Re}}\Lambda^{\bullet}_{4/3,1},\quad
  \mathbf{M}_{2,2}=-\tfrac{\bar{\mu}}{\mathrm{Re}}\Lambda^{\bullet}_{1,4/3},\\
  \mathbf{M}_{3,3}&=-\tfrac{\bar{\mu}}{\mathrm{Re}}\Lambda^{\bullet}_{1,1},\quad
  \mathbf{M}_{5,5}=-\tfrac{\gamma\bar{\mu}}{\mathrm{Pr}\,\mathrm{Re}}\Lambda^{\bullet}_{1,1},
\end{aligned}
  \label{eq:app_FCE}
\end{equation}
together with the Stokes cross-couplings
\begin{equation}
\begin{aligned}
  \mathbf{F}_{1,2}=\mathbf{F}_{2,1}&=-\tfrac{\bar{\mu}}{3\mathrm{Re}}\xi_x\xi_y,\\
  \mathbf{C}_{1,2}=\mathbf{C}_{2,1}&=-\tfrac{\bar{\mu}}{3\mathrm{Re}}\eta_x\eta_y,\\
  \mathbf{E}_{1,2}=\mathbf{E}_{2,1}&=-\tfrac{\bar{\mu}}{3\mathrm{Re}}(\xi_x\eta_y+\xi_y\eta_x).
\end{aligned}
\end{equation}

\paragraph{Matrices $\mathbf{D}$ and $\mathbf{B}$: coefficients of $\partial_\xi\hat{\mathbf{q}}$ and $\partial_\eta\hat{\mathbf{q}}$ (advection, pressure gradient, viscous-shear couplings)} Each splits as $\mathbf{D}=\mathbf{D}^{(0)}+\mathrm{i}\beta\,\mathbf{D}^{(\beta)}$ and $\mathbf{B}=\mathbf{B}^{(0)}+\mathrm{i}\beta\,\mathbf{B}^{(\beta)}$. The $\beta$-independent part of $\mathbf{D}$ has nonzero entries
\begin{align}
  \mathbf{D}^{(0)}_{1,1} &= \bar{\rho}\,\mathcal{U}_\xi - \tfrac{\bar{\mu}}{\mathrm{Re}}\Lambda^{\xi}_{4/3,1} - \tfrac{\bar{\mu}_T}{\mathrm{Re}}\!\left(\tfrac{4}{3}\xi_x\partial_x\bar{T} + \xi_y\partial_y\bar{T}\right), \nonumber \\
  \mathbf{D}^{(0)}_{1,2} &= -\tfrac{\bar{\mu}}{3\mathrm{Re}}\xi_{xy} - \tfrac{\bar{\mu}_T}{\mathrm{Re}}\!\left(\xi_x\partial_y\bar{T} - \tfrac{2}{3}\xi_y\partial_x\bar{T}\right), \nonumber \\
  \mathbf{D}^{(0)}_{1,4} &= \tfrac{\bar{T}}{\gamma\,\mathrm{Ma}^2}\,\xi_x, \nonumber \\
  \mathbf{D}^{(0)}_{1,5} &= \tfrac{\bar{\rho}}{\gamma\,\mathrm{Ma}^2}\,\xi_x - \tfrac{\bar{\mu}_T}{\mathrm{Re}}\!\left(\bar{S}_{xx}\xi_x + \bar{S}_{xy}\xi_y\right), \nonumber \\
  \mathbf{D}^{(0)}_{2,1} &= -\tfrac{\bar{\mu}}{3\mathrm{Re}}\xi_{xy} + \tfrac{\bar{\mu}_T}{\mathrm{Re}}\!\left(\tfrac{2}{3}\xi_x\partial_y\bar{T} - \xi_y\partial_x\bar{T}\right), \nonumber \\
  \mathbf{D}^{(0)}_{2,2} &= \bar{\rho}\,\mathcal{U}_\xi - \tfrac{\bar{\mu}}{\mathrm{Re}}\Lambda^{\xi}_{1,4/3} - \tfrac{\bar{\mu}_T}{\mathrm{Re}}\!\left(\xi_x\partial_x\bar{T} + \tfrac{4}{3}\xi_y\partial_y\bar{T}\right), \nonumber \\
  \mathbf{D}^{(0)}_{2,4} &= \tfrac{\bar{T}}{\gamma\,\mathrm{Ma}^2}\,\xi_y, \nonumber \\
  \mathbf{D}^{(0)}_{2,5} &= \tfrac{\bar{\rho}}{\gamma\,\mathrm{Ma}^2}\,\xi_y - \tfrac{\bar{\mu}_T}{\mathrm{Re}}\!\left(\bar{S}_{xy}\xi_x + \bar{S}_{yy}\xi_y\right), \nonumber \\
  \mathbf{D}^{(0)}_{3,3} &= \bar{\rho}\,\mathcal{U}_\xi - \tfrac{\bar{\mu}}{\mathrm{Re}}\Lambda^{\xi}_{1,1} - \tfrac{\bar{\mu}_T}{\mathrm{Re}}\!\left(\xi_x\partial_x\bar{T} + \xi_y\partial_y\bar{T}\right), \nonumber \\
  \mathbf{D}^{(0)}_{3,5} &= -\tfrac{\bar{\mu}_T}{\mathrm{Re}}\!\left(\xi_x\partial_x\bar{w} + \xi_y\partial_y\bar{w}\right), \nonumber \\
  \mathbf{D}^{(0)}_{4,1} &= \bar{\rho}\,\xi_x, \quad
  \mathbf{D}^{(0)}_{4,2} = \bar{\rho}\,\xi_y, \quad
  \mathbf{D}^{(0)}_{4,4} = \mathcal{U}_\xi, \nonumber \\
  \mathbf{D}^{(0)}_{5,1} &= (\gamma-1)\bar{\rho}\bar{T}\,\xi_x - c_{\Phi}\,\bar{\mu}\!\left(\bar{S}_{xx}\xi_x + \bar{S}_{xy}\xi_y\right), \nonumber \\
  \mathbf{D}^{(0)}_{5,2} &= (\gamma-1)\bar{\rho}\bar{T}\,\xi_y - c_{\Phi}\,\bar{\mu}\!\left(\bar{S}_{xy}\xi_x + \bar{S}_{yy}\xi_y\right), \nonumber \\
  \mathbf{D}^{(0)}_{5,3} &= -c_{\Phi}\,\bar{\mu}\!\left(\xi_x\partial_x\bar{w} + \xi_y\partial_y\bar{w}\right), \nonumber \\
  \mathbf{D}^{(0)}_{5,5} &= \bar{\rho}\,\mathcal{U}_\xi - \tfrac{\gamma\bar{\mu}}{\mathrm{Pr}\,\mathrm{Re}}\Lambda^{\xi}_{1,1} - \tfrac{2\gamma\bar{\mu}_T}{\mathrm{Pr}\,\mathrm{Re}}\!\left(\xi_x\partial_x\bar{T} + \xi_y\partial_y\bar{T}\right). \label{eq:app_D0}
\end{align}
The $\mathrm{i}\beta$-weighted streamwise-spanwise cross-viscous block has nonzero entries
\begin{equation}
  \mathbf{D}^{(\beta)}_{1,3} = \mathbf{D}^{(\beta)}_{3,1} = -\tfrac{\bar{\mu}}{3\mathrm{Re}}\,\xi_x, \qquad
  \mathbf{D}^{(\beta)}_{2,3} = \mathbf{D}^{(\beta)}_{3,2} = -\tfrac{\bar{\mu}}{3\mathrm{Re}}\,\xi_y.
  \label{eq:app_Dbeta}
\end{equation}
The matrix $\mathbf{B}$ ($\mathbf{B}^{(0)}$ and $\mathbf{B}^{(\beta)}$) has the same structure as $\mathbf{D}$, obtained by replacing every $\xi$-metric with its $\eta$-counterpart, i.e.~$\mathcal{U}_\xi\!\to\!\mathcal{U}_\eta$, $\xi_x\!\to\!\eta_x$, $\xi_y\!\to\!\eta_y$, $\xi_{xy}\!\to\!\eta_{xy}$, and $\Lambda^{\xi}_{a,b}\!\to\!\Lambda^{\eta}_{a,b}$ throughout (the $\eta$-projection of the same physical coefficients in place of the $\xi$-projection); the base-flow gradients and strains are unchanged.

\paragraph{Matrix $\mathbf{A}$: coefficient of $\hat{\mathbf{q}}$ (algebraic block)} The algebraic block lumps the no-derivative contributions of the linearized perturbation equations together with the $(-\mathrm{i}\omega)$ time-derivative, the $\mathrm{i}\beta$ spanwise terms, and the $\beta^2$ spanwise viscous diffusion. We write
\begin{equation}
  \mathbf{A} = \mathbf{A}^{(0)} + (-\mathrm{i}\omega)\,\mathbf{A}^{(\omega)} + \mathrm{i}\beta\,\mathbf{A}^{(\beta)} + \beta^2\,\mathbf{A}^{(\beta\beta)}.
  \label{eq:app_A_decomp}
\end{equation}
The time-derivative mass matrix is
\begin{equation}
  \mathbf{A}^{(\omega)} = \mathrm{diag}\!\left(\bar{\rho},\,\bar{\rho},\,\bar{\rho},\,1,\,\bar{\rho}\right).
  \label{eq:app_Aomega}
\end{equation}
The $\beta^2$ spanwise viscous block is
\begin{equation}
  \mathbf{A}^{(\beta\beta)} = \mathrm{diag}\!\left(\tfrac{\bar{\mu}}{\mathrm{Re}},\,\tfrac{\bar{\mu}}{\mathrm{Re}},\,\tfrac{4\bar{\mu}}{3\mathrm{Re}},\,0,\,\tfrac{\gamma\bar{\mu}}{\mathrm{Pr}\,\mathrm{Re}}\right).
  \label{eq:app_Abetabeta}
\end{equation}
The $\mathrm{i}\beta$ spanwise-coupling block $\mathbf{A}^{(\beta)}$ has nonzero entries
\begingroup\allowdisplaybreaks
\begin{align}
  \mathbf{A}^{(\beta)}_{1,1} &= \bar{\rho}\bar{w}, &
  \mathbf{A}^{(\beta)}_{1,3} &= \tfrac{2\bar{\mu}_T\,\partial_x\bar{T}}{3\mathrm{Re}}, \nonumber \\
  \mathbf{A}^{(\beta)}_{1,5} &= -\tfrac{\bar{\mu}_T\,\partial_x\bar{w}}{\mathrm{Re}}, &
  \mathbf{A}^{(\beta)}_{2,2} &= \bar{\rho}\bar{w}, \nonumber \\
  \mathbf{A}^{(\beta)}_{2,3} &= \tfrac{2\bar{\mu}_T\,\partial_y\bar{T}}{3\mathrm{Re}}, &
  \mathbf{A}^{(\beta)}_{2,5} &= -\tfrac{\bar{\mu}_T\,\partial_y\bar{w}}{\mathrm{Re}}, \nonumber \\
  \mathbf{A}^{(\beta)}_{3,1} &= -\tfrac{\bar{\mu}_T\,\partial_x\bar{T}}{\mathrm{Re}}, &
  \mathbf{A}^{(\beta)}_{3,2} &= -\tfrac{\bar{\mu}_T\,\partial_y\bar{T}}{\mathrm{Re}}, \nonumber \\
  \mathbf{A}^{(\beta)}_{3,3} &= \bar{\rho}\bar{w}, &
  \mathbf{A}^{(\beta)}_{3,4} &= \tfrac{\bar{T}}{\gamma\,\mathrm{Ma}^2}, \nonumber \\
  \mathbf{A}^{(\beta)}_{3,5} &= \tfrac{\bar{\rho}}{\gamma\,\mathrm{Ma}^2} + \tfrac{2\bar{\mu}_T\,\bar{D}_{\mathbf{u}}}{3\mathrm{Re}}, &
  \mathbf{A}^{(\beta)}_{4,3} &= \bar{\rho}, \nonumber \\
  \mathbf{A}^{(\beta)}_{4,4} &= \bar{w}, &
  \mathbf{A}^{(\beta)}_{5,1} &= -c_{\Phi}\,\bar{\mu}\,\bar{S}_{xz}, \nonumber \\
  \mathbf{A}^{(\beta)}_{5,2} &= -c_{\Phi}\,\bar{\mu}\,\bar{S}_{yz}, &
  \mathbf{A}^{(\beta)}_{5,3} &= (\gamma-1)\bar{\rho}\bar{T} - c_{\Phi}\,\bar{\mu}\,\bar{S}_{zz}, \nonumber \\
  \mathbf{A}^{(\beta)}_{5,5} &= \bar{\rho}\bar{w}. & \label{eq:app_Abeta}
\end{align}
\endgroup
The $\beta$-independent algebraic block $\mathbf{A}^{(0)}$ collects the linearized advective, pressure-gradient, viscous-shear, and dissipation contributions. Its nonzero entries are
\begingroup\allowdisplaybreaks
\begin{align}
  \mathbf{A}^{(0)}_{1,1} &= \bar{\rho}\,\partial_x\bar{u}, \qquad \mathbf{A}^{(0)}_{1,2} = \bar{\rho}\,\partial_y\bar{u}, \nonumber \\
  \mathbf{A}^{(0)}_{1,4} &= \bar{\mathbf{u}}\!\cdot\!\nabla\bar{u} + \tfrac{\partial_x\bar{T}}{\gamma\,\mathrm{Ma}^2}, \nonumber \\
  \mathbf{A}^{(0)}_{1,5} &= \tfrac{\partial_x\bar{\rho}}{\gamma\,\mathrm{Ma}^2} - \tfrac{\bar{\mu}_T\,\bar{\mathcal{L}}_u}{\mathrm{Re}} \nonumber \\
                         &\quad - \tfrac{\bar{\mu}_{TT}}{\mathrm{Re}}\!\left(\bar{S}_{xx}\,\partial_x\bar{T} + \bar{S}_{xy}\,\partial_y\bar{T}\right), \nonumber \\
  \mathbf{A}^{(0)}_{2,1} &= \bar{\rho}\,\partial_x\bar{v}, \qquad \mathbf{A}^{(0)}_{2,2} = \bar{\rho}\,\partial_y\bar{v}, \nonumber \\
  \mathbf{A}^{(0)}_{2,4} &= \bar{\mathbf{u}}\!\cdot\!\nabla\bar{v} + \tfrac{\partial_y\bar{T}}{\gamma\,\mathrm{Ma}^2}, \nonumber \\
  \mathbf{A}^{(0)}_{2,5} &= \tfrac{\partial_y\bar{\rho}}{\gamma\,\mathrm{Ma}^2} - \tfrac{\bar{\mu}_T\,\bar{\mathcal{L}}_v}{\mathrm{Re}} \nonumber \\
                         &\quad - \tfrac{\bar{\mu}_{TT}}{\mathrm{Re}}\!\left(\bar{S}_{yy}\,\partial_y\bar{T} + \bar{S}_{xy}\,\partial_x\bar{T}\right), \nonumber \\
  \mathbf{A}^{(0)}_{3,1} &= \bar{\rho}\,\partial_x\bar{w}, \qquad \mathbf{A}^{(0)}_{3,2} = \bar{\rho}\,\partial_y\bar{w}, \nonumber \\
  \mathbf{A}^{(0)}_{3,4} &= \bar{\mathbf{u}}\!\cdot\!\nabla\bar{w}, \nonumber \\
  \mathbf{A}^{(0)}_{3,5} &= -\tfrac{\bar{\mu}_T\,\bar{\mathcal{L}}_w}{\mathrm{Re}} \nonumber \\
                         &\quad - \tfrac{\bar{\mu}_{TT}}{\mathrm{Re}}\!\left(\partial_y\bar{w}\,\partial_y\bar{T} + \partial_x\bar{w}\,\partial_x\bar{T}\right), \nonumber \\                      
  \mathbf{A}^{(0)}_{4,1} &= \partial_x\bar{\rho}, \qquad \mathbf{A}^{(0)}_{4,2} = \partial_y\bar{\rho}, \qquad \mathbf{A}^{(0)}_{4,4} = \bar{D}_{\mathbf{u}}, \nonumber \\
  \mathbf{A}^{(0)}_{5,1} &= \bar{\rho}\,\partial_x\bar{T}, \qquad \mathbf{A}^{(0)}_{5,2} = \bar{\rho}\,\partial_y\bar{T}, \nonumber \\
  \mathbf{A}^{(0)}_{5,4} &= \bar{\mathbf{u}}\!\cdot\!\nabla\bar{T} + (\gamma-1)\bar{T}\,\bar{D}_{\mathbf{u}}, \nonumber \\
  \mathbf{A}^{(0)}_{5,5} &= (\gamma-1)\bar{\rho}\,\bar{D}_{\mathbf{u}} + \bar{\mathcal{L}}_T \nonumber \\
                         &\quad - \tfrac{\gamma(\gamma-1)\mathrm{Ma}^2}{\mathrm{Re}}\,\bar{\mu}_T\,\bar{\Phi}_0. \label{eq:app_A0}
\end{align}
\endgroup

\section{Elliptic mesh smoother}
\label{sec:app_elliptic}

The Laplace mesh equations~\eqref{eq:laplace_mesh} are discretized by second-order central differences on the uniform auxiliary grid $(\Delta\xi=\Delta\eta=1)$, with the index $i$ running along $\xi$, $j$ along $\eta$, and $X_{i,j}$, $Y_{i,j}$ the physical coordinates of the grid points. The metric coefficients of Eq.~\eqref{eq:grid_metric_coefficients} are evaluated by central differences on the current mesh, so the discrete system is nonlinear. Solving the discrete equation for the self-term gives the Gauss--Seidel update
\begin{equation}
\begin{aligned}
  X_{i,j}^{\mathrm{GS}} = & \frac{1}{2(g_{\xi\xi}+g_{\eta\eta})}\bigl[g_{\xi\xi}(X_{i,j+1}+X_{i,j-1}) \\
  & + g_{\eta\eta}(X_{i+1,j}+X_{i-1,j}) \\
  & - \frac{g_{\xi\eta}}{2}(X_{i+1,j+1}-X_{i-1,j+1}-X_{i+1,j-1}+X_{i-1,j-1}) \bigr],
\end{aligned}
\label{eq:app_gs}
\end{equation}
and analogously for $Y$. Each point is then over-relaxed as $X_{i,j}\gets(1-\omega_{\mathrm{SOR}})X_{i,j}+\omega_{\mathrm{SOR}}\,X_{i,j}^{\mathrm{GS}}$, with $\omega_{\mathrm{SOR}}=1.7$ by default. The algebraic mesh is first interpolated onto a uniform distribution of points in $\eta$, and the wall, inflow, and outflow points and the outer boundary of Sec.~\ref{subsec:elliptic} are held fixed. The sweeps are repeated until $\max|\Delta X|+\max|\Delta Y|$ falls below a prescribed tolerance or a maximum number of sweeps is reached, after which the mesh is redistributed onto the prescribed wall-normal points by shape-preserving cubic interpolation along each $\xi$-column. The near-wall clustering is therefore restored by this redistribution rather than by source terms.

\section{Summary of calculation parameters}
\label{app:params}

Tables~\ref{tab:calc_phys} and~\ref{tab:calc_num} collect the physical and numerical parameters of every calculation presented in this paper.

\begin{table*}[ht]
\centering
\caption{Physical parameters of the calculations. $R = \sqrt{\mathrm{Re}_x}$ range; $\mathrm{Pr}$ and $\gamma$ apply to the compressible cases only; $(M,N)$: number of temporal harmonics and spanwise modes; $A_0$: inflow amplitude of the imposed primary mode, given as the physical peak $2|\hat{u}|$; $F = \omega_1/\mathrm{Re}_0$ the reduced frequency.}
\label{tab:calc_phys}
\scriptsize
\setlength{\tabcolsep}{4pt}
\begin{tabular}{llcccccccccp{2.3cm}}
\toprule
Case & base flow & Ma & $\mathrm{Re}_0$ & $R$ & $\mathrm{Pr}$ & $\gamma$ & $(M,N)$ & $F\,[10^{-6}]$ & $\beta_1$ & $A_0$ & Reference \\
\midrule
\multicolumn{12}{l}{\textit{Verification}}\\
Sec.~\ref{subsec:verif_incomp} incomp.\ & Blasius & --- & 400 & 400--1067 & --- & --- & $(5,0)$ & 86 & 0 & $0.0025\sqrt{2}$ & \citet{Bertolotti1992,Westerbeek2024DeHNSSo} \\
Sec.~\ref{subsec:verif_incomp} comp.\ & Blasius & $10^{-3}$ & 400 & 400--1067 & 0.72 & 1.4 & $(5,0)$ & 86 & 0 & $0.0025\sqrt{2}$ & \citet{Bertolotti1992,Westerbeek2024DeHNSSo} \\
Sec.~\ref{subsec:verif_lumode} lu\_mode & Blasius & ---\,/\,$10^{-3}$ & 400 & 400--1067 & 0.72 & 1.4 & $(5,0)$ & 86 & 0 & $0.0025\sqrt{2}$ & --- \\
Sec.~\ref{subsec:verif_metric} metric & flat plate & 0.5 & 500 & 500--1700 & 0.70 & 1.4 & $(1,0)$ & 40 & 0 & $2\times10^{-6}$ & --- \\
Sec.~\ref{subsec:verif_metric} metric & flat plate & 4.5 & 1000 & 1000--1800 & 0.70 & 1.4 & $(1,0)$ & 153 & 0 & $2\times10^{-6}$ & --- \\
Sec.~\ref{subsec:verif_hns} nonlin.\ HNS & Blasius & --- & 400 & 400--1200 & --- & --- & $(4,0)$ & 86 & 0 & $0.004\sqrt{2}$ & \citet{Scholten2024} \\
\midrule
\multicolumn{12}{l}{\textit{Validation}}\\
Sec.~\ref{subsec:validation_lin_compr} lin.\ compr. & flat plate & 0.5 & 500 & 500--1700 & 0.70 & 1.4 & $(1,0)$ & 40 & 0 & $2\times10^{-6}$ & \citet{Zanus2021} \\
Sec.~\ref{subsec:validation_lin_compr} lin.\ compr. & flat plate & 1.6 & 400 & 400--1200 & 0.70 & 1.4 & $(1,0)$ & 40 & 0 & $2\times10^{-6}$ & \citet{Zanus2021} \\
Sec.~\ref{subsec:validation_lin_compr} lin.\ compr. & flat plate & 4.5 & 1000 & 1000--1800 & 0.70 & 1.4 & $(1,0)$ & 153 & 0 & $2\times10^{-6}$ & \citet{Zanus2021} \\
Sec.~\ref{subsec:validation_nonlin_compr} nonlin.\ compr. & flat plate & 1.6 & 750 & 750--1550 & 0.72 & 1.4 & $(4,4)$ & 20 & 0.062 & $8\times10^{-4}$ & \citet{ChangMalik1994} \\
Sec.~\ref{subsec:validation_nonlin_curved} CFI clean & M3J DNS & --- & 451 & --- & --- & --- & $(0,4)$ & 0 & 0.225 & $2.62\times10^{-3}$ & \citet{Moniripiri2026} \\
Sec.~\ref{subsec:validation_nonlin_curved} CFI hump & M3J DNS & --- & 451 & --- & --- & --- & $(0,4)$ & 0 & 0.225 & $2.62\times10^{-3}$ & \citet{Moniripiri2026} \\
Sec.~\ref{subsec:validation_resonator} resonator & Blasius & 0.06 & 371 & 371--1113 & 0.72 & 1.4 & $(1,0)$ & 64.4 & 0 & $10^{-4}$ & \citet{Michelis2023} \\
\bottomrule
\end{tabular}

\end{table*}

\begin{table*}[ht]
\centering
\caption{Numerical settings of the calculations. Entries in \textit{italic} are configurable code keywords. Solver: the Picard iteration of Sec.~\ref{subsec:operator}, with Anderson acceleration (AA, $m_{\mathrm{A}}=5$) or under-relaxation ($\theta$) where stated, the Newton continuation of Sec.~\ref{subsec:newton}, or a linear calculation. $\eta$: wall-normal scheme (\textit{FD}$_4$ = fourth-order finite difference, \textit{Ch} = Chebyshev); IC: inflow condition (\textit{CLST}, \textit{ILST} = compressible, incompressible local stability solution; \textit{PSE} = embedded parabolized stability solution; \textit{LOAD} = externally supplied profile, here the reference DNS); BC: top boundary ($n$ far-field rows of the Dirichlet condition \textit{H\_DR}) / wall thermal condition (\textit{isoth.} = isothermal); buffer: outflow treatment (\textit{para} = parabolization, \textit{on} = amplitude damping) with onset $\xi_b$ (\% of domain).}
\label{tab:calc_num}
\footnotesize
\setlength{\tabcolsep}{6pt}
\begin{tabular}{lcccccc}
\toprule
Case & $N_\xi{\times}N_\eta$ & $\eta$ & IC & BC (top / wall) & buffer ($\xi_b$) & solver \\
\midrule
\multicolumn{7}{l}{\textit{Verification}}\\
Sec.~\ref{subsec:verif_incomp} incomp.\ & $1800{\times}140$ & \textit{FD}$_4$ & \textit{ILST} & $3{\times}$\textit{H\_DR} / --- & \textit{para} (90) & Picard, AA \\
Sec.~\ref{subsec:verif_incomp} comp.\ & $1800{\times}140$ & \textit{FD}$_4$ & \textit{CLST} & $4{\times}$\textit{H\_DR} / \textit{isoth.} & \textit{para} (90) & Picard, AA \\
Sec.~\ref{subsec:verif_lumode} lu\_mode & $800{\times}40$ & \textit{FD}$_4$ & \textit{ILST}\,/\,\textit{CLST} & $3{\times}$\textit{H\_DR}\,/\,$4{\times}$\textit{H\_DR}, \textit{isoth.} & \textit{para} (90) & Picard, AA \\
Sec.~\ref{subsec:verif_metric} metric (Ma\,0.5) & $401{\times}41$ to $1201{\times}121$ & \textit{FD}$_4$ & \textit{CLST} & $4{\times}$\textit{H\_DR} / \textit{isoth.} & \textit{para} (90) & linear \\
Sec.~\ref{subsec:verif_metric} metric (Ma\,4.5) & $801{\times}101$ to $1201{\times}145$ & \textit{Ch} & \textit{CLST} & $4{\times}$\textit{H\_DR} / \textit{isoth.} & \textit{para} (90) & linear \\
Sec.~\ref{subsec:verif_hns} nonlin.\ HNS & $1050{\times}60$, $1050{\times}40$ & \textit{FD}$_4$, \textit{Ch} & \textit{ILST} & $3{\times}$\textit{H\_DR} / --- & \textit{para} (85) & Newton \\
\midrule
\multicolumn{7}{l}{\textit{Validation}}\\
Sec.~\ref{subsec:validation_lin_compr} lin.\ compr.\ (Ma\,0.5) & $800{\times}81$ & \textit{Ch} & \textit{CLST} & $4{\times}$\textit{H\_DR} / \textit{isoth.} & \textit{para} (90) & linear \\
Sec.~\ref{subsec:validation_lin_compr} lin.\ compr.\ (Ma\,1.6) & $800{\times}81$ & \textit{Ch} & \textit{CLST} & $4{\times}$\textit{H\_DR} / \textit{isoth.} & \textit{para} (90) & linear \\
Sec.~\ref{subsec:validation_lin_compr} lin.\ compr.\ (Ma\,4.5) & $800{\times}101$ & \textit{Ch} & \textit{CLST} & $4{\times}$\textit{H\_DR} / \textit{isoth.} & \textit{para} (90) & linear \\
Sec.~\ref{subsec:validation_nonlin_compr} nonlin.\ compr.\ (Ma\,1.6) & $1500{\times}60$ & \textit{Ch} & \textit{PSE} & $4{\times}$\textit{H\_DR} / \textit{isoth.} & \textit{para} (91.7) & Picard, $\theta{=}0.4$ \\
Sec.~\ref{subsec:validation_nonlin_curved} CFI clean & $2428{\times}60$ & \textit{FD}$_4$ & \textit{LOAD} (DNS) & $3{\times}$\textit{H\_DR} / --- & \textit{on} (85) & Newton \\
Sec.~\ref{subsec:validation_nonlin_curved} CFI hump & $2428{\times}60$ & \textit{FD}$_4$ & \textit{LOAD} (DNS) & $3{\times}$\textit{H\_DR} / --- & \textit{on} (85) & Newton \\
Sec.~\ref{subsec:validation_resonator} resonator & $5000{\times}40$ & \textit{FD}$_4$ & \textit{CLST} & $4{\times}$\textit{H\_DR} / \textit{isoth.} & \textit{on} (90) & linear \\
\bottomrule
\end{tabular}

\end{table*}

\section{Computational hardware}
\label{app:hardware}

Wall-clock measurements use two systems, continuing the nomenclature of \citet{Westerbeek2024DeHNSSo}. System~D is a laptop with an Apple M1~Pro processor (6 performance and 2 efficiency cores, up to $3.2$~GHz) and \SI{32}{GB} of unified memory. System~E is a workstation with an Intel Core i7-14700K processor (8 performance and 12 efficiency cores, up to $5.6$~GHz) and \SI{188}{GB} of DDR5 memory available. The thread count used for each measurement is stated in Sec.~\ref{sec:performance}. Factor sizes are identical between runs of the same calculation, whereas factorization times vary slightly between repeated runs on the same system, owing to operating-system scheduling and memory pressure.
\bibliographystyle{elsarticle-num-names} 
 \bibliography{main}

@article{Westerbeek2024DeHNSSo,
  author  = {Westerbeek, S. and Hulshoff, S. and Schuttelaars, H. and Kotsonis, M.},
  title   = {{DeHNSSo}: The {D}elft {H}armonic {N}avier-{S}tokes {S}olver for nonlinear stability problems with complex geometric features},
  journal = {Comput. Phys. Commun.},
  volume  = {302},
  pages   = {109250},
  year    = {2024},
}

@article{Michelis2023,
  author  = {Michelis, T. and De Koning, C. and Kotsonis, M.},
  title   = {On the interaction of {T}ollmien--{S}chlichting waves with a wall-embedded {H}elmholtz resonator},
  journal = {Phys. Fluids},
  volume  = {35},
  number  = {3},
  pages   = {034104},
  year    = {2023},
}

@article{Moniripiri2026,
  author  = {Moniripiri, M. and Rius-Vidales, A. F. and Kotsonis, M. and Hanifi, A.},
  title   = {Direct numerical simulation of the effects of a smooth surface hump on transition in swept-wing boundary layers},
  journal = {J. Fluid Mech.},
  volume  = {1028},
  pages   = {A47},
  year    = {2026},
}

@article{WalkerNi2011,
  author  = {Walker, H. F. and Ni, P.},
  title   = {Anderson Acceleration for Fixed-Point Iterations},
  journal = {SIAM J. Numer. Anal.},
  volume  = {49},
  number  = {4},
  pages   = {1715--1735},
  year    = {2011},
}

@article{Westerbeek2026Hump,
  author  = {Westerbeek, S. and Casacuberta, J. and Kotsonis, M.},
  title   = {Linear and nonlinear interactions between stationary cross-flow instabilities and a smooth surface hump},
  journal = {J. Fluid Mech.},
  volume  = {1026},
  pages   = {A43},
  year    = {2026},
}

@article{Malik1990,
  author  = {Malik, M. R.},
  title   = {Numerical methods for hypersonic boundary layer stability},
  journal = {J. Comput. Phys.},
  volume  = {86},
  number  = {2},
  pages   = {376--413},
  year    = {1990},
}

@article{Fornberg1998,
  author  = {Fornberg, B.},
  title   = {Calculation of weights in finite difference formulas},
  journal = {SIAM Rev.},
  volume  = {40},
  number  = {3},
  pages   = {685--691},
  year    = {1998},
}

@book{Thompson1999,
  author    = {Thompson, J. F. and Soni, B. K. and Weatherill, N. P.},
  title     = {Handbook of Grid Generation},
  publisher = {CRC Press},
  address   = {Boca Raton},
  year      = {1999},
}

@article{Scholten2024,
  author  = {Scholten, A. and Paredes, P. and Choudhari, M. M. and Li, F. and Carpenter, M. and Bailey, M.},
  title   = {Nonlinear Nonmodal Analysis of Hypersonic Flow over Blunt Cones},
  journal = {AIAA J.},
  volume  = {62},
  number  = {9},
  pages   = {3271--3283},
  year    = {2024},
}

@article{Saric1994,
  author  = {Saric, W. S.},
  title   = {G\"ortler vortices},
  journal = {Annu. Rev. Fluid Mech.},
  volume  = {26},
  pages   = {379--409},
  year    = {1994},
}

@article{Fedorov2011,
  author  = {Fedorov, A. V.},
  title   = {Transition and stability of high-speed boundary layers},
  journal = {Annu. Rev. Fluid Mech.},
  volume  = {43},
  pages   = {79--95},
  year    = {2011},
}

@inproceedings{Franco2018,
  author    = {Franco, J. A. and Hein, S. and Valero, E.},
  title     = {Effect of humps and indentations on boundary-layer transition of compressible flows using the {AHLNS} methodology},
  booktitle = {6th European Conference on Computational Mechanics (ECCM 6) / 7th European Conference on Computational Fluid Dynamics (ECFD 7)},
  address   = {Glasgow, UK},
  year      = {2018},
}

@article{FrancoSumariva2020,
  author  = {Franco Sumariva, J. A. and Hein, S. and Valero, E.},
  title   = {On the influence of two-dimensional hump roughness on laminar--turbulent transition},
  journal = {Phys. Fluids},
  volume  = {32},
  number  = {3},
  pages   = {034102},
  year    = {2020},
}

@article{Zhao2019,
  author  = {Zhao, L. and Dong, M. and Yang, Y.},
  title   = {Harmonic linearized {N}avier--{S}tokes equation on describing the effect of surface roughness on hypersonic boundary-layer transition},
  journal = {Phys. Fluids},
  volume  = {31},
  number  = {3},
  pages   = {034108},
  year    = {2019},
}

@inproceedings{Gonzalez2024,
  author    = {Gonzalez, C. A. and Harris, S. R. and Moin, P.},
  title     = {{glimPSE}: an open-source incompressible nonlinear parabolized stability equation solver},
  booktitle = {Annual Research Briefs},
  publisher = {Center for Turbulence Research, Stanford University},
  pages     = {231--243},
  year      = {2024},
}

@article{Bertolotti1992,
  author  = {Bertolotti, F. P. and Herbert, T. and Spalart, P. R.},
  title   = {Linear and nonlinear stability of the {B}lasius boundary layer},
  journal = {J. Fluid Mech.},
  volume  = {242},
  pages   = {441--474},
  year    = {1992},
}

@phdthesis{Paredes2014,
  author  = {Paredes, P.},
  title   = {Advances in Global Instability Computations: From Incompressible to Hypersonic Flow},
  school  = {Technical University of Madrid},
  address = {Madrid, Spain},
  year    = {2014},
}

@article{Amestoy2001,
  author  = {Amestoy, P. R. and Duff, I. S. and L'Excellent, J.-Y. and Koster, J.},
  title   = {A fully asynchronous multifrontal solver using distributed dynamic scheduling},
  journal = {SIAM J. Matrix Anal. Appl.},
  volume  = {23},
  number  = {1},
  pages   = {15--41},
  year    = {2001},
}

@article{Amestoy2006,
  author  = {Amestoy, P. R. and Guermouche, A. and L'Excellent, J.-Y. and Pralet, S.},
  title   = {Hybrid scheduling for the parallel solution of linear systems},
  journal = {Parallel Comput.},
  volume  = {32},
  number  = {2},
  pages   = {136--156},
  year    = {2006},
}

@inproceedings{Anderson1990,
  author    = {Anderson, E. and Bai, Z. and Dongarra, J. and Greenbaum, A. and McKenney, A. and Du Croz, J. and Hammarling, S. and Demmel, J. and Bischof, C. and Sorensen, D.},
  title     = {{LAPACK}: A portable linear algebra library for high-performance computers},
  booktitle = {Proceedings of the 1990 ACM/IEEE Conference on Supercomputing},
  publisher = {IEEE Computer Society Press},
  year      = {1990},
  pages     = {2--11}
}

@incollection{Keller1977,
  author    = {Keller, H. B.},
  title     = {Numerical Solution of Bifurcation and Nonlinear Eigenvalue Problems},
  booktitle = {Applications of Bifurcation Theory},
  editor    = {Rabinowitz, P. H.},
  publisher = {Academic Press},
  address   = {New York},
  year      = {1977},
  pages     = {359--384}
}

@article{Joslin1998,
  author  = {Joslin, R. D.},
  title   = {Aircraft laminar flow control},
  journal = {Annu. Rev. Fluid Mech.},
  volume  = {30},
  number  = {1},
  pages   = {1--29},
  year    = {1998},
}

@article{Schrauf2005,
  author  = {Schrauf, G.},
  title   = {Status and perspectives of laminar flow},
  journal = {Aeronaut. J.},
  volume  = {109},
  number  = {1102},
  pages   = {639--644},
  year    = {2005},
}

@article{Saric2003,
  author  = {Saric, W. S. and Reed, H. L. and White, E. B.},
  title   = {Stability and transition of three-dimensional boundary layers},
  journal = {Annu. Rev. Fluid Mech.},
  volume  = {35},
  number  = {1},
  pages   = {413--440},
  year    = {2003},
}

@article{Holmes1985,
  author  = {Holmes, B. J. and Obara, C. J. and Martin, G. L. and Domack, C. S.},
  title   = {Manufacturing tolerances for natural laminar flow airframe surfaces},
  journal = {SAE Trans.},
  volume  = {94},
  pages   = {522--531},
  year    = {1985},
}

@article{Worner2003,
  author  = {W\"orner, A. and Rist, U. and Wagner, S.},
  title   = {Humps/steps influence on stability characteristics of two-dimensional laminar boundary layer},
  journal = {AIAA J.},
  volume  = {41},
  number  = {2},
  pages   = {192--197},
  year    = {2003},
}

@inproceedings{Crouch2006,
  author    = {Crouch, J. and Kosorygin, V. and Ng, L.},
  title     = {Modeling the effects of steps on boundary-layer transition},
  booktitle = {IUTAM Symposium on Laminar-Turbulent Transition},
  editor    = {Govindarajan, R.},
  publisher = {Springer},
  address   = {Dordrecht},
  pages     = {37--44},
  year      = {2006},
}

@article{WangGaster2005,
  author  = {Wang, Y. and Gaster, M.},
  title   = {Effect of surface steps on boundary layer transition},
  journal = {Exp. Fluids},
  volume  = {39},
  pages   = {679--686},
  year    = {2005},
}

@article{RiusVidales2021,
  author  = {Rius-Vidales, A. F. and Kotsonis, M.},
  title   = {Impact of a forward-facing step on the development of crossflow instability},
  journal = {J. Fluid Mech.},
  volume  = {924},
  pages   = {A34},
  year    = {2021},
}

@article{RiusVidales2025Hump,
  author  = {Rius-Vidales, A. F. and Morais, L. and Westerbeek, S. and Casacuberta, J. and Soyler, M. and Kotsonis, M.},
  title   = {Delay of swept-wing transition using a surface hump},
  journal = {J. Fluid Mech.},
  volume  = {1014},
  pages   = {A35},
  year    = {2025},
}

@article{Herbert1997,
  author  = {Herbert, T.},
  title   = {Parabolized stability equations},
  journal = {Annu. Rev. Fluid Mech.},
  volume  = {29},
  number  = {1},
  pages   = {245--283},
  year    = {1997},
}

@inproceedings{Cooke2019,
  author    = {Cooke, E. E. and Mughal, M. S. and Sherwin, S. and Ashworth, R. and Rolston, S.},
  title     = {Destabilisation of stationary and travelling crossflow disturbances due to steps over a swept wing},
  booktitle = {AIAA Aviation 2019 Forum},
  note      = {AIAA Paper 2019-3533},
  year      = {2019},
}

@article{Casacuberta2022,
  author  = {Casacuberta, J. and Hickel, S. and Westerbeek, S. and Kotsonis, M.},
  title   = {Direct numerical simulation of interaction between a stationary crossflow instability and forward-facing steps},
  journal = {J. Fluid Mech.},
  volume  = {943},
  pages   = {A46},
  year    = {2022},
}

@inproceedings{Streett1998,
  author    = {Streett, C. L.},
  title     = {Direct harmonic linear {N}avier--{S}tokes methods for efficient simulation of wave packets},
  booktitle = {36th AIAA Aerospace Sciences Meeting and Exhibit},
  note      = {AIAA Paper 98-0784},
  year      = {1998},
}

@article{Towne2019,
  author  = {Towne, A. and Rigas, G. and Colonius, T.},
  title   = {A critical assessment of the parabolized stability equations},
  journal = {Theor. Comput. Fluid Dyn.},
  volume  = {33},
  pages   = {359--382},
  year    = {2019},
}

@article{Rigas2021,
  author  = {Rigas, G. and Sipp, D. and Colonius, T.},
  title   = {Nonlinear input/output analysis: application to boundary layer transition},
  journal = {J. Fluid Mech.},
  volume  = {911},
  pages   = {A15},
  year    = {2021},
}

@article{BertolottiHerbert1991,
  author  = {Bertolotti, F. P. and Herbert, T.},
  title   = {Analysis of the linear stability of compressible boundary layers using {PSE}},
  journal = {Theor. Comput. Fluid Dyn.},
  volume  = {3},
  number  = {2},
  pages   = {117--124},
  year    = {1991},
}

@inproceedings{ChangMalik1991,
  author    = {Chang, C.-L. and Malik, M. R. and Erlebacher, G. and Hussaini, M. Y.},
  title     = {Compressible stability of growing boundary layers using parabolized stability equations},
  booktitle = {22nd AIAA Fluid Dynamics, Plasma Dynamics and Lasers Conference},
  note      = {AIAA Paper 91-1636},
  year      = {1991},
}

@article{ChangMalik1994,
  author  = {Chang, C.-L. and Malik, M. R.},
  title   = {Oblique-mode breakdown and secondary instability in supersonic boundary layers},
  journal = {J. Fluid Mech.},
  volume  = {273},
  pages   = {323--360},
  year    = {1994},
}

@article{Paredes2019,
  author  = {Paredes, P. and Choudhari, M. M. and Li, F. and Jewell, J. S. and Kimmel, R. L.},
  title   = {Nonmodal growth of traveling waves on blunt cones at hypersonic speeds},
  journal = {AIAA J.},
  volume  = {57},
  number  = {11},
  pages   = {4738--4749},
  year    = {2019},
}

@article{Lacombe2022,
  author  = {Lacombe, F. and Hickey, J.-P.},
  title   = {{Krypton}: Nonlinear Parabolized Stability Equation solver for transonic flow in curvilinear coordinates},
  journal = {SoftwareX},
  volume  = {20},
  pages   = {101206},
  year    = {2022},
}

@article{Barahona2026,
  author  = {Barahona, M. and Chen, Y. and Rius-Vidales, A. F. and Kotsonis, M.},
  title   = {Impact of wall-heating on the development of crossflow instability},
  journal = {Exp. Therm. Fluid Sci.},
  volume  = {175},
  pages   = {111743},
  year    = {2026},
}

@phdthesis{Zanus2021,
  author = {Zanus, L.},
  title  = {Numerical Study of Chemically Reacting Hypersonic Boundary Layers by Means of Non-linear Parabolized Stability Equations},
  school = {Universit\'e Libre de Bruxelles and von Karman Institute for Fluid Dynamics},
  year   = {2021},
}

@article{Paredes2024,
  author  = {Paredes, P. and Choudhari, M. M. and Carpenter, M. H. and Li, F.},
  title   = {The harmonic linearized {N}avier--{S}tokes equations for transition prediction in three-dimensional flows},
  journal = {Comput. Fluids},
  volume  = {273},
  pages   = {106206},
  year    = {2024},
}

@article{Cook2026,
  author  = {Cook, D. A. and Bugel, J. G. and Nichols, J. W.},
  title   = {A graph-based rank-reduced interface preconditioner for the harmonic linearized {N}avier--{S}tokes equations},
  journal = {Theor. Comput. Fluid Dyn.},
  volume  = {40},
  number  = {4},
  pages   = {16},
  year    = {2026},
}

@article{ZhaoDong2025,
  author  = {Zhao, L. and Dong, M.},
  title   = {Excitation of non-modal perturbations in hypersonic boundary layers by free stream forcing: shock-fitting harmonic linearised {N}avier--{S}tokes approach},
  journal = {J. Fluid Mech.},
  volume  = {1013},
  pages   = {A44},
  year    = {2025},
}

@article{Sleeman2025,
  author  = {Sleeman, M. K. and Colonius, T. and Lakebrink, M. T.},
  title   = {Boundary-layer stability analysis using the nonlinear one-way {N}avier--{S}tokes approach},
  journal = {AIAA J.},
  volume  = {63},
  number  = {8},
  pages   = {3145--3159},
  year    = {2025},
}

@article{TowneColonius2015,
  author  = {Towne, A. and Colonius, T.},
  title   = {One-way spatial integration of hyperbolic equations},
  journal = {J. Comput. Phys.},
  volume  = {300},
  pages   = {844--861},
  year    = {2015},
}

@article{ZhuTowne2023,
  author  = {Zhu, M. and Towne, A.},
  title   = {Recursive one-way {N}avier--{S}tokes equations with {PSE}-like cost},
  journal = {J. Comput. Phys.},
  volume  = {473},
  pages   = {111744},
  year    = {2023},
}

@article{BadcockMughal2026,
  author  = {Badcock, E. J. and Mughal, S.},
  title   = {A carrier-wave factored one-way {N}avier--{S}tokes method for boundary-layer instability modelling},
  journal = {Theor. Comput. Fluid Dyn.},
  volume  = {40},
  number  = {4},
  pages   = {19},
  year    = {2026},
}

@article{LiMalik1996,
  author  = {Li, F. and Malik, M. R.},
  title   = {On the nature of {PSE} approximation},
  journal = {Theor. Comput. Fluid Dyn.},
  volume  = {8},
  number  = {4},
  pages   = {253--273},
  year    = {1996},
}

@techreport{ChangMalik1993,
  author      = {Chang, C.-L. and Malik, M. R. and Erlebacher, G. and Hussaini, M. Y.},
  title       = {Linear and nonlinear {PSE} for compressible boundary layers},
  institution = {NASA Langley Research Center},
  number      = {NASA CR-191537, ICASE Report 93-70},
  year        = {1993},
}

@article{Dijkstra2014,
  author  = {Dijkstra, H. A. and Wubs, F. W. and Cliffe, A. K. and Doedel, E. and
             Dragomirescu, I. F. and Eckhardt, B. and Gelfgat, A. Yu. and Hazel, A. L. and
             Lucarini, V. and Salinger, A. G. and Phipps, E. T. and Sanchez-Umbria, J. and
             Schuttelaars, H. and Tuckerman, L. S. and Thiele, U.},
  title   = {Numerical bifurcation methods and their application to fluid dynamics: analysis beyond simulation},
  journal = {Commun. Comput. Phys.},
  volume  = {15},
  number  = {1},
  pages   = {1--45},
  year    = {2014},
}

@book{GolubVanLoan2013,
  author    = {Golub, G. H. and Van Loan, C. F.},
  title     = {Matrix Computations},
  edition   = {4th},
  publisher = {Johns Hopkins University Press},
  address   = {Baltimore},
  year      = {2013},
}

@article{Vanderstraeten1998,
  author  = {Vanderstraeten, D. and Keunings, R.},
  title   = {A parallel solver based on the dual {S}chur decomposition of general finite element matrices},
  journal = {Int. J. Numer. Methods Fluids},
  volume  = {28},
  number  = {1},
  pages   = {23--46},
  year    = {1998},
}

@inproceedings{CuthillMcKee1969,
  author    = {Cuthill, E. and McKee, J.},
  title     = {Reducing the bandwidth of sparse symmetric matrices},
  booktitle = {Proceedings of the 24th National Conference of the ACM},
  publisher = {Association for Computing Machinery},
  address   = {New York},
  year      = {1969},
  pages     = {157--172},
}

@article{Davis2004Colamd,
  author  = {Davis, T. A. and Gilbert, J. R. and Larimore, S. I. and Ng, E. G.},
  title   = {A column approximate minimum degree ordering algorithm},
  journal = {ACM Trans. Math. Softw.},
  volume  = {30},
  number  = {3},
  pages   = {353--376},
  year    = {2004},
}

@article{Busquet2021,
  author  = {Busquet, D. and Marquet, O. and Richez, F. and Juniper, M. and Sipp, D.},
  title   = {Bifurcation scenario for a two-dimensional static airfoil exhibiting trailing edge stall},
  journal = {J. Fluid Mech.},
  volume  = {928},
  pages   = {A3},
  year    = {2021},
}

@article{Beaume2017,
  author  = {Beaume, C.},
  title   = {Adaptive {S}tokes preconditioning for steady incompressible flows},
  journal = {Commun. Comput. Phys.},
  volume  = {22},
  number  = {2},
  pages   = {494--516},
  year    = {2017},
}

@inproceedings{Benkert2007,
  author    = {Benkert, K. and Fischer, R.},
  title     = {An efficient implementation of the {T}homas-algorithm for block penta-diagonal systems on vector computers},
  booktitle = {Computational Science -- {ICCS} 2007},
  series    = {Lecture Notes in Computer Science},
  volume    = {4487},
  pages     = {144--151},
  publisher = {Springer},
  address   = {Berlin, Heidelberg},
  year      = {2007},
}

@article{Colella2011,
  author  = {Colella, P. and Dorr, M. R. and Hittinger, J. A. F. and Martin, D. F.},
  title   = {High-order, finite-volume methods in mapped coordinates},
  journal = {J. Comput. Phys.},
  volume  = {230},
  number  = {8},
  pages   = {2952--2976},
  year    = {2011},
}

@article{GrimmStrele2014,
  author  = {Grimm-Strele, H. and Kupka, F. and Muthsam, H. J.},
  title   = {Curvilinear grids for {WENO} methods in astrophysical simulations},
  journal = {Comput. Phys. Commun.},
  volume  = {185},
  number  = {3},
  pages   = {764--776},
  year    = {2014},
}

@article{Boldini2025CUBENS,
  author  = {Boldini, P. C. and Hirai, R. and Costa, P. and Peeters, J. W. R. and Pecnik, R.},
  title   = {{CUBENS}: A {GPU}-accelerated high-order solver for wall-bounded flows with non-ideal fluids},
  journal = {Comput. Phys. Commun.},
  volume  = {309},
  pages   = {109507},
  year    = {2025},
}

@article{DiRenzo2020HTR,
  author  = {Di Renzo, M. and Fu, L. and Urzay, J.},
  title   = {{HTR} solver: An open-source exascale-oriented task-based multi-{GPU} high-order code for hypersonic aerothermodynamics},
  journal = {Comput. Phys. Commun.},
  volume  = {255},
  pages   = {107262},
  year    = {2020},
}

@article{LuchiniBottaro2014,
  author  = {Luchini, P. and Bottaro, A.},
  title   = {Adjoint equations in stability analysis},
  journal = {Annu. Rev. Fluid Mech.},
  volume  = {46},
  pages   = {493--517},
  year    = {2014},
}

@article{Poulain2023,
  author  = {Poulain, A. and Content, C. and Sipp, D. and Rigas, G. and Garnier, E.},
  title   = {{BROADCAST}: A high-order compressible {CFD} toolbox for stability and sensitivity using Algorithmic Differentiation},
  journal = {Comput. Phys. Commun.},
  volume  = {283},
  pages   = {108557},
  year    = {2023}
}

@article{MartiniSchmidt2024,
  author  = {Martini, E. and Schmidt, O. T.},
  title   = {{Linstab2D}: stability and resolvent analysis of compressible viscous flows in {MATLAB}},
  journal = {Theor. Comput. Fluid Dyn.},
  volume  = {38},
  pages   = {665--685},
  year    = {2024}
}

@article{Kloker1993,
  author  = {Kloker, M. and Konzelmann, U. and Fasel, H.},
  title   = {Outflow boundary conditions for spatial {Navier--Stokes} simulations of transition boundary layers},
  journal = {AIAA J.}, 
  volume = {31}, 
  number = {4}, 
  pages = {620--628}, 
  year = {1993}}

@inproceedings{LiuLiu1993,
  author    = {Liu, C. and Liu, Z.},
  title     = {Multigrid methods and high order finite difference for flow in transition: effects of isolated and distributed roughness elements},
  booktitle = {11th AIAA Computational Fluid Dynamics Conference},
  note      = {AIAA Paper 1993-3354}, 
  year = {1993}}

@article{Avallone2026,
  author  = {Avallone, F. and Bosia, F. and Chen, Y. and others},
  title   = {Metamaterials and fluid flows},
  journal = {Nat. Commun.},
  volume  = {17},
  pages   = {4144},
  year    = {2026}
}

\end{document}